\documentclass[11pt]{article}%
\usepackage{amsfonts}
\usepackage{amssymb}
\usepackage{amsmath}
\usepackage{graphicx}
\usepackage{amscd}%
\providecommand{\U}[1]{\protect\rule{.1in}{.1in}}
\newtheorem{theorem}{Theorem}

\newtheorem{proposition}[theorem]{Proposition}

\begin{document}

\title{The Objective Indefiniteness Interpretation\\of Quantum Mechanics}
\author{David Ellerman\\University of California at Riverside\\Draft Version 3 (not for quotation)}
\maketitle

\begin{abstract}
The common-sense view of reality is expressed logically in Boolean subset
logic (each element is either definitely in or not in a subset, i.e., either
definitely has or does not have a property). But quantum mechanics does not
agree with this "properties all the way down" picture of micro-reality. Are
there other coherent alternative views of reality? A logic of partitions, dual
to the Boolean logic of subsets (partitions are dual to subsets), was recently
developed along with a logical version of information theory. In view of the
subset-partition duality, partition logic is \textit{the} alternative to
Boolean subset logic and thus it abstractly describes the alternative dual
view of micro-reality. Perhaps QM is compatible with this dual view? Indeed,
when the mathematics of partitions using sets is "lifted" from sets to vector
spaces, then it yields the mathematics and relations of quantum mechanics.
Thus the vision of micro-reality abstractly characterized by partition logic
matches that described by quantum mechanics. The key concept explicated by
partition logic is the old idea of "objective indefiniteness" (emphasized by
Shimony). Thus partition logic, logical information theory, and the lifting
program provide the back story so that the old idea then yields the
\textit{objective indefiniteness interpretation} of quantum mechanics.

\end{abstract}
\tableofcontents

\section{Introduction: the back story for objective indefiniteness}

Classical physics is compatible with the common-sense view of reality that is
expressed at the logical level in Boolean subset logic. Each element in the
Boolean universe set is either definitely in or not in a subset, i.e., each
element either definitely has or does not have a property. Each element is
characterized by a full set of properties, a view that might be referred to as
"definite properties all the way down."

It is now rather widely accepted that this common-sense view of reality is not
compatible with quantum mechanics (QM). If we think in terms of only two
positions, \textit{here} and \textit{there}, then in classical physics a
particle is either definitely \textit{here} or \textit{there}, while in QM,
the particle can be "neither definitely here nor there."\cite[p.
144]{weinberg:dreams}\footnote{This is usually misrepresented in the popular
literature as the particle being "both \textit{here} and \textit{there} at the
same time." Weinberg also mentions a particle "spinning neither definitely
clockwise nor counterclockwise" and then notes that for elementary particles,
"it is possible to have a particle in a state in which it is neither
definitely an electron nor definitely a neutrino until we measure some
property that would distinguish the two, like the electric charge."\cite[pp.
144-145 (thanks to Noson Yanofsky for this reference)]{weinberg:dreams}} This
is not an epistemic or subjective indefiniteness of location; it is an
ontological or objective indefiniteness. The notion of \textit{objective
indefiniteness} in QM has been most emphasized by Abner Shimony
(\cite{shim:reality},\cite{shim:concept}).

\begin{quotation}
From these two basic ideas alone -- indefiniteness and the superposition
principle -- it should be clear already that quantum mechanics conflicts
sharply with common sense. If the quantum state of a system is a complete
description of the system, then a quantity that has an indefinite value in
that quantum state is objectively indefinite; its value is not merely unknown
by the scientist who seeks to describe the system. ...Classical physics did
not conflict with common sense in these fundamental ways.\cite[p.
47]{shim:reality}
\end{quotation}

\noindent Other quantum philosophers have used similar concepts. For instance,
in his discussion of Heisenberg's uncertainty\footnote{Heisenberg's German
word was "Unbestimmtheit" which could well be translated as "indefiniteness"
or "indeterminateness" rather than "uncertainty."} principle, Paul Feyerabend
asserted that "inherent indefiniteness is a universal and objective property
of matter."\cite[p. 202]{feyer:micro} Thus one path to arrive at the notion of
"inherent indefiniteness" is to understand that Heisenberg's indefiniteness
principle is \textit{not} about the clumsiness of instruments in
simultaneously measuring incompatible observables that always have definite values.

Two questions arise:

\begin{itemize}
\item What is the logic of objective indefiniteness that plays the role
analogous to Boolean subset logic (the logic of definite properties "all the
way down")?

\item And given such a logic, how would one fill in the gap between the
austere level of logic and the rich mathematical framework of quantum mechanics?
\end{itemize}

These questions can now be answered. The logic of objective indefiniteness
that plays the role analogous to subset logic is the recently developed dual
logic of partitions.\cite{ell:partitions} Partition logic is not just "another
alternative" logic; it is the unique mathematical dual (explained below) to
subset logic. Hence if subset logic is not the logic for quantum mechanics,
then the natural next step to check is if the dual logic of partitions fits
QM. The point of this paper is to show that it fits perfectly.

Moreover, Boole developed a logical finite probability theory out of his logic
of subsets \cite{boole:lot}, and the analogous theory developed out of the
logic of partitions is the logical version of information
theory.\cite{ell:distinctions} This logical information theory generalized to
the density matrices of quantum mechanics describes and interprets the changes
made in a quantum measurement.

The concepts and operations of partition logic and logical information theory
are developed in the rather austere set-theoretic context; they needed to be
"lifted" to the richer environment of vector spaces. This lifting program from
sets to vector spaces is part of the mathematical folklore (e.g., used
intuitively by von Neumann). When applied to the concepts and operations of
partition mathematics, the lifting program indeed yields the mathematics of
quantum mechanics. This corroborates that the vision of micro-reality provided
by \textit{the} dual form of logic (i.e., partition logic rather than subset
logic) is, in fact, the micro-reality described by QM. Thus the development of
the logic of partitions, logical information theory, and the lifting program
provides the back story to the notion of objective indefiniteness. The result
is the objective indefiniteness interpretation of quantum mechanics.

\section{The logic of partitions}

\subsection{From "propositional" logic to subset logic}

Our treatment of partition logic here will only develop the basic concepts
necessary to be lifted to the vector spaces of quantum mechanics.\footnote{See
\cite{ell:partitions} for a detailed development from the basic concepts up
through the correctness and completeness theorems for a tableau system of
partition logic.} But first it might be helpful to explain why it has taken so
long for partition logic to be developed as the dual of subset logic, and to
explain the duality.

George Boole \cite{boole:lot} originally developed his logic as the logic of
subsets. As noted by Alonzo Church:

\begin{quotation}
\noindent The algebra of logic has its beginning in 1847, in the publications
of Boole and De Morgan. This concerned itself at first with an algebra or
calculus of classes,\ldots\ a true propositional calculus perhaps first
appeared\ldots in 1877.\cite[pp. 155-156]{church:ml}
\end{quotation}

In the logic of subsets, a \textit{tautology} is defined as a formula such
that no matter what subsets of the given universe $U$ are substituted for the
variables, when the set-theoretic operations are applied, then the whole
formula evaluates to $U$. Boole noted that to determine these valid formulas,
it suffices to take the special case of $U=1$ which has only two subsets
$0=\emptyset$ and $1$. Thus what was later called the "truth table"
characterization of a tautology was a theorem, not a
definition.\footnote{Alfred Renyi \cite{renyi:prin} gave a generalization of
the theorem to probability theory.}

But over the years, the whole became identified with the special case. The
Boolean logic of subsets was reconceptualized as "propositional logic." The
truth-table characterization of a tautology became the \textit{definition} of
a tautology in propositional logic rather than a theorem in subset logic. This
facilitated the further analysis of the propositional atoms into statements
with quantifiers and the development of model theory. But the restricted
notion of "propositional" logic also had a downside; it hid the idea of a dual
logic since propositions don't have duals.

Subsets and partitions (or equivalence relations or quotient sets) are dual in
the category-theoretic sense of the duality between monomorphisms and
epimorphisms. This duality is familiar in abstract algebra in the interplay of
subobjects (e.g., subgroups, subrings, etc.) and quotient objects. William
Lawvere calls the general category-theoretic notion of a subobject a
\textit{part}, and then he notes: "The dual notion (obtained by reversing the
arrows) of `part' is the notion of partition."\cite[p. 85]{law:sets} The image
of monomorphic or injective map between sets is a subset of the codomain, and
dually the inverse-image of an epimorphic or surjective map between sets is a
partition of the domain. The development of the dual logic of partitions was
long delayed by the conceptualization of subset logic as "propositional" logic.

\subsection{Basic concepts of partition logic}

In the Boolean logic of subsets, the basic algebraic structure is the Boolean
lattice $\wp\left(  U\right)  $ of subsets of a universe set $U$ enriched by
the implication $A\Rightarrow B=A^{c}\cup B$ to form the Boolean algebra of
subsets of $U$. In a similar manner, we form the lattice of partitions on $U$
enriched by the partition operation of implication and other partition operations.

Given a universe set $U$, a \textit{partition} $\pi$ of $U$ is a set of
non-empty subsets or blocks $\left\{  B\right\}  $ of $U$ that are pairwise
disjoint and whose union is $U$. In category-theoretic terms, a partition is a
direct sum decomposition of a set, and that concept will lift, in the
sets-to-vector-spaces lifting program, to the concept of a direct sum
decomposition of a vector space.

Given two partitions $\pi=\left\{  B\right\}  $ and $\sigma=\left\{
C\right\}  $ of the same universe $U$, the partition $\sigma$ is
\textit{refined} by $\pi$, written by $\sigma\preceq\pi$, if for every block
$B\in\pi$, there is a block $C\in\sigma$ such that $B\subseteq C$. Given the
two partitions of the same universe, their \textit{join} $\pi\vee\sigma$ is
the partition whose blocks are the non-empty intersections $B\cap C$. The join
of two partitions of the same set will lift to the join of two direct sum
decompositions that are in a certain sense "compatible."

The \textit{top} of the lattice is the \textit{discrete partition}
$\mathbf{1}=\left\{  \left\{  u\right\}  :u\in U\right\}  $ whose blocks are
all the singletons, and the \textit{bottom} is the \textit{indiscrete
partition} (nicknamed the "blob") $\mathbf{0}=\left\{  \left\{  U\right\}
\right\}  $ whose only block is all of $U$. Together with the meet
operation,\footnote{To define the meet $\pi\wedge\sigma$, consider an
undirected graph on $U$ where there is a link between any two elements
$u,u^{\prime}\in U$ if they are in the same block of $\pi$ or the same block
of $\sigma$. Then the blocks of $\pi\wedge\sigma$ are the connected components
of that graph.} this defines the \textit{lattice of partitions} $\prod(U)$ on
$U$.\footnote{For anything worthy to be called "partition logic," an operation
of implication would be needed if not partition versions of all the sixteen
binary subset operations. Given $\pi=\left\{  B\right\}  $ and $\sigma
=\left\{  C\right\}  $, the \textit{implication} $\sigma\Rightarrow\pi$ is the
partition whose blocks are like the blocks of $\pi$ except that whenever a
block $B$ is contained in some block $C\in\sigma$, then $B$ is discretized,
i.e., replaced by the singletons of its elements. If we think of a whole block
$B$ as a mini-$\mathbf{0}$ and a discretized $B$ as a mini-$\mathbf{1}$, then
the implication $\sigma\Rightarrow\pi$ is just the indicator function for the
inclusion of the $\pi$-blocks in the $\sigma$-blocks. In the Boolean algebra
$\wp\left(  U\right)  $, the implication is related to the partial order by
the relation, $A\Rightarrow B=U$ iff $A\subseteq B$, and we immediately see
that the corresponding relation $\sigma\Rightarrow\pi=\mathbf{1}$ holds iff
$\sigma\preceq\pi$.}

We will use the representation of the partition lattice $\prod\left(
U\right)  $ as a lattice of subsets of $U\times U$. Given a partition
$\pi=\left\{  B\right\}  $ on $U$, the \textit{distinctions} or \textit{dits}
of $\pi$ are the ordered pairs $\left(  u,u^{\prime}\right)  $ where $u$ and
$u^{\prime}$ are in distinct blocks of $\pi$, and $\operatorname*{dit}\left(
\pi\right)  $ is the \textit{set of distinctions} or \textit{dit set }of $\pi
$. Similarly, an \textit{indistinction} or \textit{indit} of $\pi$ is an
ordered pair $\left(  u,u^{\prime}\right)  $ where $u$ and $u^{\prime}$ are in
the same block of $\pi$, and $\operatorname*{indit}\left(  \pi\right)  $ is
the \textit{indit set} of $\pi$. Of course, $\operatorname*{indit}\left(
\pi\right)  $ is just the equivalence relation determined by $\pi$, and it is
the complement of $\operatorname*{dit}\left(  \pi\right)  $ in $U\times U$.

The complement of an equivalence relation is properly called a
\textit{partition relation }[also an "apartness relation"]. An equivalence
relation is reflexive, symmetric, and transitive, so a partition relation is
irreflexive [i.e., contains no self-pairs $\left(  u,u\right)  $ from the
diagonal $\Delta_{U}$], symmetric, and anti-transitive, where a binary
relation $R$ is \textit{anti-transitive} if for any $\left(  u,u^{\prime
\prime}\right)  \in R$, and any other element $u^{\prime}\in U$, then either
$\left(  u,u^{\prime}\right)  \in R$ or $\left(  u^{\prime},u^{\prime\prime
}\right)  \in R$. Otherwise both pairs would be in the complement
$R^{c}=U\times U-R$ which is transitive so $\left(  u,u^{\prime\prime}\right)
\in R^{c}$ contrary to the assumption.

Every subset $S\subseteq U\times U$ has a reflexive-symmetric-transitive
\textit{closure} $\overline{S}$ which is the smallest equivalence relation
containing $S$. Hence we can define an \textit{interior} operation as the
complement of the closure of the complement, i.e., $\operatorname*{int}\left(
S\right)  =\left(  \overline{S^{c}}\right)  ^{c}$, which is the largest
partition relation included in $S$. While some motivation might be supplied by
thinking of the partition relations as "open" subsets and the equivalence
relations as "closed" subsets, they do not form a topology. The closure
operation is not a topological closure operation since the union of two closed
subsets is not necessarily closed, and the intersection of two open subsets is
not necessarily open.

Every partition $\pi$ is represented by its dit set $\operatorname*{dit}%
\left(  \pi\right)  $. The refinement relation between partitions,
$\sigma\preceq\pi$ is represented by the inclusion relation between dit sets,
i.e., $\sigma\preceq\pi$ iff $\operatorname*{dit}\left(  \sigma\right)
\subseteq\operatorname*{dit}\left(  \pi\right)  $.\footnote{Unfortunately in
much of the literature of combinatorial theory, the refinement partial
ordering is written the other way around (so Gian-Carlo Rota sometimes called
it "unrefinement"), and thus the "join" and "meet" are reversed, and the
lattice of partitions is then "upside-down." That upside-down representation
of the "lattice of partitions" uses the indit sets so it is actually the
lattice of equivalence relations rather than the lattice of partition
relations.} The join $\pi\vee\sigma$ is represented in $U\times U$ by the
union of the dit sets, i.e., $\operatorname*{dit}\left(  \pi\vee\sigma\right)
=\operatorname*{dit}\left(  \pi\right)  \cup\operatorname*{dit}\left(
\sigma\right)  $.\footnote{But the intersection of two dit sets is not
necessarily a dit set so to find the dit set of the meet $\pi\wedge\sigma$, we
have to take the interior of the intersection of their dit sets, i.e.,
$\operatorname*{dit}\left(  \pi\wedge\sigma\right)  =\operatorname*{int}%
\left(  \operatorname*{dit}\left(  \pi\right)  \cap\operatorname*{dit}\left(
\sigma\right)  \right)  $. These equations for the dit sets of the join and
meet are theorems, not definitions, since the join and meet were already
defined above. The general algorithm to represent a partition operation is to
apply the corresponding set operation to the dit sets and then apply the
interior to the result (if it is not already a partition relation). Thus, for
instance,
\par
\begin{center}
$\operatorname*{dit}\left(  \sigma\Rightarrow\pi\right)  =\operatorname*{int}%
\left(  \operatorname*{dit}\left(  \sigma\right)  ^{c}\cup\operatorname*{dit}%
\left(  \pi\right)  \right)  $.
\end{center}
\par
\noindent It is a striking fact (see \cite{ell:partitions} for a proof) that
$\operatorname*{int}\left(  \operatorname*{dit}\left(  \sigma\right)  ^{c}%
\cup\operatorname*{dit}\left(  \pi\right)  \right)  $ is the dit set of
$\sigma\Rightarrow\pi$ previously defined as the indicator function for the
inclusion of $\pi$-blocks in $\sigma$-blocks.} In this manner, the lattice of
partitions $\prod\left(  U\right)  $ enriched by implication and other
partition operations can be represented by the lattice of partition relations
$\mathcal{O}\left(  U\times U\right)  $ on $U\times U$.

\begin{center}%
\begin{tabular}
[c]{|c|c|c|}\hline
Representation & $\prod\left(  U\right)  $ & $\mathcal{O}\left(  U\times
U\right)  $\\\hline\hline
Partition & $\pi$ & $\operatorname*{dit}\left(  \pi\right)  $\\\hline
Refinement order & $\sigma\preceq\pi$ & $\operatorname*{dit}\left(
\sigma\right)  \subseteq\operatorname*{dit}\left(  \pi\right)  $\\\hline
Top & $\mathbf{1}=\left\{  \{u\right\}  :u\in U\}$ & $\operatorname*{dit}%
\left(  \mathbf{1}\right)  =U\times U-\Delta_{U}$ all dits\\\hline
Bottom & $\mathbf{0}=\left\{  \left\{  U\right\}  \right\}  $ &
$\operatorname*{dit}\left(  \mathbf{0}\right)  =\emptyset$ no dits\\\hline
Join & $\pi\vee\sigma$ & $\operatorname*{dit}\left(  \pi\vee\sigma\right)
=\operatorname*{dit}\left(  \pi\right)  \cup\operatorname*{dit}\left(
\sigma\right)  $\\\hline
Meet & $\pi\wedge\sigma$ & $\operatorname*{dit}\left(  \pi\wedge\sigma\right)
=\operatorname*{int}\left(  \operatorname*{dit}\left(  \pi\right)
\cap\operatorname*{dit}\left(  \sigma\right)  \right)  $\\\hline
Implication & $\sigma\Rightarrow\pi$ & $\operatorname*{dit}\left(
\sigma\Rightarrow\pi\right)  =\operatorname*{int}\left(  \operatorname*{dit}%
\left(  \sigma\right)  ^{c}\cup\operatorname*{dit}\left(  \pi\right)  \right)
$\\\hline
Any logical op. $\#$ & $\sigma\#\pi$ & Int. of subset op. $\#$ applied to dit
sets\\\hline
\end{tabular}

Lattice of partitions $\prod\left(  U\right)  $ represented as lattice of
partition relations $\mathcal{O}\left(  U\times U\right)  $.
\end{center}

\subsection{Analogies between subset logic and partition logic}

The development of partition logic was guided by some basic analogies between
the two dual forms of logic. The most basic analogy is that a distinction or
dit of a partition is the analogue of an element of a subset:

\begin{center}
a subset $S$ of $U$ contains an element $u$ $\approx$ a partition $\pi$ on $U$
distinguishes a pair $\left(  u,u^{\prime}\right)  $.
\end{center}

\noindent The top of the subset lattice is the universe set $U$ of all
possible elements and the top of the partition lattice is the partition
$\mathbf{1}$ with all possible distinctions $\operatorname*{dit}%
(\mathbf{1)}=U\times U-\Delta_{U}$ (all the ordered pairs minus the diagonal
self-pairs which can never be distinctions). The bottoms of the lattices are
the null subset $\emptyset$ of no elements and the indiscrete partition
$\mathbf{0}$ of no distinctions. The partial orders in the lattices are the
inclusion of elements $S\subseteq T$, and the inclusion of distinctions
$\operatorname*{dit}\left(  \sigma\right)  \subseteq\operatorname*{dit}\left(
\pi\right)  $.

\ Intuitively, a \textit{property} on $U$ is something that each element has
or does not have (like a person being female or not), while intuitively an
attribute on $U$ is something that each element has but with various values
(like the weight or height of a person). The subsets of $U$ can be thought of
as abstract versions of properties of the elements of $U$ while the partitions
on $U$ are abstract versions of the attributes on $U$ where the different
blocks of a partition represent the different values of the attribute.
Technically, an \textit{attribute} is given by a function $f:U\rightarrow%
%TCIMACRO{\U{211d} }%
%BeginExpansion
\mathbb{R}
%EndExpansion
$ (for some value set which we might take as the reals $%
%TCIMACRO{\U{211d} }%
%BeginExpansion
\mathbb{R}
%EndExpansion
$) and the partition induced by the attribute is the inverse image partition
$\left\{  f^{-1}\left(  r\right)  \neq\emptyset:r\in%
%TCIMACRO{\U{211d} }%
%BeginExpansion
\mathbb{R}
%EndExpansion
\right\}  $. A real-valued attribute $f:U\rightarrow%
%TCIMACRO{\U{211d} }%
%BeginExpansion
\mathbb{R}
%EndExpansion
$ will lift to a Hermitian operator so that the attribute's inverse image
partition $\left\{  f^{-1}\left(  r\right)  \neq\emptyset:r\in%
%TCIMACRO{\U{211d} }%
%BeginExpansion
\mathbb{R}
%EndExpansion
\right\}  $ lifts to the direct sum decomposition of the operator's
eigenspaces, and the attribute's values $r$ lift to the operator's eigenvalues.%

%TCIMACRO{\FRAME{dtbpF}{5.4233in}{3.3855in}{0in}{}{}{fig1-twologicstable.eps}%
%width 5.4233in;  height 3.3855in;  depth 0in;  original-width 7.4753in;
%original-height 4.6517in;  cropleft "0";  croptop "1";  cropright "1";
%cropbottom "0";  filename 'fig1-twologicstable.eps';file-properties "XNPEU";}}
%}%
%BeginExpansion
\begin{center}
\includegraphics[
height=3.3855in,
width=5.4233in
]%
{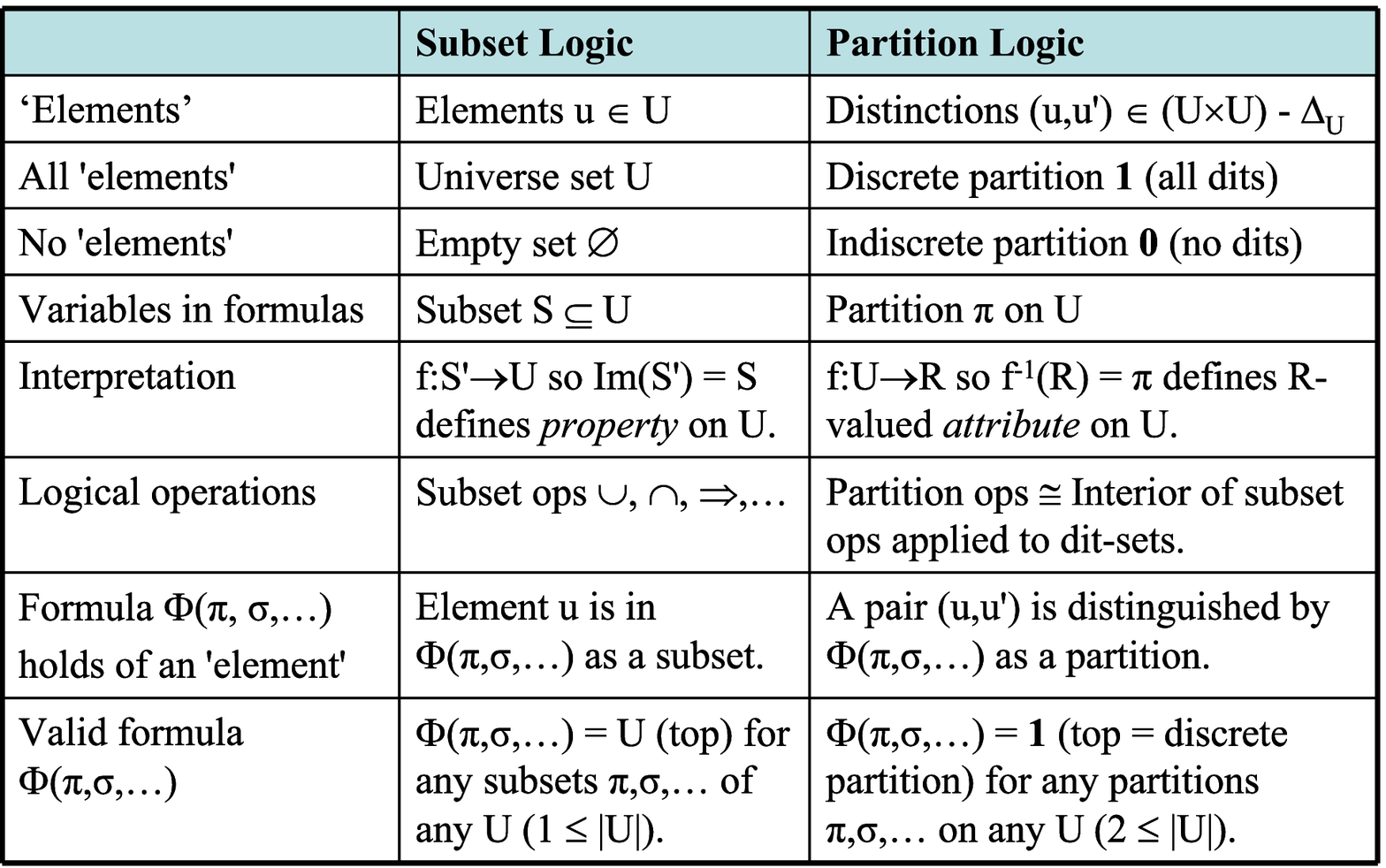}%
\end{center}
%EndExpansion

\begin{center}
Figure 1: Table of analogies between dual logics of subsets and partitions.
\end{center}

\section{Logical information theory}

We have so far made no assumptions about the finitude of the universe $U$. For
a finite universe $U$, Boole developed the "logical" version of finite
probability theory by assigning the normalized counting measure $\Pr\left(
S\right)  =\frac{|S|}{\left\vert U\right\vert }$ to each subset which can be
interpreted as a probability under the Laplacian assumption of equiprobable
elements. Using the elements-distinctions analogy, we can assign the analogous
normalized counting measure of the dit set of a partition $h\left(
\pi\right)  =\frac{\left\vert \operatorname*{dit}\left(  \pi\right)
\right\vert }{\left\vert U\times U\right\vert }$ to each partition which can
be interpreted as the \textit{logical information content} or \textit{logical
entropy }of the partition. Under the assumption of equiprobable elements, the
logical entropy of a partition can be interpreted as the probability that two
drawings from $U$ (with replacement) will give a distinction of the partition.%

%TCIMACRO{\FRAME{dtbpF}{6.0908in}{2.0806in}{0in}{}{}%
%{fig2-logicalprobsandentropy.eps}{\special{ language "Scientific Word";
%type "GRAPHIC";  maintain-aspect-ratio TRUE;  display "USEDEF";
%valid_file "F";  width 6.0908in;  height 2.0806in;  depth 0in;
%original-width 7.4753in;  original-height 2.5318in;  cropleft "0";
%filename 'fig2-logicalprobsandentropy.eps';file-properties "XNPEU";}} }%
%BeginExpansion
\begin{center}
\includegraphics[
height=2.0806in,
width=6.0908in
]%
{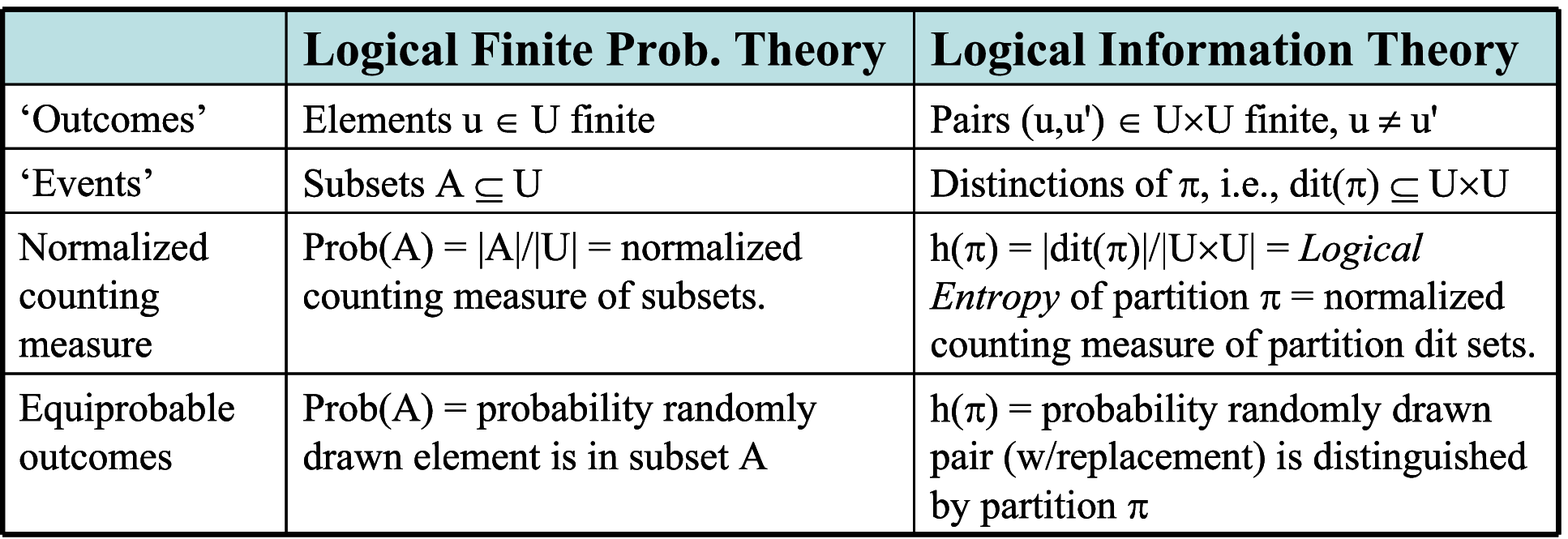}%
\end{center}
%EndExpansion

\begin{center}
Figure 2: Logical probability theory is to subset logic

as logical information theory is to partition logic
\end{center}

The probability of drawing an element from a block $B\in\pi$ is $p_{B}%
=\frac{\left\vert B\right\vert }{\left\vert U\right\vert }$ so the logical
entropy of a partition can be written in terms of these block probabilities
since $\left\vert \operatorname*{dit}\left(  \pi\right)  \right\vert
=\sum_{B\neq B^{\prime}\in\pi}\left\vert B\times B^{\prime}\right\vert
=\left\vert U\right\vert ^{2}-\sum_{B\in\pi}\left\vert B\right\vert ^{2}$. Hence:

\begin{center}
$h\left(  \pi\right)  =\frac{\left\vert \operatorname*{dit}\left(  \pi\right)
\right\vert }{\left\vert U\times U\right\vert }=\frac{\left\vert U\right\vert
^{2}-\sum_{B\in\pi}\left\vert B\right\vert ^{2}}{\left\vert U\right\vert ^{2}%
}=1-\sum_{B\in\pi}p_{B}^{2}$.
\end{center}

\noindent This formula has a long history (see \cite{ell:distinctions}) and is
usually called the \textit{Gini-Simpson diversity index} in the biological
literature \cite{rao:div}. For instance, if we partition animals by species,
then it is the probability in two independent samples that we will find
animals of different species.

This version of the logical entropy formula also makes clear the
generalization path to define the logical entropy of any finite probability
distribution $p=(p_{1},...,p_{n})$:

\begin{center}
$h\left(  p\right)  =1-\sum_{i}p_{i}^{2}$.\footnote{In the general case, the
$p_{i}$ becomes a probability density function and the summation an integral.}
\end{center}

\noindent C. R. Rao \cite{rao:div} has defined a general notion of quadratic
entropy in terms of a distance function $d(u,u^{\prime})$ between the elements
of $U$. In the most general "logical" case, the natural logical distance
function is:

\begin{center}
$d\left(  u,u^{\prime}\right)  =1-\delta\left(  u,u^{\prime}\right)  =\left\{
\begin{array}
[c]{c}%
1\text{ if }u\neq u^{\prime}\\
0\text{ if }u=u^{\prime}%
\end{array}
\right.  $
\end{center}

\noindent and, in that case, the quadratic entropy is just the logical entropy.

Further details about logical information theory and the relationship with the
usual notion of Shannon entropy can be found in \cite{ell:distinctions}. For
our purposes here, the important thing is the lifting of logical entropy to
the context of vector spaces and quantum mathematics where for any density
matrix $\rho$, the logical entropy $h\left(  \rho\right)
=1-\operatorname*{tr}\left[  \rho^{2}\right]  $ allows us to directly measure
and interpret the changes made in a measurement.

\section{Partitions and objective indefiniteness}

\subsection{Representing objective indistinctness}

It has already been emphasized how Boolean subset logic captures at the
logical level the common sense vision of reality where an entity definitely
has or does not have any property. We can now describe how the dual logic of
partitions captures at the logical level a vision of reality with objectively
indefinite (or indistinct)\footnote{The adjectives "indefinite" and
"indistinct" will be used interchangeably as synonyms. The word
"indefiniteness" is more common in the QM literature, but "indistinctness" has
a better noun form as "indistinctions" (with the opposite as "distinctions").}
entities. The key step is to:

\begin{quote}
interpret a subset $S$ as a \textit{single objectively indistinct element}
that, with further distinctions, could become any of the fully distinct
elements $u\in S$.
\end{quote}

To anticipate the lifted concepts in vector spaces, the fully distinct
elements $u\in U$ might be called "eigen-elements" and the single indistinct
element $S$ is a "superposition" of the eigen-elements $u\in S$ (thinking of
the collecting together $\left\{  u,u^{\prime},...\right\}  =S$ of the
elements of $S$ as their "superposition"). With distinctions, the indistinct
element $S$ might be refined into one of the singletons $\left\{  u\right\}  $
for $u\in S$ [where $\left\{  u\right\}  $ is the "superposition" consisting
of a single eigen-element so it just denotes that element $u$].

Abner Shimony (\cite{shim:reality} and \cite{shim:concept}), in his
description of a superposition state as being objectively indefinite, adopted
Heisenberg's \cite{heisen:phy-phil} language of "potentiality" and "actuality"
to describe the relationship of the eigenstates that are superposed to give an
objectively indefinite superposition. This terminology could be adapted to the
case of the sets. The elements $u\in S$ are "potential" in the objectively
indefinite "superposition" $S$, and, with further distinctions, the indefinite
element $S$ might "actualize" to $\left\{  u\right\}  $ for one of the
"potential" $u\in S$. Starting with $S$, the other $u\notin S$ are not
"potentialities" that could be "actualized" with further distinctions.

This terminology is, however, somewhat misleading since the indefinite element
$S$ is perfectly actual; it is only the multiple eigen-elements $u\in S$ that
are "potential" until "actualized" by some further distinctions. In a
"measurement," a single actual indefinite element becomes a single actual
definite element. Since the "measurement" goes from actual indefinite to
actual definite, the potential-to-actual language of Heisenberg should only be
used with proper care--if at all.

Consider a three-element universe $U=\left\{  a,b,c\right\}  $ and a partition
$\pi=\left\{  \left\{  a\right\}  ,\left\{  b,c\right\}  \right\}  $. The
block $S=\left\{  b,c\right\}  $ is objectively indefinite between $\left\{
b\right\}  $ and $\left\{  c\right\}  $ so those singletons are its
"potentialities" in the sense that a distinction could result in either
$\left\{  b\right\}  $ or $\left\{  c\right\}  $ being "actualized." However
$\left\{  a\right\}  $ is not a "potentiality" when one is starting with the
indefinite element $\left\{  b,c\right\}  $.

Note that this objective indefiniteness is not well-described as saying that
indefinite pre-distinction element is "simultaneously both $b$ and $c$"; it is
indefinite between $b$ and $c$. That is, a "superposition" should \textit{not}
be thought of like a double exposure photograph which has two fully definite
images. That imagery is a holdover from classical wave imagery (e.g., in
Fourier analysis) where definite eigen-waveforms are superposed to give a
superposition waveform. Instead, the objectively indistinct element is like an
out-of-focus photograph that with some sharpening could be resolved into one
of two or more definite images. Yet one needs some way to indicate what are
the definite eigen-elements that could be "actualized" from a single
indefinite element $S$, and that is the role in the set case of
conceptualizing $S$ as a collecting together or a "superposition" of certain
"potential" eigen-elements $u$.

The following is another attempt to clarify the imagery.%

%TCIMACRO{\FRAME{dtbpF}{3.4223in}{2.676in}{0pt}{}{}{fig3-guymasks.eps}%
%width 3.4223in;  height 2.676in;  depth 0pt;  original-width 7.3957in;
%original-height 5.7721in;  cropleft "0";  croptop "1";  cropright "1";
%cropbottom "0";  filename 'fig3-guymasks.eps';file-properties "XNPEU";}} }%
%BeginExpansion
\begin{center}
\includegraphics[
height=2.676in,
width=3.4223in
]%
{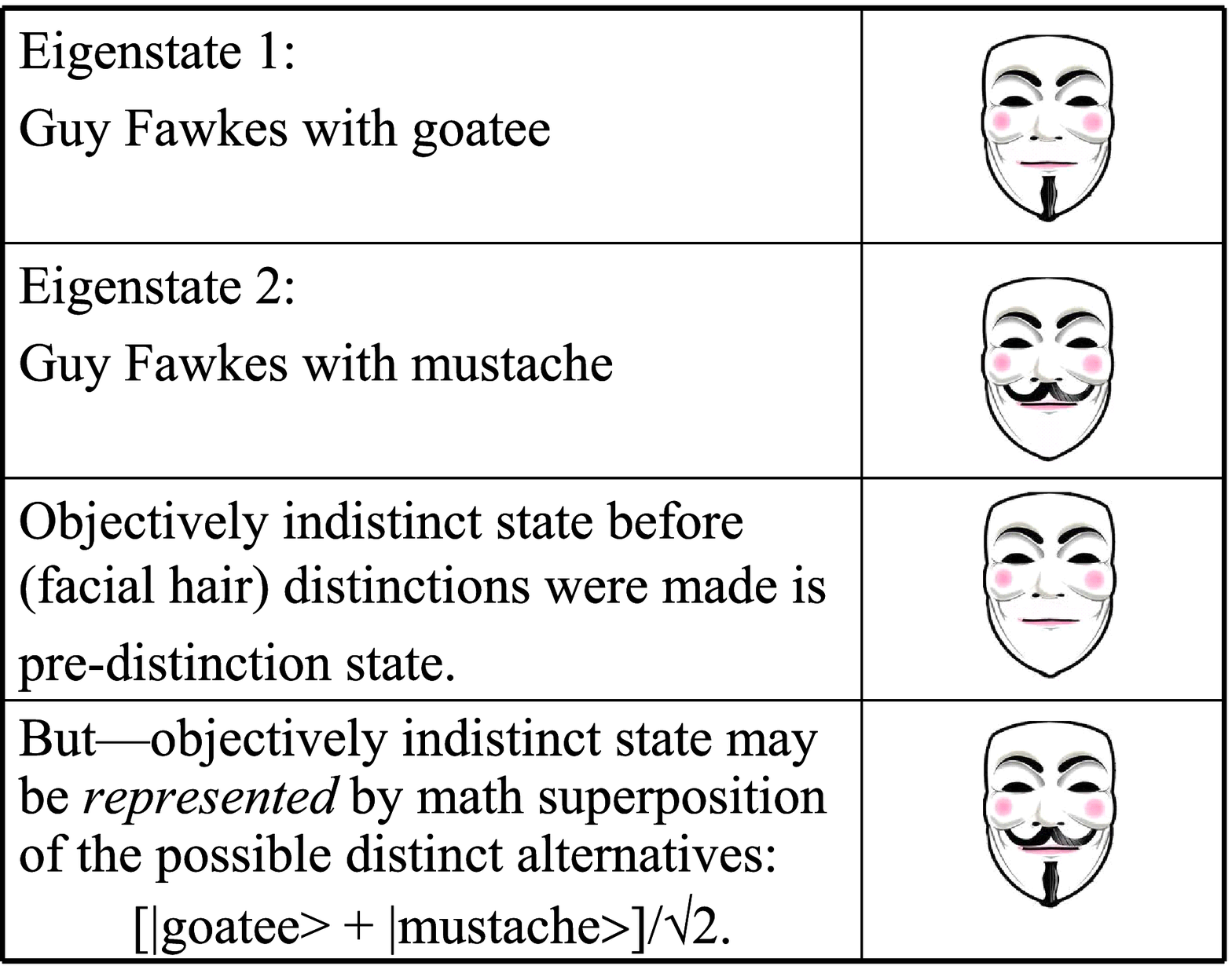}%
\end{center}
%EndExpansion

\begin{center}
Figure 3: Indistinct pre-distinction state \textit{represented} as superposition
\end{center}

The following table gives yet another attempt at visualization by contrasting
a classical picture and an objectively indefinite (or "quantum") picture of a
"particle" getting from $A$ to $B$.%

%TCIMACRO{\FRAME{dtbpF}{3.7561in}{2.7037in}{0pt}{}{}{fig4-trajectorytable.eps}%
%width 3.7561in;  height 2.7037in;  depth 0pt;  original-width 6.9093in;
%original-height 4.9629in;  cropleft "0";  croptop "1";  cropright "1";
%cropbottom "0";  filename 'fig4-trajectorytable.eps';file-properties "XNPEU";}%
%} }%
%BeginExpansion
\begin{center}
\includegraphics[
height=2.7037in,
width=3.7561in
]%
{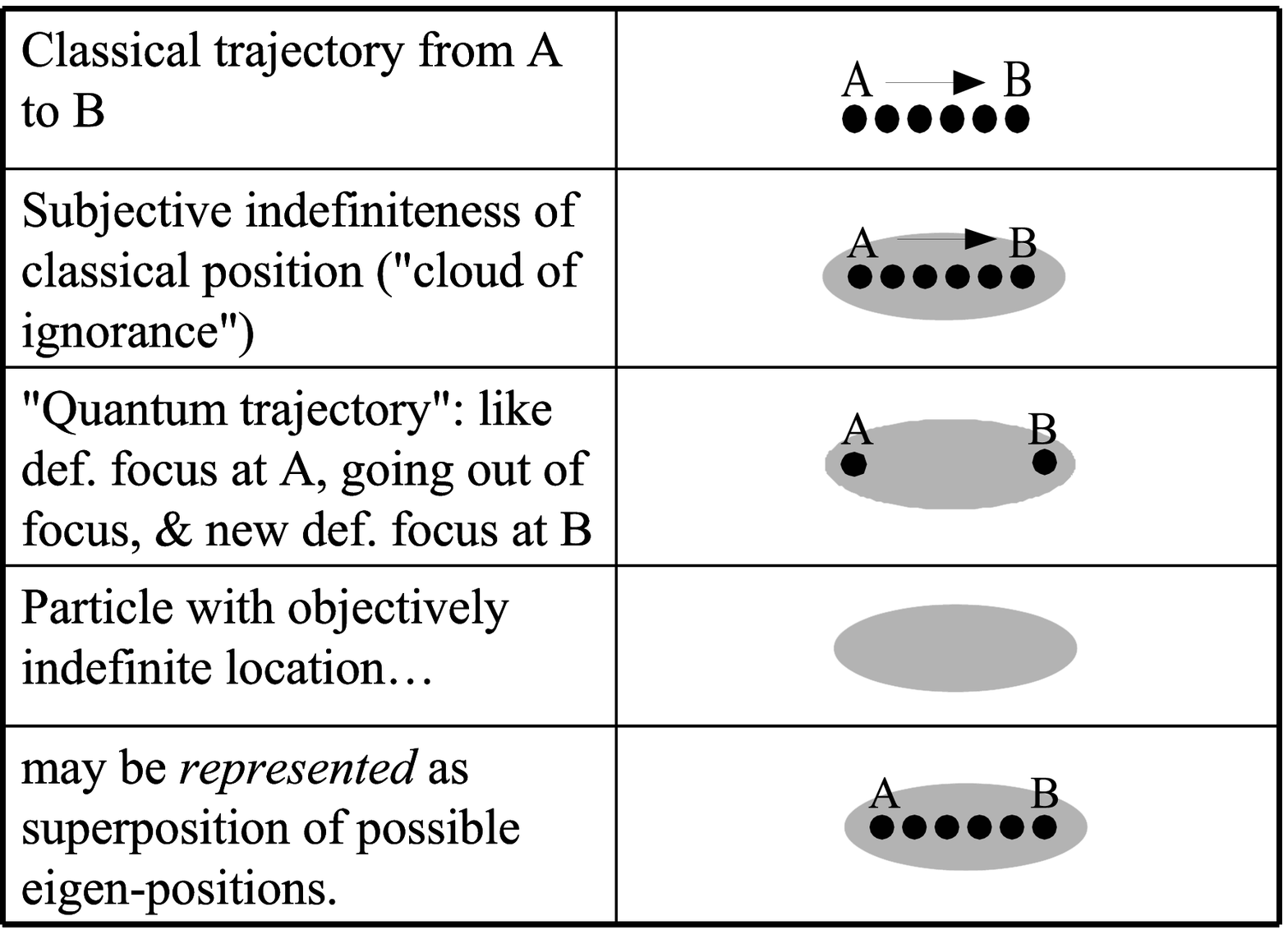}%
\end{center}
%EndExpansion

\begin{center}
Figure 4: Getting from $A$ to $B$ in classical and quantum ways
\end{center}

The classical trajectory is a sequence of definite positions. A state of
subjective indefiniteness is compatible with a classical trajectory when we
have a "cloud of ignorance" about the actual definite location of the
particle. The "quantum trajectory" might be envisaged as starting with a
definite focus or location at $A$, then evolving to an objectively indefinite
state (with the various positions as potentialities), and then finally another
"look" or measurement that achieves a definite focus at location $B$. The
particle in its objectively indefinite position state is represented as the
superposition of the possible definite position states.

\subsection{The conceptual duality between the two lattices}

The conceptual duality between the lattice of subsets and the lattice of
partitions could be described using the rather meta-physical notions of
substance and form. Consider what happens when one starts at the bottom of
each lattice and moves towards the top.%

%TCIMACRO{\FRAME{dtbpFU}{5.4384in}{1.9355in}{0pt}{\Qcb{Figure 5: Conceptual
%duality between the two lattices}}{}{fig5-twolattices.eps}%
%width 5.4384in;  height 1.9355in;  depth 0pt;  original-width 5.279in;
%original-height 1.8617in;  cropleft "0";  croptop "1";  cropright "1";
%cropbottom "0";  filename 'fig5-twolattices.eps';file-properties "XNPEU";}} }%
%BeginExpansion
\begin{center}
\includegraphics[
height=1.9355in,
width=5.4384in
]%
{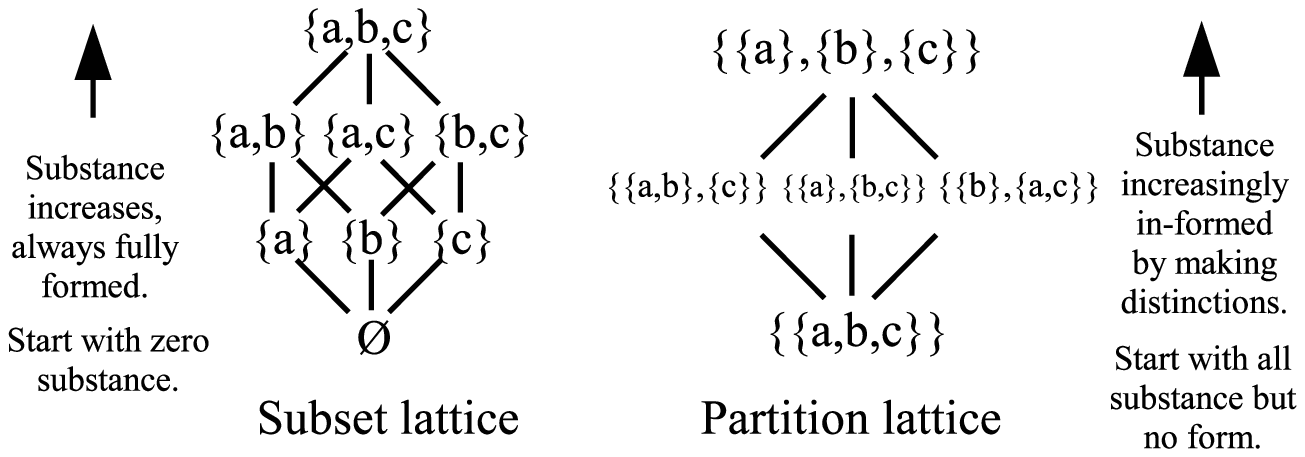}%
\\
Figure 5: Conceptual duality between the two lattices
\end{center}
%EndExpansion

At the bottom of the Boolean lattice is the empty set $\emptyset$ which
represents no substance. As one moves up the lattice, new fully propertied
elements of substance appear until finally one reaches the top, the universe
$U$. Thus new substance is created but each element is fully formed and
distinguished in terms of its properties.

At the bottom of the partition lattice is the blob $\mathbf{0}$ which
represents all the substance but with no distinctions to in-form the
substance. As one moves up the lattice, no new substance appears but
distinctions objectively in-form the indistinct elements as they become more
and more distinct, until one finally reaches the top, the discrete partition
$\mathbf{1}$, where all the eigen-elements of $U$ have been fully
distinguished from each other. Thus one ends up at the same place either way,
but by two totally different but dual ways.

The notion of logical entropy expresses this idea of objective in-formation as
the normalized count of the informing distinctions. For instance, in the
partition lattice on a three element set pictured above, the logical entropy
of the blob is always $h\left(  \mathbf{0}\right)  =0$ since there are no
distinctions. For a middle partition such as $\pi=\left\{  \left\{  a\right\}
,\left\{  b,c\right\}  \right\}  $, the distinctions are $\left(  a,b\right)
$, $\left(  b,a\right)  $, $\left(  a,c\right)  $, and $\left(  c,a\right)  $
for a total of $4$ where $\left\vert U\right\vert ^{2}=3^{2}=9$ so the logical
entropy is $h\left(  \pi\right)  =\frac{\left\vert \operatorname*{dit}\left(
\pi\right)  \right\vert }{\left\vert U\times U\right\vert }=\frac{4}{9}$. For
the discrete partition, there are all possible distinctions for a total of
$\left\vert U\right\vert ^{2}-\left\vert \Delta_{U}\right\vert =9-3=6$ so the
logical entropy is $h\left(  \mathbf{1}\right)  =1-\frac{1}{\left\vert
U\right\vert }=\frac{6}{9}$. In each case, the logical entropy of a partition
is the probability that two independent draws from $U$ will yield a
distinction of the partition.

The progress from bottom to top of the two lattices could also be described as
two creation stories.

\begin{itemize}
\item \noindent\textit{Subset creation story}: \textquotedblleft In the
Beginning was the Void\textquotedblright, and then elements are created, fully
propertied and distinguished from one another, until finally reaching all the
elements of the universe set $U$.

\item \noindent\textit{Partition creation story}: \textquotedblleft In the
Beginning was the Blob\textquotedblright, which is an undifferentiated
\textquotedblleft substance,\textquotedblright\ and then there is a "Big Bang"
where elements (\textquotedblleft its\textquotedblright) are created by being
objectively in-formed (objective "dits") by the making of distinctions (e.g.,
breaking symmetries) until the result is finally the singletons which
designate the elements of the universe $U$.\footnote{Heisenberg identifies the
"substance" with energy.
\par
\begin{quotation}
\noindent Energy is in fact the substance from which all elementary particles,
all atoms and therefore all things are made, and energy is that which moves.
Energy is a substance, since its total amount does not change, and the
elementary particles can actually be made from this substance as is seen in
many experiments on the creation of elementary particles.\cite[p.
63]{heisen:phy-phil}
\end{quotation}
\par
In his sympathetic interpretation of Aristotle's treatment of substance and
form, Heisenberg refers to the substance as: "a kind of indefinite corporeal
substratum, embodying the possibility of passing over into actuality by means
of the form."\cite[p. 148]{heisen:phy-phil} It was previously noted that
Heisenberg's "potentiality" "passing over into actuality by means of the form"
should be seen as the actual indefinite "passing over into" the actual
definite by being objectively in-formed through the making of distinctions.}
\end{itemize}

These two creation stories might also be illustrated as follows.%

%TCIMACRO{\FRAME{dtbpFU}{5.7864in}{1.3787in}{0pt}{\Qcb{Figure 6: Two ways to
%create a universe $U$}}{}{fig6-twocreatons.eps}%
%width 5.7864in;  height 1.3787in;  depth 0pt;  original-width 7.191in;
%original-height 1.694in;  cropleft "0";  croptop "1";  cropright "1";
%cropbottom "0";  filename 'fig6-twocreatons.eps';file-properties "XNPEU";}} }%
%BeginExpansion
\begin{center}
\includegraphics[
height=1.3787in,
width=5.7864in
]%
{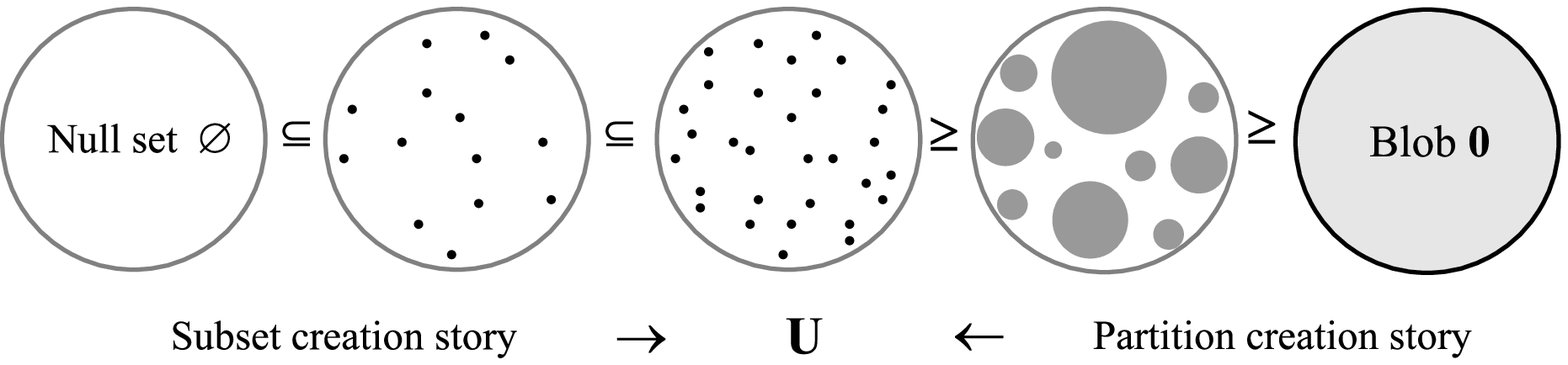}%
\\
Figure 6: Two ways to create a universe $U$%
\end{center}
%EndExpansion

One might think of the universe $U$ (in the middle of the above picture) as
the macroscopic world of fully definite entities that we ordinarily
experience. Common sense and classical physics assumes, as it were, the subset
creation story on the left. But \textit{a priori,} it could just as well have
been the dual story, the partition creation story pictured on the right, that
leads to the \textit{same} macro-picture $U$. And, as we will see, that is
indeed the message of quantum mechanics.

\section{The Lifting Program}

\subsection{From sets to vector spaces}

We have so far outlined the mathematics of set partitions such as the
representation of an indefinite element as a (non-singleton) block in a
partition and carving out the fully distinct eigen-elements by making more
distinctions, e.g., joining together the distinctions of different partitions
(on the same universe). The lifting program lifts these set-based concepts to
the much richer environment of vector spaces.

Why vector spaces? Dirac \cite{dirac:principles} noted that the notion of
superposition was basic to and characteristic of quantum mechanics. At the
level of sets, there is only a very simple and austere notion of
"superposition," namely collecting together definite eigen-elements into one
subset interpreted as one indefinite element (indistinct between the
"superposed" eigen-elements). In a vector space, superposition is represented
by a weighted vector sum with weights drawn from the base field. Thus the
lifting of set concepts to vector spaces (Hilbert spaces in particular) gives
a much richer version of partition mathematics, and, as we will see, the
lifting gives the mathematics of quantum mechanics.

The lifting program is not an algorithm but has the guiding:

\begin{center}
\textbf{Basis Principle:} \textit{Apply the set concept to a basis set and
then generate the lifted vector space concept.}
\end{center}

\noindent For instance, what is the vector space lift of the set concept of
cardinality? We apply the set concept of cardinality to a basis set of a
vector space where it yields the notion of \textit{dimension} of the vector
space (after checking that all bases have equal cardinality). Thus the lift of
set-cardinality is not the cardinality of a vector space but its
dimension.\footnote{In QM, the extension of concepts on finite dimensional
Hilbert space to infinite dimensional ones is well-known. Since our expository
purpose is conceptual rather than mathematical, we will stick to finite
dimensional spaces.} Thus the null set $\emptyset$ with cardinality $0$ lifts
to the trivial zero vector space with dimension $0$.

It is often convenient to refer to a set concept in terms of its lifted vector
space concept. This will be done by using the name of the vector space concept
enclosed in scare quotes, e.g., the cardinality of a set is its "dimension."

\subsection{Lifting set partitions}

To lift the mathematics of set partitions to vector spaces, the first question
is the lift of a set partition. In the category of sets, the direct sum is the
disjoint union, and the union of the blocks in a partition is a disjoint
union. Hence a set partition is a direct sum decomposition of the universe
set, so one might expect the corresponding vector space concept to be a direct
sum decomposition of the space (where "direct sum" is defined in the category
of vector spaces over some base field). That answer is immediately obtained by
applying the set concept of a partition to a basis set and then seeing what it
generates. Each block $B$ of the set partition of a basis set generates a
subspace $W_{B}\subseteq V$, and the subspaces together form a \textit{direct
sum decomposition}: $V=\sum_{B}\oplus W_{B}$. Thus the proper lifted notion of
a partition for a vector space is \textit{not} a set partition of the space,
e.g., defined by a subspace $W\subseteq V$ where $v\thicksim v^{\prime}$ if
$v-v^{\prime}\in W$, but is a direct sum decomposition of the vector
space.\footnote{The usual quantum logic approach to define a `propositional'
logic for QM focused on the question of whether or not a vector was in a
subspace, which in turn lead to a misplaced focus on the set equivalence
relations defined by the subspaces, equivalence relations that have a special
property of being \textit{commuting} \cite{fmr:commuting}. If "quantum logic"
is to be the logic that is to QM as Boolean subset logic is to classical
mechanics, then that is partition logic.}

\subsection{Lifting partition joins}

The main partition operation that we need to lift to vector spaces is the join
operation. Two set partitions cannot be joined unless they are
\textit{compatible} in the sense of being defined on the same universe set.
This notion of compatibility lifts to vector spaces by defining two vector
space partitions $\omega=\left\{  W_{\lambda}\right\}  $ and $\xi=\{X_{\mu}\}$
on $V$ as being \textit{compatible} if there is a basis set for $V$ so that
the two vector space partitions arise from two set partitions of that common
basis set.

If two set partitions $\pi=\left\{  B\right\}  $ and $\sigma=\left\{
C\right\}  $ are compatible, then their \textit{join} $\pi\vee\sigma$ is
defined as the set partition whose blocks are the non-empty intersections
$B\cap C$. Similarly the lifted concept is that if two vector space partitions
$\omega=\left\{  W_{\lambda}\right\}  $ and $\xi=\{X_{\mu}\}$ are compatible,
then their \textit{join} $\omega\vee\xi$ is defined as the vector space
partition whose subspaces are the non-zero intersections $W_{\lambda}\cap
X_{\mu}$. And by the definition of compatibility, we could generate the
subspaces of the join $\omega\vee\xi$ by the blocks in the join of the two set
partitions of the common basis set.

\subsection{Lifting attributes}

A set partition might be seen as an abstract rendition of the inverse image
partition $\left\{  f^{-1}\left(  r\right)  \right\}  $ defined by some
concrete attribute $f:U\rightarrow%
%TCIMACRO{\U{211d} }%
%BeginExpansion
\mathbb{R}
%EndExpansion
$ on $U$. What is the lift of an attribute? At first glance, the basis
principle would seem to imply: define a set attribute on a basis set (with
values in the base field) and then linearly generate a functional from the
vector space to the base field. But a functional does not define a vector
space partition; it only defines the set partition of the vector space
compatible with the vector space operations that is determined by the kernel
of the functional. Hence we need to try a more careful application of the
basis principle.

It is helpful to first give a suggestive reformulation of a set attribute
$f:U\rightarrow%
%TCIMACRO{\U{211d} }%
%BeginExpansion
\mathbb{R}
%EndExpansion
$. If $f$ is constant on a subset $S\subseteq U$ with a value $r$, then we
might symbolize this as:

\begin{center}
$f\upharpoonright S=rS$
\end{center}

\noindent and suggestively call $S$ an "eigenvector" and $r$ an "eigenvalue."
\noindent For any "eigenvalue" $r$, define $f^{-1}\left(  r\right)  $ =
"eigenspace of $r$" as the union of all the "eigenvectors" with that
"eigenvalue." Since the "eigenspaces" span the set $U$, the attribute
$f:U\rightarrow%
%TCIMACRO{\U{211d} }%
%BeginExpansion
\mathbb{R}
%EndExpansion
$ can be represented by:

\begin{center}
$f=\sum_{r}r\chi_{f^{-1}\left(  r\right)  }$

"Spectral decomposition" of set attribute $f:U\rightarrow%
%TCIMACRO{\U{211d} }%
%BeginExpansion
\mathbb{R}
%EndExpansion
$
\end{center}

\noindent\lbrack where $\chi_{f^{-1}\left(  r\right)  }$ is the characteristic
function for the "eigenspace" $f^{-1}\left(  r\right)  $]. Thus a set
attribute determines a set partition and has a constant value on the blocks of
the set partition, so by the basis principle, that lifts to a vector space
concept that determines a vector space partition and has a constant value on
the blocks of the vector space partition.

The suggestive terminology gives the proper lift. The lift of
$f\upharpoonright S=rS$ is the eigenvector equation $Lv=\lambda v$ where $L$
is a linear operator on $V$. In particular, if $Lv_{1}=\lambda v_{1}$ and
$Lv_{2}=\lambda v_{2}$ for two basis vectors $v_{1}$ and $v_{2}$, then
$Lv=\lambda v$ for all $v\in\left[  v_{1},v_{2}\right]  $ (the subspace
generated by $v_{1}$ and $v_{2}$). The lift of an "eigenspace" $f^{-1}\left(
r\right)  $ is the eigenspace $W_{\lambda}$ of an eigenvalue $\lambda$. The
lift of the simplest attributes, which are the characteristic functions
$\chi_{f^{-1}\left(  r\right)  }$, are the projection operators $P_{\lambda}$
that project to the eigenspaces $W_{\lambda}$. The characteristic property of
the characteristic functions $\chi:U\rightarrow%
%TCIMACRO{\U{211d} }%
%BeginExpansion
\mathbb{R}
%EndExpansion
$ is that they are idempotent in the sense that $\chi\left(  u\right)
\chi\left(  u\right)  =\chi\left(  u\right)  $ for all $u\in U$, and the
lifted characteristic property of the projection operators $P:V\rightarrow V$
is that they are idempotent in the sense that $P^{2}:V\rightarrow V\rightarrow
V=P:V\rightarrow V$. Finally, the "spectral decomposition" of a set attribute
lifts to the spectral decomposition of a \textit{vector space attribute}:

\begin{center}
$f=\sum_{r}r\chi_{f^{-1}\left(  r\right)  }$ lifts to $L=\sum_{\lambda}\lambda
P_{\lambda}$.

Lift of a set attribute to a vector space attribute
\end{center}

\noindent Thus a vector space attribute is just a linear operator whose
eigenspaces span the whole space which is called a \textit{diagonalizable}
\textit{linear operator} \cite{hk:la}. Then we see that the proper lift of a
set attribute using the basis principle does indeed define a vector space
partition, namely that of the eigenspaces of a diagonalizable linear operator,
and that the values of the attribute are constant on the blocks of the vector
space partition--as desired. To keep the eigenvalues of the linear operator
real, quantum mechanics restricts the vector space attributes to
\textit{Hermitian} (or \textit{self-adjoint}) linear operators, which
represent \textit{observables}, on a Hilbert space.%

%TCIMACRO{\FRAME{dtbpFU}{5.0182in}{2.8781in}{0pt}{\Qcb{Figure 7: Set attributes
%lift to linear operators}}{}{fig7-liftattributetable.eps}%
%width 5.0182in;  height 2.8781in;  depth 0pt;  original-width 7.4753in;
%original-height 4.2727in;  cropleft "0";  croptop "1";  cropright "1";
%filename 'fig7-liftattributetable.eps';file-properties "XNPEU";}} }%
%BeginExpansion
\begin{center}
\includegraphics[
height=2.8781in,
width=5.0182in
]%
{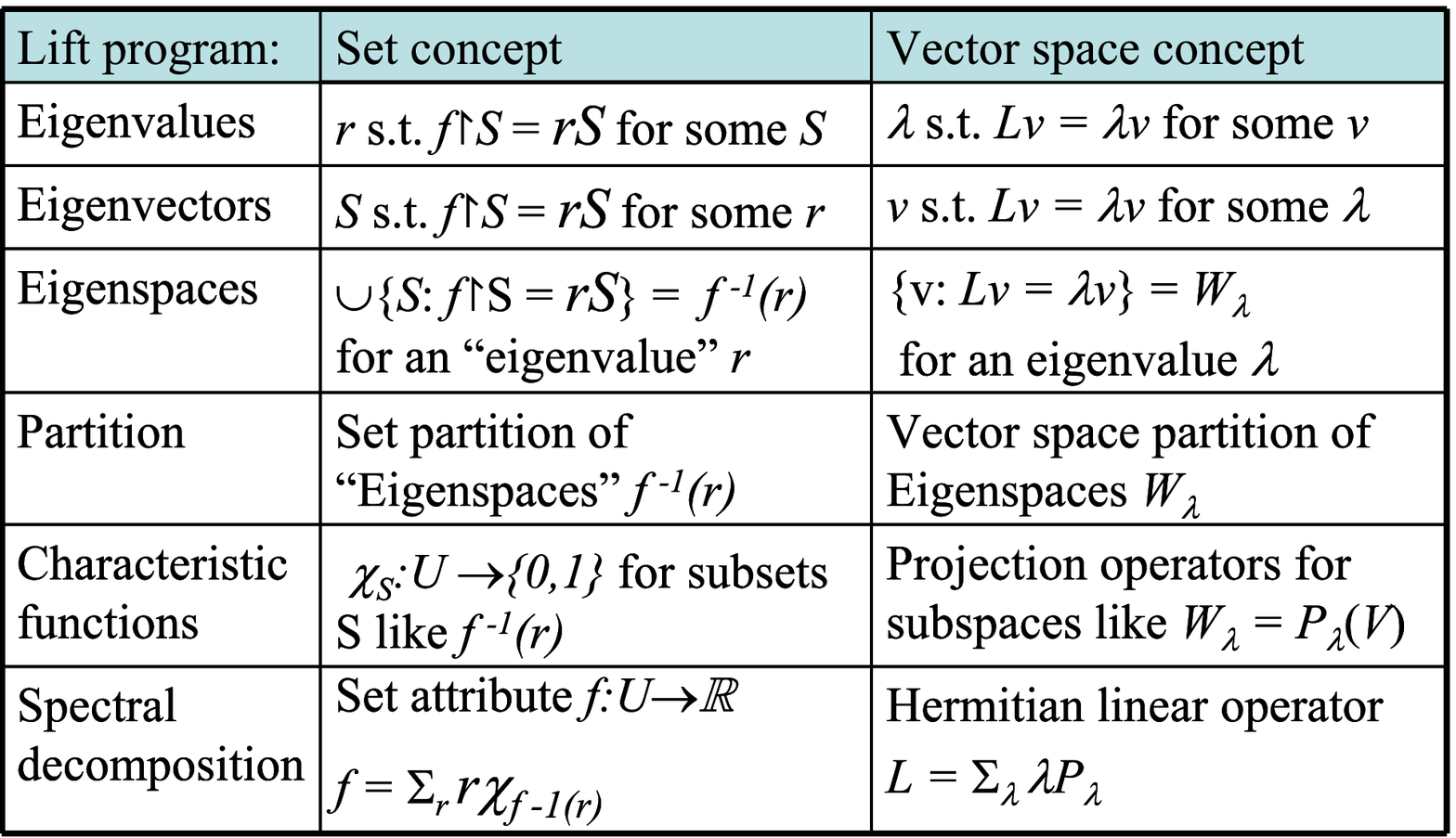}%
\\
Figure 7: Set attributes lift to linear operators
\end{center}
%EndExpansion

One of the mysteries of quantum mechanics is that the set attributes such as
position or momentum on the phase spaces of classical physics become linear
operators on the state spaces of QM. The lifting program "explains" that mystery.

\subsection{Lifting compatible attributes}

Since two set attributes $f:U\rightarrow%
%TCIMACRO{\U{211d} }%
%BeginExpansion
\mathbb{R}
%EndExpansion
$ and $g:U^{\prime}\rightarrow%
%TCIMACRO{\U{211d} }%
%BeginExpansion
\mathbb{R}
%EndExpansion
$ define two inverse image partitions $\left\{  f^{-1}\left(  r\right)
\right\}  $ and $\left\{  g^{-1}\left(  s\right)  \right\}  $ on their
domains, we need to extend the concept of compatible partitions to the
attributes that define the partitions. That is, two attributes $f:U\rightarrow%
%TCIMACRO{\U{211d} }%
%BeginExpansion
\mathbb{R}
%EndExpansion
$ and $g:U^{\prime}\rightarrow%
%TCIMACRO{\U{211d} }%
%BeginExpansion
\mathbb{R}
%EndExpansion
$ are \textit{compatible} if they have the same domain $U=U^{\prime}%
$.\footnote{This simplified definition is justified by the later treatment of
compatible attributes in the context of "quantum mechanics" on sets.} We have
previously lifted the notion of compatible set partitions to compatible vector
space partitions. Since real-valued set attributes lift to Hermitian linear
operators, the notion of compatible set attributes just defined would lift to
two linear operators being \textit{compatible} if their eigenspace partitions
are compatible. It is a standard fact of the QM literature (e.g., \cite[pp.
102-3]{hughes:interp} or \cite[p. 177]{hk:la}) that two (Hermitian) linear
operators $L,M:V\rightarrow V$ are compatible if and only if they commute,
$LM=ML$. Hence the \textit{commutativity} of linear operators is the lift of
the compatibility (i.e., defined on the same set) of set attributes.

Given two compatible set attributes $f:U\rightarrow%
%TCIMACRO{\U{211d} }%
%BeginExpansion
\mathbb{R}
%EndExpansion
$ and $g:U\rightarrow%
%TCIMACRO{\U{211d} }%
%BeginExpansion
\mathbb{R}
%EndExpansion
$, the join of their "eigenspace" partitions has as blocks the non-empty
intersections $f^{-1}\left(  r\right)  \cap g^{-1}\left(  s\right)  $. Each
block in the join of the "eigenspace" partitions could be characterized by the
ordered pair of "eigenvalues" $\left(  r,s\right)  $. An "eigenvector" of $f$,
$S\subseteq f^{-1}\left(  r\right)  $, and of $g$, $S\subseteq g^{-1}\left(
s\right)  $, would be a "simultaneous eigenvector": $S\subseteq f^{-1}\left(
r\right)  \cap g^{-1}\left(  s\right)  $.

In the lifted case, two commuting Hermitian linear operator $L$ and $M$ have
compatible eigenspace partitions $W_{L}=\left\{  W_{\lambda}\right\}  $ (for
the eigenvalues $\lambda$ of $L$) and $W_{M}=\left\{  W_{\mu}\right\}  $ (for
the eigenvalues $\mu$ of $M$). The blocks in the join $W_{L}\vee W_{M}$ of the
two compatible eigenspace partitions are the non-zero subspaces $\left\{
W_{\lambda}\cap W_{\mu}\right\}  $ which can be characterized by the ordered
pairs of eigenvalues $\left(  \lambda,\mu\right)  $. The nonzero vectors $v\in
W_{\lambda}\cap W_{\mu}$ are \textit{simultaneous eigenvectors} for the two
commuting operators, and there is a basis for the space consisting of
simultaneous eigenvectors.\footnote{One must be careful not to assume that the
simultaneous eigenvectors are the eigenvectors for the operator $LM=ML$ due to
the problem of degeneracy.}

A set of compatible set attributes is said to be \textit{complete} if the join
of their partitions is discrete, i.e., the blocks have cardinality $1$. Each
element of $U$ is then characterized by the ordered $n$-tuple $\left(
r,...,s\right)  $ of attribute values.

In the lifted case, a set of commuting linear operators is said to be
\textit{complete} if the join of their eigenspace partitions is nondegenerate,
i.e., the blocks have dimension $1$. The eigenvectors that generate those
one-dimensional blocks of the join are characterized by the ordered $n$-tuples
$\left(  \lambda,...,\mu\right)  $ of eigenvalues so the eigenvectors are
usually denoted as the eigenkets $\left\vert \lambda,...,\mu\right\rangle $ in
the Dirac notation. These \textit{Complete Sets of Commuting Operators} are
Dirac's CSCOs \cite{dirac:principles}.

\subsection{Summary of lifting program}

The lifting program so far is summarized in the following table.%

%TCIMACRO{\FRAME{dtbpFU}{4.6903in}{2.9351in}{0pt}{\Qcb{Figure 8: Summary of
%Lifting Program}}{}{fig8-liftingtable.eps}%
%width 4.6903in;  height 2.9351in;  depth 0pt;  original-width 7.4753in;
%original-height 4.6643in;  cropleft "0";  croptop "1";  cropright "1";
%cropbottom "0";  filename 'fig8-liftingtable.eps';file-properties "XNPEU";}}
%}%
%BeginExpansion
\begin{center}
\includegraphics[
height=2.9351in,
width=4.6903in
]%
{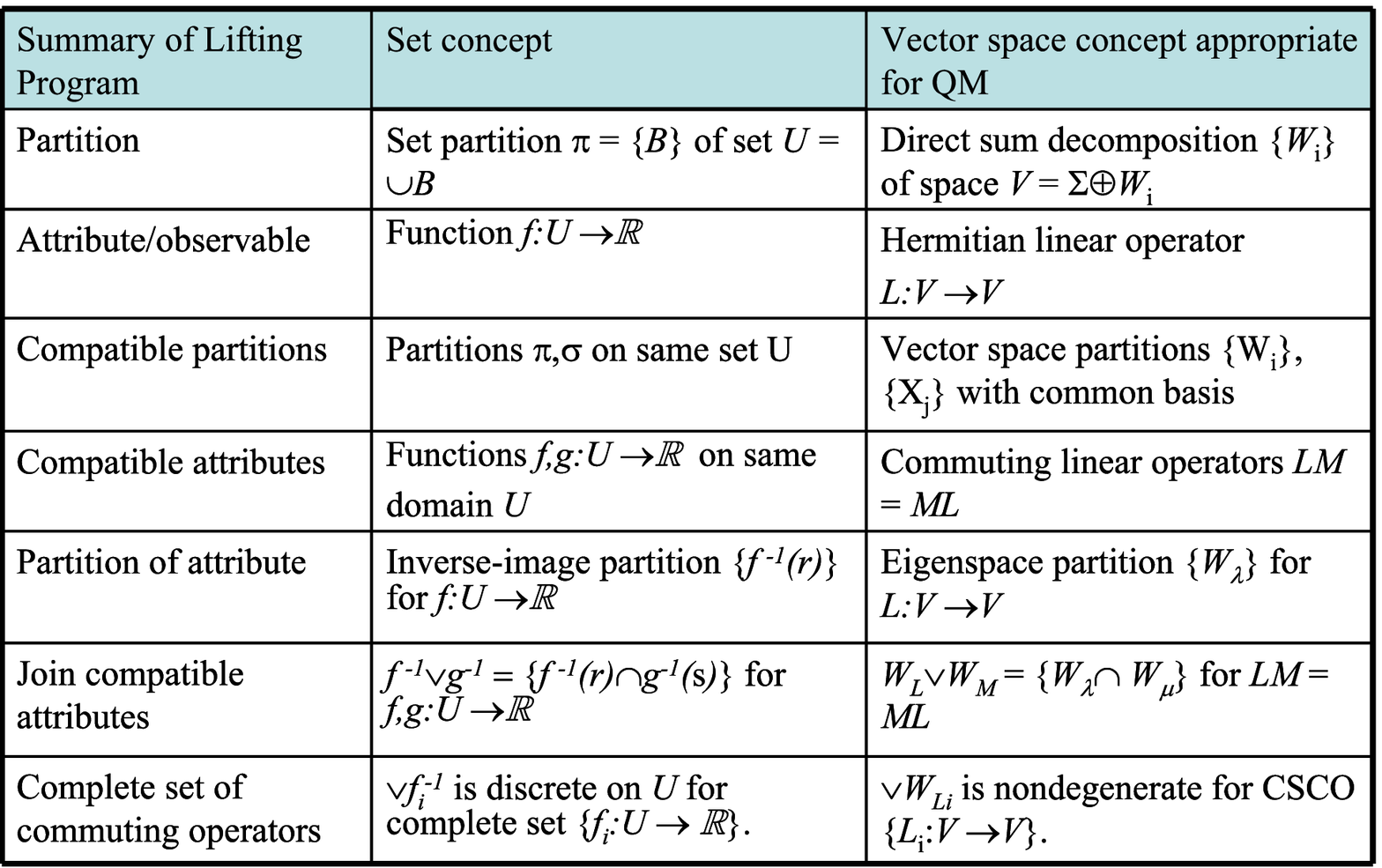}%
\\
Figure 8: Summary of Lifting Program
\end{center}
%EndExpansion

\subsection{Some subtleties of the lifting program}

The relation between set concepts and the lifted vector space concepts is not
a one-to-one mapping. For instance, the same subset $S=f^{-1}\left(  r\right)
$ appears both as an "eigenvector" $S$ such that $f\upharpoonright S=rS$ and
as an "eigenspace"--which are two very different vector space concepts. The
two-dimensional space $\left[  a,b\right]  $ generated by vectors $a$ and $b$
is quite different from the vector $a+b$, but at the austere level of sets,
they are both $\left\{  a,b\right\}  $. Thus the same set concept of a subset
$\left\{  a,b\right\}  $ (depending on whether it is viewed as $\left\{
a,b\right\}  =f^{-1}\left(  r\right)  $ or as $f\upharpoonright\left\{
a,b\right\}  =r\left\{  a,b\right\}  $) lifts to quite different vector space
concepts: the subspace $\left[  a,b\right]  $ or the vector $a+b$. This is one
of the reasons that the lifting program cannot be reduced to a simple mapping.

Moreover, the same vector space concept, viewed from different angles, may
"delift" to quite different set concepts. Consider the vector space concept of
a projection operator $P:V\rightarrow V$ that projects to the subspace
$P\left(  V\right)  =W$. As a linear operator with the eigenvalues $0$ and
$1$, a projection operator is the lift of a characteristic function $\chi
_{S}:U\rightarrow%
%TCIMACRO{\U{211d} }%
%BeginExpansion
\mathbb{R}
%EndExpansion
$ as an attribute. The projection operator assigns the eigenvalues $1$ and $0$
to the two blocks $P\left(  V\right)  $ and $\ker\left(  P\right)  $ of its
eigenspace partition, just as the attribute $\chi_{S}$ assigns the two values
to the two blocks $\chi_{S}\left(  1\right)  $ and $\chi_{S}\left(  0\right)
$ of its set partition. But a projection operator also serves to project an
arbitrary vector $v\in V$ to the part of $v$, namely $P\left(  v\right)  $,
that is in the range-space $W$. Since the delift of vectors $v\in V$ are
subsets $S\subseteq U$ (viewed as single indefinite elements), the delift of
the projecting operation would be a mapping from arbitrary subsets to the part
of each subset that is in the "range eigenspace" $\chi_{S}^{-1}\left(
1\right)  $. That "projection" is the idempotent mapping:

\begin{center}
$\chi_{S}^{-1}\left(  1\right)  \cap\left(  {}\right)  :\wp\left(  U\right)
\rightarrow\wp\left(  U\right)  $.
\end{center}

\noindent Thus the same vector space concept of a projection operator delifts
to two quite different set concepts: the set attribute $\chi_{S}:U\rightarrow%
%TCIMACRO{\U{211d} }%
%BeginExpansion
\mathbb{R}
%EndExpansion
$ and the subset operator $\chi_{S}^{-1}\left(  1\right)  \cap\left(
{}\right)  :\wp\left(  U\right)  \rightarrow\wp\left(  U\right)  $.

The subset operator treatment of a projection allows another type of "spectral
decomposition" associated with an attribute $f:U\rightarrow%
%TCIMACRO{\U{211d} }%
%BeginExpansion
\mathbb{R}
%EndExpansion
$. The previous statement for $S\subseteq f^{-1}\left(  r\right)  $ that
$f\upharpoonright S=rS$ can now be written $r\left[  f^{-1}\left(  r\right)
\cap S\right]  =rS$ so that the action of $f$ on subsets can be symbolically
represented as:

\begin{center}
$f\upharpoonright\left(  {}\right)  =\sum_{r}r\left[  f^{-1}\left(  r\right)
\cap\left(  {}\right)  \right]  $
\end{center}

\noindent that identifies the "eigenvectors" and "eigenvalues" in the set case
and thus could be taken as the set operator analogue of $L=\sum_{\lambda
}\lambda P_{\lambda}$.

\section{The Delifting Program: "Quantum mechanics" on sets}

\subsection{Probabilities in "quantum mechanics" on sets}

The lifting program establishes a relationship between concepts and operations
for sets and those for vector spaces. We have so far started with set
concepts, like the concept of a set partition, and then developed the
corresponding concept for vector spaces (direct sum decomposition). However
the relation between set and vector space concepts can also be established by
going the other way, by delifting quantum mechanical concepts from vector
spaces to sets. By delifting QM concepts to sets, we can develop a toy model
called \textit{"quantum mechanics"} \textit{on sets}--which shows the logical
structure of QM in a pedagogically simple and understandable
context.\footnote{Recall that a delifted vector space concept is indicated by
the concept's name in scare quotes.}

The connection between sets and the complex vector spaces of QM can be
facilitated by considering an intermediate stage. A power set $\wp\left(
U\right)  $ can be considered as a vector space over $%
%TCIMACRO{\U{2124} }%
%BeginExpansion
\mathbb{Z}
%EndExpansion
_{2}=\left\{  0,1\right\}  $ with the \textit{symmetric difference} of
subsets, i.e., $S\Delta T=S\cup T-S\cap T$ for $S,T\subseteq U$, as the vector
addition operation. Thus set concepts can be first translated into
sets-as-vectors concepts for vector spaces over $%
%TCIMACRO{\U{2124} }%
%BeginExpansion
\mathbb{Z}
%EndExpansion
_{2}$ and then lifted to vector spaces over $%
%TCIMACRO{\U{2102} }%
%BeginExpansion
\mathbb{C}
%EndExpansion
$ (or vice-versa for delifting). A vector in $%
%TCIMACRO{\U{2124} }%
%BeginExpansion
\mathbb{Z}
%EndExpansion
_{2}^{n}$ is specified in the $U$-basis $\left\{  \left\{  u_{1}\right\}
,\left\{  u_{2}\right\}  ,...,\left\{  u_{n}\right\}  \right\}  $ by its
characteristic function $\chi_{S}:U\rightarrow%
%TCIMACRO{\U{2124} }%
%BeginExpansion
\mathbb{Z}
%EndExpansion
_{2}$, and a vector $v$ in $%
%TCIMACRO{\U{2102} }%
%BeginExpansion
\mathbb{C}
%EndExpansion
^{n}$ is specified in terms of an orthonormal basis $\left\{  \left\vert
v_{i}\right\rangle \right\}  $ by a function $\left\langle \_|v\right\rangle
:\left\{  \left\vert v_{i}\right\rangle \right\}  \rightarrow%
%TCIMACRO{\U{2102} }%
%BeginExpansion
\mathbb{C}
%EndExpansion
$ assigning a complex amplitude $\left\langle v_{i}|v\right\rangle $ to each
basis vector. One of the key pieces of machinery in QM, namely the inner
product, does not exist in vector spaces over finite fields but a
basis-dependent "bracket" can be defined and a norm can be defined to play a
similar role in the probability algorithm of "quantum mechanics" on sets.

Seeing $\wp\left(  U\right)  $ as the vector space $%
%TCIMACRO{\U{2124} }%
%BeginExpansion
\mathbb{Z}
%EndExpansion
_{2}^{|U|}$ allows different bases in which the vectors can be expressed (as
well as the basis-free notion of a vector as a ket). Consider the simple case
of $U=\left\{  a,b,c\right\}  $ where the $U$-basis is $\left\{  a\right\}  $,
$\left\{  b\right\}  $, and $\left\{  c\right\}  $. But the three subsets
$\left\{  a,b\right\}  $, $\left\{  b,c\right\}  $, and $\left\{
a,b,c\right\}  $ also form a basis since: $\left\{  a,b\right\}  +\left\{
a,b,c\right\}  =\left\{  c\right\}  $; $\left\{  b,c\right\}  +\left\{
c\right\}  =\left\{  b\right\}  $; and $\left\{  a,b\right\}  +\left\{
b\right\}  =\left\{  a\right\}  $. These new basis vectors could be considered
as the basis-singletons in another equicardinal universe $U^{\prime}=\left\{
a^{\prime},b^{\prime},c^{\prime}\right\}  $ where $a^{\prime}=\left\{
a,b\right\}  $, $b^{\prime}=\left\{  b,c\right\}  $, and $c^{\prime}=\left\{
a,b,c\right\}  $. In the following \textit{ket table}, each row is a ket of
$V=%
%TCIMACRO{\U{2124} }%
%BeginExpansion
\mathbb{Z}
%EndExpansion
_{2}^{3}$ expressed in the $U$-basis and the $U^{\prime}$-basis.

\begin{center}%
\begin{tabular}
[c]{|c|c|}\hline
$U=\left\{  a,b,c\right\}  $ & $U^{\prime}=\left\{  a^{\prime},b^{\prime
},c^{\prime}\right\}  $\\\hline\hline
$\left\{  a,b,c\right\}  $ & $\left\{  c^{\prime}\right\}  $\\\hline
$\left\{  a,b\right\}  $ & $\left\{  a^{\prime}\right\}  $\\\hline
$\left\{  b,c\right\}  $ & $\left\{  b^{\prime}\right\}  $\\\hline
$\left\{  a,c\right\}  $ & $\left\{  a^{\prime},b^{\prime}\right\}  $\\\hline
$\left\{  a\right\}  $ & $\left\{  b^{\prime},c^{\prime}\right\}  $\\\hline
$\left\{  b\right\}  $ & $\left\{  a^{\prime},b^{\prime},c^{\prime}\right\}
$\\\hline
$\left\{  c\right\}  $ & $\left\{  a^{\prime},c^{\prime}\right\}  $\\\hline
$\emptyset$ & $\emptyset$\\\hline
\end{tabular}

Vector space isomorphism (i.e., preserves $+$) $%
%TCIMACRO{\U{2124} }%
%BeginExpansion
\mathbb{Z}
%EndExpansion
_{2}^{3}\cong\wp\left(  U\right)  \cong\wp\left(  U^{\prime}\right)  $: row = ket.
\end{center}

In a Hilbert space, the inner product is used to define the amplitudes
$\left\langle v_{i}|v\right\rangle $ and the norm $\left\Vert v\right\Vert
=\sqrt{\left\langle v|v\right\rangle }$, and the probability algorithm can be
formulated using this norm. In a vector space over $%
%TCIMACRO{\U{2124} }%
%BeginExpansion
\mathbb{Z}
%EndExpansion
_{2}$, the Dirac notation can still be used but in a basis-dependent form
(like matrices as opposed to operators) that defines a real-valued norm even
though there is no inner product. The kets $\left\vert S\right\rangle $ for
$S\in\wp\left(  U\right)  $ are basis-free but the corresponding bras are
basis-dependent. For $u\in U$, the "\textit{bra"} $\left\langle \left\{
u\right\}  \right\vert _{U}:\wp\left(  U\right)  \rightarrow%
%TCIMACRO{\U{211d} }%
%BeginExpansion
\mathbb{R}
%EndExpansion
$ is defined by the "\textit{bracket"}:

\begin{center}
$\left\langle \left\{  u\right\}  |_{U}S\right\rangle =\left\{
\begin{array}
[c]{c}%
1\text{ if }u\in S\\
0\text{ if }u\notin S
\end{array}
\right.  .$
\end{center}

\noindent Then $\left\langle \left\{  {}\right\}  |_{U}S\right\rangle
=\chi_{S}:U\rightarrow%
%TCIMACRO{\U{2124} }%
%BeginExpansion
\mathbb{Z}
%EndExpansion
_{2}$ is the delift of $\left\langle \_|v\right\rangle :\left\{  \left\vert
v_{i}\right\rangle \right\}  \rightarrow%
%TCIMACRO{\U{2102} }%
%BeginExpansion
\mathbb{C}
%EndExpansion
$. Assuming a finite $U$, the "bracket" linearly extends to the more general
basis-dependent form:

\begin{center}
$\left\langle T|_{U}S\right\rangle =\left\vert T\cap S\right\vert $ for
$T,S\subseteq U$.
\end{center}

\noindent Note that for $u,u^{\prime}\in U$, $\left\langle \left\{  u^{\prime
}\right\}  |_{U}\left\{  u\right\}  \right\rangle =\delta_{u^{\prime}u}$
taking the distinct elements of $U$ as being paired with the vectors in an
orthonormal basis in the lift-delift relationship. In fact, this delifting of
the Dirac bracket is motivated by the basis principle in reverse. Consider an
orthonormal basis set $\left\{  \left\vert v_{i}\right\rangle \right\}  $ in a
finite dimensional Hilbert space. Given two subsets $T,S\subseteq\left\{
\left\vert v_{i}\right\rangle \right\}  $, consider the unnormalized vector
$\psi_{T}=\sum_{\left\vert v_{i}\right\rangle \in T}\left\vert v_{i}%
\right\rangle $ and similarly for $\psi_{S}$. Then their inner product in the
Hilbert space is $\left\langle \psi_{T}|\psi_{S}\right\rangle =\left\vert
T\cap S\right\vert $, which "delifts" (running the basis principle in reverse)
to $\left\langle T|_{U}S\right\rangle =\left\vert T\cap S\right\vert $ for
subsets $T,S\subseteq U$.

The basis-dependent "\textit{ket-bra}" $\left\vert u\right\rangle \left\langle
u\right\vert _{U}:\wp\left(  U\right)  \rightarrow\wp\left(  U\right)  $ is
defined as the "one-dimensional" projection operator $\left\{  u\right\}
\cap():\wp\left(  U\right)  \rightarrow\wp\left(  U\right)  $ and the "ket-bra
identity" holds as usual:

\begin{center}
$\sum_{u\in U}\left\vert u\right\rangle \left\langle u\right\vert _{U}%
=\Delta_{u\in U}\left(  \left\{  u\right\}  \cap()\right)  =I:\wp\left(
U\right)  \rightarrow\wp\left(  U\right)  $
\end{center}

\noindent where the summation is the symmetric difference of sets in $%
%TCIMACRO{\U{2124} }%
%BeginExpansion
\mathbb{Z}
%EndExpansion
_{2}^{n}$.

Then the (basis-dependent) $U$\textit{-norm} $\left\Vert S\right\Vert _{U}%
:\wp\left(  U\right)  \rightarrow%
%TCIMACRO{\U{211d} }%
%BeginExpansion
\mathbb{R}
%EndExpansion
$ is defined, as usual, as the square root of the bracket:

\begin{center}
$\left\Vert S\right\Vert _{U}=\sqrt{\left\langle S|_{U}S\right\rangle }%
=\sqrt{|S|}$
\end{center}

\noindent for $S\in\wp\left(  U\right)  $ which is the delift of the
basis-free norm $\left\Vert \psi\right\Vert =\sqrt{\left\langle \psi
|\psi\right\rangle }$ (since the inner product does not depend on the basis).
Note that a ket has to be expressed in the $U$-basis to apply the
basis-dependent definition so in the above example, $\left\Vert \left\{
a^{\prime}\right\}  \right\Vert _{U}=\sqrt{2}$ since $\left\{  a^{\prime
}\right\}  =\left\{  a,b\right\}  $ in the $U$-basis.

For a specific basis $\left\{  \left\vert v_{i}\right\rangle \right\}  $ and
for any nonzero vector $v$ in a finite dimensional vector space, $\left\Vert
v\right\Vert =\sqrt{\sum_{i}\left\langle v_{i}|v\right\rangle \left\langle
v_{i}|v\right\rangle ^{\ast}}$ whose delifted version would be: $\left\Vert
S\right\Vert _{U}=\sqrt{\sum_{u\in U}\left\langle \left\{  u\right\}
|_{U}S\right\rangle ^{2}}$. Thus squaring both sides, we also have:

\begin{center}
$\sum_{i}\frac{\left\langle v_{i}|v\right\rangle \left\langle v_{i}%
|v\right\rangle ^{\ast}}{\left\Vert v\right\Vert ^{2}}=1$ and $\sum_{u}%
\frac{\left\langle \left\{  u\right\}  |_{U}S\right\rangle ^{2}}{\left\Vert
S\right\Vert _{U}^{2}}=\sum_{u}\frac{\left\vert \left\{  u\right\}  \cap
S\right\vert }{\left\vert S\right\vert }=1$
\end{center}

\noindent where $\frac{\left\langle v_{i}|v\right\rangle \left\langle
v_{i}|v\right\rangle ^{\ast}}{\left\Vert v\right\Vert ^{2}}$ is a `mysterious'
quantum probability while $\frac{\left\vert \left\{  u\right\}  \cap
S\right\vert }{\left\vert S\right\vert }$ is the unmysterious probability
$\Pr\left(  \left\{  u\right\}  |S\right)  $ of getting $u$ when sampling $S$
(equiprobable elements of $U$). We previously saw that a subset $S\subseteq U$
as a block in a partition could be interpreted as a single indefinite element
rather than a subset of definite elements. In like manner, we can interpret a
subset of outcomes (an event) in a finite probability space as a single
indefinite outcome where the conditional probability $\Pr\left(  \left\{
u\right\}  |S\right)  $ is the objective probability of a "$U$-measurement" of
$S$ yielding the definite outcome $\left\{  u\right\}  $.

An observable, i.e., a Hermitian operator, on a Hilbert space determines its
home basis set of orthonormal eigenvectors. In a similar manner, an attribute
$f:U\rightarrow%
%TCIMACRO{\U{211d} }%
%BeginExpansion
\mathbb{R}
%EndExpansion
$ defined on $U$ has the $U$-basis as its "home basis set." Then given a
Hermitian operator $L=\sum_{\lambda}\lambda P_{\lambda}$ and a $U$-attribute
$f:U\rightarrow%
%TCIMACRO{\U{211d} }%
%BeginExpansion
\mathbb{R}
%EndExpansion
$, we have:

\begin{center}
$\left\Vert v\right\Vert =\sqrt{\sum_{\lambda}\left\Vert P_{\lambda}\left(
v\right)  \right\Vert ^{2}}$ and $\left\Vert S\right\Vert _{U}=\sqrt{\sum
_{r}\left\Vert f^{-1}\left(  r\right)  \cap S\right\Vert _{U}^{2}}$
\end{center}

\noindent where $f^{-1}\left(  r\right)  \cap S$ is the "projection operator"
$f^{-1}\left(  r\right)  \cap\left(  {}\right)  $ applied to $S$, the delift
of applying the projection operator $P_{\lambda}$ to $v$.\footnote{Since
$\wp\left(  U\right)  $ is now interpreted as a vector space, it should be
noted that the projection operator $S\cap():\wp\left(  U\right)
\rightarrow\wp\left(  U\right)  $ is linear, i.e., $\left(  S\cap
S_{1}\right)  \Delta(S\cap S_{2})=S\cap\left(  S_{1}\Delta S_{2}\right)  $.
Indeed, this is the distributive law when $\wp\left(  U\right)  $ is
interpreted as a Boolean ring.} This can also be written as:

\begin{center}
$\sum_{\lambda}\frac{\left\Vert P_{\lambda}\left(  v\right)  \right\Vert ^{2}%
}{\left\Vert v\right\Vert ^{2}}=1$ and $\sum_{r}\frac{\left\Vert f^{-1}\left(
r\right)  \cap S\right\Vert _{U}^{2}}{\left\Vert S\right\Vert _{U}^{2}}%
=\sum_{r}\frac{\left\vert f^{-1}\left(  r\right)  \cap S\right\vert
}{\left\vert S\right\vert }=1$
\end{center}

\noindent where $\frac{\left\Vert P_{\lambda}\left(  v\right)  \right\Vert
^{2}}{\left\Vert v\right\Vert ^{2}}$ is the quantum probability of getting
$\lambda$ in an $L$-measurement of $v$ while $\frac{\left\vert f^{-1}\left(
r\right)  \cap S\right\vert }{\left\vert S\right\vert }$ has the rather
unmysterious interpretation of the probability $\Pr\left(  r|S\right)  $ of
the random variable $f$ $:U\rightarrow%
%TCIMACRO{\U{211d} }%
%BeginExpansion
\mathbb{R}
%EndExpansion
$ having the value $r$ when sampling $S\subseteq U$.\footnote{"Quantum
mechanics" on sets is just a noncommutative version of the logical finite
probability theory developed by Boole (finite set of equiprobable outcomes).}
Under the set version of the objective indefiniteness interpretation, i.e.,
"quantum mechanics" on sets, the indefinite element $S$ is being "measured"
using the "observable" $f$ and the probability $\Pr\left(  r|S\right)  $ of
getting the "eigenvalue" $r$ is $\frac{\left\vert f^{-1}\left(  r\right)  \cap
S\right\vert }{\left\vert S\right\vert }$ with the "projected resultant state"
as $f^{-1}\left(  r\right)  \cap S$.

These delifts are summarized in the following table for a finite $U$ and a
finite dimensional Hilbert space $V$.

\begin{center}%
\begin{tabular}
[c]{|c|c|}\hline
Set Case & Vector space case\\\hline\hline
Projection $B\cap\left(  {}\right)  :\wp\left(  U\right)  \rightarrow
\wp\left(  U\right)  $ & Projection $P:V\rightarrow V$\\\hline
$f\upharpoonright\left(  {}\right)  =\sum_{r}r\left(  f^{-1}\left(  r\right)
\cap\left(  {}\right)  \right)  $ & Herm. $L=\sum_{\lambda}\lambda P_{\lambda
}$\\\hline
$\Delta_{B\in\pi}B\cap\left(  {}\right)  =I:\wp\left(  U\right)
\rightarrow\wp\left(  U\right)  $ & $\sum_{\lambda}P_{\lambda}=I$\\\hline
$\left\langle S|_{U}T\right\rangle =\left\vert S\cap T\right\vert $ where
$S,T\subseteq U$ & $\left\langle \psi|\varphi\right\rangle =$ "overlap" of
$\psi$ and $\varphi$\\\hline
$\left\Vert S\right\Vert _{U}=\sqrt{\left\langle S|_{U}S\right\rangle }%
=\sqrt{\left\vert S\right\vert }$ where $S\subseteq U$ & $\left\Vert
\psi\right\Vert =\sqrt{\left\langle \psi|\psi\right\rangle }$\\\hline
$\left\Vert S\right\Vert _{U}=\sqrt{\sum_{u\in U}\left\langle \left\{
u\right\}  |_{U}S\right\rangle ^{2}}$ & $\left\Vert \psi\right\Vert
=\sqrt{\sum_{i}\left\langle v_{i}|\psi\right\rangle \left\langle v_{i}%
|\psi\right\rangle ^{\ast}}$\\\hline
$S\neq\emptyset$, $\sum_{u\in U}\frac{\left\langle \left\{  u\right\}
|_{U}S\right\rangle ^{2}}{\left\Vert S\right\Vert _{U}^{2}}=\sum_{u\in S}%
\frac{1}{\left\vert S\right\vert }=1$ & $\left\vert \psi\right\rangle \neq0$,
$\sum_{i}\frac{\left\langle v_{i}|\psi\right\rangle \left\langle v_{i}%
|\psi\right\rangle ^{\ast}}{\left\Vert \psi\right\Vert ^{2}}=1$\\\hline
$\left\Vert S\right\Vert _{U}=\sqrt{\sum_{r}\left\Vert f^{-1}\left(  r\right)
\cap S\right\Vert _{U}^{2}}$ & $\left\Vert \psi\right\Vert =\sqrt
{\sum_{\lambda}\left\Vert P_{\lambda}\left(  \psi\right)  \right\Vert ^{2}}%
$\\\hline
$S\neq\emptyset$, $\sum_{r}\frac{\left\Vert f^{-1}\left(  r\right)  \cap
S\right\Vert _{U}^{2}}{\left\Vert S\right\Vert _{U}^{2}}=\sum_{r}%
\frac{\left\vert f^{-1}\left(  r\right)  \cap S\right\vert }{\left\vert
S\right\vert }=1$ & $\left\vert \psi\right\rangle \neq0$, $\sum_{\lambda}%
\frac{\left\Vert P_{\lambda}\left(  \psi\right)  \right\Vert ^{2}}{\left\Vert
\psi\right\Vert ^{2}}=1$\\\hline
Given $S$, prob. of $r$ is $\frac{\left\Vert f^{-1}\left(  r\right)  \cap
S\right\Vert _{U}^{2}}{\left\Vert S\right\Vert _{U}^{2}}=\frac{\left\vert
f^{-1}\left(  r\right)  \cap S\right\vert }{\left\vert S\right\vert }$ & Given
$\psi$, prob. of $\lambda$ is $\frac{\left\Vert P_{\lambda}\left(
\psi\right)  \right\Vert ^{2}}{\left\Vert \psi\right\Vert ^{2}}$\\\hline
\end{tabular}

Demystifying quantum probabilities using "quantum mechanics" on sets
\end{center}

\subsection{Measurement in "QM" on sets}

Certainly the notion of measurement is one of the most opaque notions of QM so
let's consider a set version of (projective) measurement starting at some
block (the "state") in a partition in a partition lattice. In the simple
example illustrated below we start at the one block or "state" of the
indiscrete partition or blob which is the completely indistinct element
$\left\{  a,b,c\right\}  $. A measurement always uses some attribute that
defines an inverse-image partition on $U=\left\{  a,b,c\right\}  $. In the
case at hand, there are "essentially" four possible attributes that could be
used to "measure" the indefinite element $\left\{  a,b,c\right\}  $ (since
there are four partitions that refine the blob).

For an example of a "nondegenerate measurement," consider any attribute
$f:U\rightarrow%
%TCIMACRO{\U{211d} }%
%BeginExpansion
\mathbb{R}
%EndExpansion
$ which has the discrete partition as its inverse image, such as the ordinal
number of a letter in the alphabet: $f\left(  a\right)  =1$, $f\left(
b\right)  =2$, and $f\left(  c\right)  =3$. This attribute or "observable" has
three "eigenvectors": $f\upharpoonright\left\{  a\right\}  =1\left\{
a\right\}  $, $f\upharpoonright\left\{  b\right\}  =2\left\{  b\right\}  $,
and $f\upharpoonright\left\{  c\right\}  =3\left\{  c\right\}  $ with the
corresponding "eigenvalues." The "eigenspaces" in the inverse image are also
$\left\{  a\right\}  $, $\left\{  b\right\}  $, and $\left\{  c\right\}  $,
the blocks in the discrete partition of $U$ all of which have "dimension"
(i.e., cardinality) one. Starting in the "state" $S=\left\{  a,b,c\right\}  $,
a $U$-measurement with this observable would yield the "eigenvalue" $r$ with
the probability of $\Pr\left(  r|S\right)  =\frac{\left\vert f^{-1}\left(
r\right)  \cap S\right\vert }{\left\vert S\right\vert }=\frac{1}{3}$. A
"projective measurement" makes distinctions in the measured "state" that are
sufficient to induce the "quantum jump" or "projection" to the "eigenvector"
associated with the observed "eigenvalue." If the observed "eigenvalue" was
$3$, then the "state" $\left\{  a,b,c\right\}  $ "projects" to $f^{-1}\left(
3\right)  \cap\left\{  a,b,c\right\}  =\left\{  c\right\}  \cap\left\{
a,b,c\right\}  =\left\{  c\right\}  $ as pictured below.%

%TCIMACRO{\FRAME{dtbpF}{2.2324in}{1.4541in}{0in}{}{}{fig9-measpathnd.eps}%
%width 2.2324in;  height 1.4541in;  depth 0in;  original-width 2.151in;
%original-height 1.3921in;  cropleft "0";  croptop "1";  cropright "1";
%cropbottom "0";  filename 'fig9-measpathnd.eps';file-properties "XNPEU";}} }%
%BeginExpansion
\begin{center}
\includegraphics[
height=1.4541in,
width=2.2324in
]%
{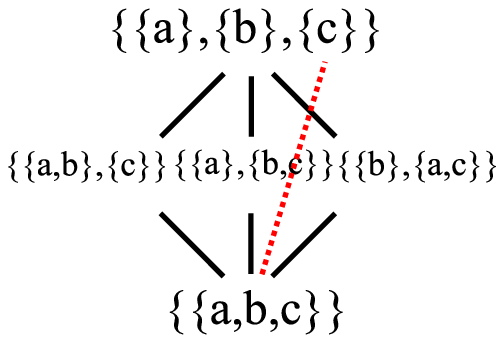}%
\end{center}
%EndExpansion

\begin{center}
Figure 9: "Nondegenerate measurement"
\end{center}

It might be emphasized that this is an objective state reduction (or "collapse
of the wave packet") from the single indefinite element $\left\{
a,b,c\right\}  $ to the single definite element $\left\{  c\right\}  $, not a
subjective removal of ignorance as if the "state" had all along been $\left\{
c\right\}  $. For instance, Pascual Jordan in 1934 argued that:

\begin{quotation}
\noindent the electron is forced to a decision. We compel it to assume a
definite position; previously, in general, it was neither here nor there; it
had not yet made its decision for a definite position... . ... [W]e ourselves
produce the results of the measurement. (quoted in \cite[p. 161]{jammer:phil})
\end{quotation}

It may be useful (for later purposes) to formulate this measurement using
density matrices and logical entropy. Given a partition $\pi=\left\{
B\right\}  $ on $U=\left\{  1,2,...,n\right\}  $. If the points had the
probabilities $p=\left\{  p_{1},...,p_{n}\right\}  ,$the $n\times n$ "density
matrix" $\rho^{\pi}$ representing $\pi$ has the entries:

\begin{center}
$\rho_{ij}^{\pi}=\left\{
\begin{array}
[c]{c}%
\sqrt{p_{i}p_{j}}\text{ if }\left(  i,j\right)  \in\operatorname*{indit}%
\left(  \pi\right)  \\
0\text{ if }\left(  i,j\right)  \in\operatorname*{dit}\left(  \pi\right)
\text{.}%
\end{array}
\right.  $
\end{center}

\noindent All the entries are "amplitudes" whose squares are probabilities.
Since the diagonal pairs $\left(  i,i\right)  $ are always indits of a
partition, the probabilities $p_{i}$ are the diagonal entries. To foreshadow
the quantum case, the non-zero off-diagonal entries $\sqrt{p_{i}p_{j}}$
indicate that $i$ and $j$ "cohere" together in a block of the partition. After
interchanging some rows and the corresponding columns, the density matrix
would be a block-diagonal matrix with the blocks corresponding to the blocks
$B$ of the partition $\pi$.

The \textit{quantum logical entropy} of a density matrix $\rho$ is: $h\left(
\rho\right)  =1-\operatorname*{tr}\left[  \rho^{2}\right]  $, and the logical
entropy of a set partition $\pi$ with point probabilities $p$ would be:
$h\left(  \rho^{\pi}\right)  =1-\operatorname*{tr}\left[  \left(  \rho^{\pi
}\right)  ^{2}\right]  $--which generalizes $h\left(  p\right)  =1-\sum
_{i=1}^{n}p_{i}^{2}$ where $\pi$ is the discrete partition on $U$. In the case
at hand, all the points are considered equiprobable with $p_{i}=\frac{1}{n}$.
Then a little calculation shows that: 

\begin{center}
$h\left(  \rho^{\pi}\right)  =1-\operatorname*{tr}\left[  \left(  \rho^{\pi
}\right)  ^{2}\right]  =1-\sum_{B\in\pi}p_{B}^{2}=h\left(  \pi\right)  $
\end{center}

\noindent where $p_{B}=\frac{\left\vert B\right\vert }{\left\vert U\right\vert
}$.

For $U=\left\{  a,b,c\right\}  $ (equiprobable points), the "density
matrix"(in the $U$-basis) of the indiscrete partition $\mathbf{0}$ is the
constant matrix:

\begin{center}
$\rho^{\mathbf{0}}=%
\begin{bmatrix}
\frac{1}{3} & \frac{1}{3} & \frac{1}{3}\\
\frac{1}{3} & \frac{1}{3} & \frac{1}{3}\\
\frac{1}{3} & \frac{1}{3} & \frac{1}{3}%
\end{bmatrix}
$. 
\end{center}

\noindent While all the entries are the same, they have quite different
meanings. The diagonal entries are the point probabilities. The off-diagonal
entries $\rho_{ij}^{\mathbf{0}}$ are the square roots $\sqrt{\frac{1}{3}%
\frac{1}{3}}$ of the probabilities of drawing the $\left(  i,j\right)  $ pair
(in two independent draws) if that pair is an indistinction of the partition,
and otherwise the entry is $0$. Since all pairs are indistinctions of the
indiscrete partition (i.e., cohere together in this indiscrete = "pure"
state), all the off-diagonal elements of $\rho^{\mathbf{0}}$ are $\sqrt
{\frac{1}{3}\frac{1}{3}}=\frac{1}{3}$. In general, $\operatorname*{tr}\left[
\left(  \rho^{\pi}\right)  ^{2}\right]  $ will be the probability of drawing
an indistinction of the partition, so $h\left(  \rho^{\pi}\right)  $ is the
probability of drawing a distinction of $\pi$.

Then, with computations in the reals, 

\begin{center}
$h\left(  \rho^{\mathbf{0}}\right)  =1-\operatorname*{tr}\left[  \left(
\rho^{\mathbf{0}}\right)  ^{2}\right]  =1-\left(  \frac{1}{3}+\frac{1}%
{3}+\frac{1}{3}\right)  =0$
\end{center}

\noindent as we expect since there are no distinctions in the indiscrete
partition (everything "coheres" together), and the interpretation of the
logical entropy is the probability of drawing a distinction.

In the above example of a non-degenerate measurement, we assumed a particular
outcome of the eigenvalue $3$ and the eigenstate $\left\{  c\right\}  $. But
the density matrix allows the more general formulation that the measurement
turns the "pure state" $\rho^{\mathbf{0}}$ into a "mixed state" $\hat{\rho
}^{\mathbf{0}}$ without specifying the particular outcome. Since the
measurement was non-degenerate, the "mixed state density matrix" is diagonal:

\begin{center}
$\hat{\rho}^{\mathbf{0}}=%
\begin{bmatrix}
\frac{1}{3} & 0 & 0\\
0 & \frac{1}{3} & 0\\
0 & 0 & \frac{1}{3}%
\end{bmatrix}
$
\end{center}

\noindent which indicates that the measurement could have produced $\left\{
a\right\}  $, $\left\{  b\right\}  $, or $\left\{  c\right\}  $ each with
probability $\frac{1}{3}$. Each of the off-diagonal terms was "decohered" by
the non-degenerate measurement so the post-measurement "amplitude" of $\left(
i,j\right)  $ still "cohering" is $0$. 

The general result is that the logical entropy increase resulting from a
measurement is the sum of the new distinction probabilities, which means the
squared amplitudes of the off-diagonal indistinction amplitudes that were
zeroed or "decohered" by the measurement. In this case, the six off-diagonal
amplitudes of $\frac{1}{3}$ were all zeroed so the change in logical entropy
is: $6\times\left(  \frac{1}{3}\right)  ^{2}=\frac{6}{9}=\frac{2}{3}$. Since
the initial logical entropy was $0$, that is also just the logical entropy of
the discrete partition: $h\left(  \mathbf{1}\right)  =1-\sum_{i=1}^{3}\left(
\frac{1}{3}\right)  ^{2}=1-\frac{3}{9}=\frac{2}{3}$.

For an example of a "degenerate measurement," we choose an attribute with a
non-discrete inverse-image partition such as $\left\{  \left\{  a\right\}
,\left\{  b,c\right\}  \right\}  $, which could, for instance, just be the
characteristic function $\chi_{\left\{  b,c\right\}  }$ with the two
"eigenspaces" $\left\{  a\right\}  $ and $\left\{  b,c\right\}  $ and the two
"eigenvalues" $0$ and $1$ respectively. Since one of the two "eigenspaces" is
not a singleton of an eigen-element, the "eigenvalue" of $1$ is a set version
of a "degenerate eigenvalue." This attribute $\chi_{\left\{  b,c\right\}  }$
has four "eigenvectors": $\chi_{\left\{  b,c\right\}  }\upharpoonright\left\{
b,c\right\}  =1\left\{  b,c\right\}  $, $\chi_{\left\{  b,c\right\}
}\upharpoonright\left\{  b\right\}  =1\left\{  b\right\}  $, $\chi_{\left\{
b,c\right\}  }\upharpoonright\left\{  c\right\}  =1\left\{  c\right\}  $, and
$\chi_{\left\{  b,c\right\}  }\upharpoonright\left\{  a\right\}  =0\left\{
a\right\}  $.

The "measuring apparatus" makes distinctions that further distinguishes the
indefinite element $S=\left\{  a,b,c\right\}  $ but the measurement returns
one of "eigenvalues" with certain probabilities:

\begin{center}
$\Pr(0|S)=\frac{\left\vert \left\{  a\right\}  \cap\left\{  a,b,c\right\}
\right\vert }{\left\vert \left\{  a,b,c\right\}  \right\vert }=\frac{1}{3}$
and $\Pr\left(  1|S\right)  =\frac{\left\vert \left\{  b,c\right\}
\cap\left\{  a,b,c\right\}  \right\vert }{\left\vert \left\{  a,b,c\right\}
\right\vert }=\frac{2}{3}$.
\end{center}

Suppose it returns the "eigenvalue" $1$. Then the indefinite element $\left\{
a,b,c\right\}  $ "jumps" to the "projection" $\chi_{\left\{  b,c\right\}
}^{-1}\left(  1\right)  \cap\left\{  a,b,c\right\}  =\left\{  b,c\right\}  $
of the "state" $\left\{  a,b,c\right\}  $ to that "eigenspace" \cite[p.
221]{cohen-t:QM1}.

In the density matrix treatment of this degenerate measurement, the result is
the mixed state with the density matrix corresponding to the partition
$\left\{  \left\{  a\right\}  ,\left\{  b,c\right\}  \right\}  $ which is:

\begin{center}
$\hat{\rho}^{\mathbf{0}}=%
\begin{bmatrix}
\frac{1}{3} & 0 & 0\\
0 & \frac{1}{3} & \frac{1}{3}\\
0 & \frac{1}{3} & \frac{1}{3}%
\end{bmatrix}
$.
\end{center}

\noindent The logical entropy increase is the sum of off-diagonal terms
squared that were zeroed or "decohered" by the measurement which in this case
is: $4\times\left(  \frac{1}{3}\right)  ^{2}=\frac{4}{9}$. This is also
obtained as:

\begin{center}
$h\left(  \hat{\rho}^{\mathbf{0}}\right)  =1-\operatorname*{tr}\left[  \left(
\hat{\rho}^{0}\right)  ^{2}\right]  =1-\operatorname*{tr}\allowbreak%
\begin{bmatrix}
\frac{1}{9} & 0 & 0\\
0 & \frac{2}{9} & \frac{2}{9}\\
0 & \frac{2}{9} & \frac{2}{9}%
\end{bmatrix}
=1-\frac{5}{9}=\frac{4}{9}$.
\end{center}

\noindent In simpler terms, the new dits created by the measurement that took
the indiscrete partition $\left\{  \left\{  a,b,c\right\}  \right\}  $ to the
partition $\left\{  \left\{  a\right\}  ,\left\{  b,c\right\}  \right\}  $ are
$\left\{  \left(  a,b\right)  ,\left(  a,c\right)  ,\left(  b,a\right)
,\left(  c,a\right)  \right\}  $ (which correspond to the four zeroed
off-diagonal terms) so the change in the normalized count of the dit sets is:

\begin{center}
$\frac{\left\vert \left\{  \left(  a,b\right)  ,\left(  a,c\right)  ,\left(
b,a\right)  ,\left(  c,a\right)  \right\}  \right\vert }{\left\vert U\times
U\right\vert }=\frac{4}{9}$.
\end{center}

Since this is a "degenerate" result (i.e., the "eigenspace" don't all have
"dimension" one), another measurement is needed to make more distinctions.
Measurements by attributes that give either of the other two middle
partitions, $\left\{  \left\{  a,b\right\}  ,\{c\right\}  \}$ or $\left\{
\left\{  b\right\}  ,\left\{  a,c\right\}  \right\}  $, suffice to distinguish
$\left\{  b,c\right\}  $ into $\left\{  b\right\}  $ or $\left\{  c\right\}
$, so either attribute together with the attribute $\chi_{\left\{
b,c\right\}  }$ would form a \textit{complete set of compatible attributes}
(i.e., the set version of a CSCO). The join of the two attributes' partitions
gives the discrete partition. Taking the other attribute as $\chi_{\left\{
a,b\right\}  }$, the join of the two attributes' "eigenspace" partitions is discrete:

\begin{center}
$\left\{  \left\{  a\right\}  ,\left\{  b,c\right\}  \right\}  \vee\left\{
\left\{  a,b\right\}  ,\{c\right\}  \}=\left\{  \left\{  a\right\}  ,\left\{
b\right\}  ,\left\{  c\right\}  \right\}  =\mathbf{1}$.
\end{center}

\noindent Hence all the singletons can be characterized by the ordered pairs
of the "eigenvalues" of these two attributes: $\left\{  a\right\}  =\left\vert
0,1\right\rangle $, $\left\{  b\right\}  =\left\vert 1,1\right\rangle $, and
$\left\{  c\right\}  =\left\vert 1,0\right\rangle $ (using Dirac's kets to
give the ordered pairs).

The second "projective measurement" of the indefinite "superposition" element
$\left\{  b,c\right\}  $ using the attribute $\chi_{\left\{  a,b\right\}  }$
with the "eigenspace" partition $\left\{  \left\{  a,b\right\}  ,\{c\right\}
\}$ would induce a jump to either $\left\{  b\right\}  $ or $\left\{
c\right\}  $ with the probabilities:

\begin{center}
$\Pr\left(  1|\left\{  b,c\right\}  \right)  =\frac{\left\vert \left\{
a,b\right\}  \cap\left\{  b,c\right\}  \right\vert }{\left\vert \left\{
b,c\right\}  \right\vert }=\frac{1}{2}$ and $\Pr\left(  0|\left\{
b,c\right\}  \right)  =\frac{\left\vert \left\{  c\right\}  \cap\left\{
b,c\right\}  \right\vert }{\left\vert \left\{  b,c\right\}  \right\vert
}=\frac{1}{2}$.
\end{center}

If the measured "eigenvalue" is $0$, then the "state" $\left\{  b,c\right\}  $
"projects" to $\chi_{\left\{  a,b\right\}  }^{-1}\left(  0\right)
\cap\left\{  b,c\right\}  =\left\{  c\right\}  $ as pictured below.%

%TCIMACRO{\FRAME{dtbpF}{2.2324in}{1.4541in}{0in}{}{}{fig10-measpath.eps}%
%width 2.2324in;  height 1.4541in;  depth 0in;  original-width 2.151in;
%original-height 1.3921in;  cropleft "0";  croptop "1";  cropright "1";
%cropbottom "0";  filename 'fig10-measpath.eps';file-properties "XNPEU";}} }%
%BeginExpansion
\begin{center}
\includegraphics[
height=1.4541in,
width=2.2324in
]%
{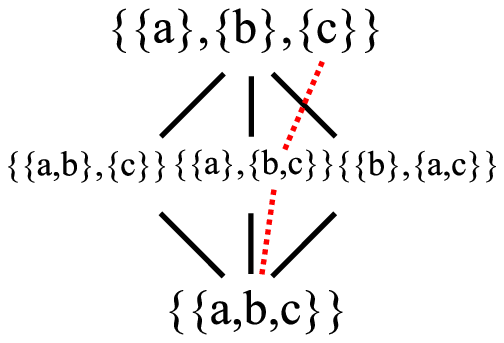}%
\end{center}
%EndExpansion

\begin{center}
Figure 10: "Degenerate measurement"
\end{center}

\noindent The two "projective measurements" of $\left\{  a,b,c\right\}  $
using the complete set of compatible (both defined on $U$) attributes
$\chi_{\left\{  b,c\right\}  }$ and $\chi_{\left\{  a,b\right\}  }$ produced
the respective "eigenvalues" $1$ and $0$, and the resulting "eigenstate" was
characterized by the "eigenket" $\left\vert 1,0\right\rangle =\{c\}$.

In terms of density matrices, the second measurement makes the following change:

\begin{center}
$%
\begin{bmatrix}
\frac{1}{3} & 0 & 0\\
0 & \frac{1}{3} & \frac{1}{3}\\
0 & \frac{1}{3} & \frac{1}{3}%
\end{bmatrix}
\Rightarrow%
\begin{bmatrix}
\frac{1}{3} & 0 & 0\\
0 & \frac{1}{3} & 0\\
0 & 0 & \frac{1}{3}%
\end{bmatrix}
$.
\end{center}

\noindent The off-diagonal terms zeroed by that measurement were the $\left(
b,c\right)  $ and $\left(  c,b\right)  $ terms (also the new dits) so the
increase in logical entropy is $2\times\left(  \frac{1}{3}\right)  ^{2}%
=\frac{2}{9}$, which can also be computed as:

\begin{center}
$\left\{  1-\operatorname*{tr}%
\begin{bmatrix}
\frac{1}{9} & 0 & 0\\
0 & \frac{1}{9} & 0\\
0 & 0 & \frac{1}{9}%
\end{bmatrix}
\right\}  -\left\{  1-\operatorname*{tr}\allowbreak%
\begin{bmatrix}
\frac{1}{9} & 0 & 0\\
0 & \frac{2}{9} & \frac{2}{9}\\
0 & \frac{2}{9} & \frac{2}{9}%
\end{bmatrix}
\right\}  =\left(  1-\frac{3}{9}\right)  -\left(  1-\frac{5}{9}\right)
=\frac{2}{9}$.
\end{center}

\noindent The normalized dit count in going from the indiscrete to the
discrete partition is same regardless of going all in one non-degenerate
measurement or in two or more steps: $\frac{2}{3}=\frac{4}{9}+\frac{2}{9}$.

In this manner, the toy model of "quantum mechanics" on sets provides a set
version of:

\begin{itemize}
\item a "nondegenerate measurement" by an "observable," 

\item "degenerate measurement" by "observables," 

\item "projections" associated with "eigenvalues" that "project" to
"eigenvectors," 

\item characterizations of "eigenvectors" by "eigenkets" of "eigenvalues" and

\item "density matrix" treatments of the "measurements" and the corresponding
changes in logical entropy.
\end{itemize}

\noindent This all shows the bare-bones logical structure of QM measurement in
the simple context of "QM" on sets.

\subsection{The indeterminacy principle in "QM" on sets}

Behind Heisenberg's indeterminacy principle, the basic idea (not the numerical
formula) is that a vector space can have quite different bases so that a ket
that is a definite state in one basis is an indefinite superposition in
another basis. And that basic idea can be well illustrated at the set level by
interpreting $\wp\left(  U\right)  $ as a vector space $%
%TCIMACRO{\U{2124} }%
%BeginExpansion
\mathbb{Z}
%EndExpansion
_{2}^{n}$ (where $\left\vert U\right\vert =n$) which has many bases. In our
previous (simplified) treatment of attributes $f:U\rightarrow%
%TCIMACRO{\U{211d} }%
%BeginExpansion
\mathbb{R}
%EndExpansion
$ and $g:U^{\prime}\rightarrow%
%TCIMACRO{\U{211d} }%
%BeginExpansion
\mathbb{R}
%EndExpansion
$ not using $%
%TCIMACRO{\U{2124} }%
%BeginExpansion
\mathbb{Z}
%EndExpansion
_{2}^{n}$, the attributes were compatible if $U=U^{\prime}$. Now we can give a
more sophisticated treatment of the set case using $%
%TCIMACRO{\U{2124} }%
%BeginExpansion
\mathbb{Z}
%EndExpansion
_{2}^{n}$, but with the similar result that attributes are compatible, i.e.,
"commute," if and only if there is a common basis set of "simultaneous
eigenvectors" on which both attributes can be defined. The lifted version is
the same; two observable operators are compatible if there is a basis of
simultaneous eigenvectors, and that holds if and only if the operators
commute--which is also equivalent to all the projection operators in the two
spectral decompositions commuting.

We are given two basis sets $\left\{  \left\{  a\right\}  ,\left\{  b\right\}
,...\mid a,b,...\in U\right\}  $ and $\left\{  \left\{  a^{\prime}\right\}
,\left\{  b^{\prime}\right\}  ,...\mid a^{\prime},b^{\prime},...\in U^{\prime
}\right\}  $ for $%
%TCIMACRO{\U{2124} }%
%BeginExpansion
\mathbb{Z}
%EndExpansion
_{2}^{n}$ such as in the previous example where $n=3$ and the $U^{\prime}%
$-basis was the three kets $\left\{  a^{\prime}\right\}  =\left\{
a,b\right\}  $, $\left\{  b^{\prime}\right\}  =\left\{  b,c\right\}  $, and
$\left\{  c^{\prime}\right\}  =\left\{  a,b,c\right\}  $. Then we have two
real-valued set attributes defined on the different bases, $f:U\rightarrow%
%TCIMACRO{\U{211d} }%
%BeginExpansion
\mathbb{R}
%EndExpansion
$ and $g:U^{\prime}\rightarrow%
%TCIMACRO{\U{211d} }%
%BeginExpansion
\mathbb{R}
%EndExpansion
$, and we want to investigate their compatibility.

The set attributes define set partitions $\left\{  f^{-1}\left(  r\right)
\right\}  $ and $\left\{  g^{-1}\left(  s\right)  \right\}  $ respectively on
$U$ and $U^{\prime}$. These set partitions on the basis sets define, as usual,
vector space partitions $\left\{  \wp\left(  f^{-1}\left(  r\right)  \right)
\right\}  $ and $\left\{  \wp\left(  g^{-1}\left(  s\right)  \right)
\right\}  $ on $%
%TCIMACRO{\U{2124} }%
%BeginExpansion
\mathbb{Z}
%EndExpansion
_{2}^{n}$. But those vector space partitions cannot in general be obtained as
the eigenspace partitions of Hermitian operators on $%
%TCIMACRO{\U{2124} }%
%BeginExpansion
\mathbb{Z}
%EndExpansion
_{2}^{n}$ since the only available eigenvalues are $0$ and $1$. But any set
attribute that is the characteristic function $\chi_{S}:U\rightarrow\left\{
0,1\right\}  \subseteq%
%TCIMACRO{\U{211d} }%
%BeginExpansion
\mathbb{R}
%EndExpansion
$ of a subset $S\subseteq U$ can represented by an operator, indeed a
projection operator, whose action on $\wp\left(  U\right)  \cong%
%TCIMACRO{\U{2124} }%
%BeginExpansion
\mathbb{Z}
%EndExpansion
_{2}^{n}$ is given by the "projection operator" $S\cap\left(  {}\right)
:\wp\left(  U\right)  \rightarrow\wp\left(  U\right)  $, and similarly for
$U^{\prime}$. The properties of the real-valued attributes $f$ and $g$ can
then stated in terms of these projection operators for subsets $S=f^{-1}%
\left(  r\right)  \subseteq U$ and $S^{\prime}=g^{-1}\left(  s\right)
\subseteq U^{\prime}$.

Consider first the above example and the simple case where the attributes are
just characteristic functions $f=\chi_{\left\{  b,c\right\}  }:U\rightarrow
\left\{  0,1\right\}  \subseteq%
%TCIMACRO{\U{211d} }%
%BeginExpansion
\mathbb{R}
%EndExpansion
$ so $f^{-1}\left(  1\right)  =\left\{  b,c\right\}  $ and $g=\chi_{\left\{
a^{\prime},b^{\prime}\right\}  }:U^{\prime}\rightarrow\left\{  0,1\right\}
\subseteq%
%TCIMACRO{\U{211d} }%
%BeginExpansion
\mathbb{R}
%EndExpansion
$ so $g^{-1}\left(  1\right)  =\left\{  a^{\prime},b^{\prime}\right\}  $. The
two projection operators are $\left\{  b,c\right\}  \cap\left(  {}\right)
:\wp\left(  U\right)  \rightarrow\wp\left(  U\right)  $ and $\left\{
a^{\prime},b^{\prime}\right\}  \cap\left(  {}\right)  :\wp\left(  U^{\prime
}\right)  \rightarrow\wp\left(  U^{\prime}\right)  $. Note that this
representation of the projection operators is basis-dependent. For instance,
$\left\{  a^{\prime},b^{\prime}\right\}  =\left\{  a,c\right\}  $ but the
operator $\left\{  a,c\right\}  \cap\left(  {}\right)  $ operating on
$\wp\left(  U\right)  $ is a very different operator than $\left\{  a^{\prime
},b^{\prime}\right\}  \cap()$ operating on $\wp\left(  U^{\prime}\right)  $.
The following ket table computes the two projection operators and checks if
they commute.

\begin{center}%
\begin{tabular}
[c]{|c|c|c|c|c|c|}\hline
$U$ & $U^{\prime}$ & $f\upharpoonright=\left\{  b,c\right\}  \cap()$ &
$g\upharpoonright=\left\{  a^{\prime},b^{\prime}\right\}  \cap()$ &
$g\upharpoonright f\upharpoonright$ & $f\upharpoonright g\upharpoonright
$\\\hline\hline
$\left\{  a,b,c\right\}  $ & $\left\{  c^{\prime}\right\}  $ & $\left\{
b,c\right\}  $ & $\emptyset$ & $\left\{  b,c\right\}  $ & $\emptyset$\\\hline
$\left\{  a,b\right\}  $ & $\left\{  a^{\prime}\right\}  $ & $\left\{
b\right\}  $ & $\left\{  a^{\prime}\right\}  =\left\{  a,b\right\}  $ &
$\left\{  a,c\right\}  $ & $\left\{  b\right\}  $\\\hline
$\left\{  b,c\right\}  $ & $\left\{  b^{\prime}\right\}  $ & $\left\{
b,c\right\}  $ & $\left\{  b^{\prime}\right\}  =\left\{  b,c\right\}  $ &
$\left\{  b,c\right\}  $ & $\left\{  b,c\right\}  $\\\hline
$\left\{  a,c\right\}  $ & $\left\{  a^{\prime},b^{\prime}\right\}  $ &
$\left\{  c\right\}  $ & $\left\{  a^{\prime},b^{\prime}\right\}  =\left\{
a,c\right\}  $ & $\left\{  a,b\right\}  $ & $\left\{  c\right\}  $\\\hline
$\left\{  a\right\}  $ & $\left\{  b^{\prime},c^{\prime}\right\}  $ &
$\emptyset$ & $\left\{  b^{\prime}\right\}  =\left\{  b,c\right\}  $ &
$\emptyset$ & $\left\{  b,c\right\}  $\\\hline
$\left\{  b\right\}  $ & $\left\{  a^{\prime},b^{\prime},c^{\prime}\right\}  $
& $\left\{  b\right\}  $ & $\left\{  a^{\prime},b^{\prime}\right\}  =\left\{
a,c\right\}  $ & $\left\{  a,c\right\}  $ & $\left\{  a,c\right\}  $\\\hline
$\left\{  c\right\}  $ & $\left\{  a^{\prime},c^{\prime}\right\}  $ &
$\left\{  c\right\}  $ & $\left\{  a^{\prime}\right\}  =\left\{  a,b\right\}
$ & $\left\{  a,b\right\}  $ & $\left\{  b\right\}  $\\\hline
$\emptyset$ & $\emptyset$ & $\emptyset$ & $\emptyset$ & $\emptyset$ &
$\emptyset$\\\hline
\end{tabular}

Non-commutativity of the projections $\left\{  b,c\right\}  \cap\left(
{}\right)  $ and $\left\{  a^{\prime},b^{\prime}\right\}  \cap\left(
{}\right)  $.
\end{center}

We can move even closer to QM mathematics by using matrices in $%
%TCIMACRO{\U{2124} }%
%BeginExpansion
\mathbb{Z}
%EndExpansion
_{2}^{n}$ to represent the operators. The $U$-basis vectors $\left\{
a\right\}  $, $\left\{  b\right\}  $, and $\left\{  c\right\}  $ are
represented by the respective column vectors:

\begin{center}
$%
\begin{bmatrix}
1\\
0\\
0
\end{bmatrix}
_{U}$, $%
\begin{bmatrix}
0\\
1\\
0
\end{bmatrix}
_{U}$, and $%
\begin{bmatrix}
0\\
0\\
1
\end{bmatrix}
_{U}$
\end{center}

\noindent where the subscripts indicate the basis. The projection operator
$\left\{  b,c\right\}  \cap\left(  {}\right)  $ would be represented by the
matrix whose columns give the result of applying the operator to the basis vectors:

\begin{center}
$%
\begin{bmatrix}
0 & 0 & 0\\
0 & 1 & 0\\
0 & 0 & 1
\end{bmatrix}
_{U}$

$\left\{  b,c\right\}  \cap\left(  {}\right)  $ projection matrix in $U$-basis.
\end{center}

In the $U^{\prime}$-basis (with the corresponding basis vectors using the
$U^{\prime}$ subscript), the $\left\{  a^{\prime},b^{\prime}\right\}
\cap\left(  {}\right)  $ projection operator is represented by the projection matrix:

\begin{center}
$%
\begin{bmatrix}
1 & 0 & 0\\
0 & 1 & 0\\
0 & 0 & 0
\end{bmatrix}
_{U^{\prime}}$

$\left\{  a^{\prime},b^{\prime}\right\}  \cap\left(  {}\right)  $ projection
matrix in $U^{\prime}$-basis.
\end{center}

These matrices cannot be meaningfully multiplied since they are in different
bases but we can convert them into the same basis to see if they commute.
Since $\left\{  a^{\prime}\right\}  =\left\{  a,b\right\}  $, $\left\{
b^{\prime}\right\}  =\left\{  b,c\right\}  $, and $\left\{  c^{\prime
}\right\}  =\left\{  a,b,c\right\}  $, the conversion matrix $\mathcal{C}%
_{U\leftarrow U^{\prime}}$ to convert $U^{\prime}$-basis vectors to $U$-basis
vectors is given by the entries such as $\left\langle \left\{  a\right\}
|_{U}\left\{  a^{\prime}\right\}  \right\rangle =1$:

\begin{center}
$\mathcal{C}_{U\leftarrow U^{\prime}}=%
\begin{bmatrix}
\left\langle \left\{  a\right\}  |_{U}\left\{  a^{\prime}\right\}
\right\rangle  & \left\langle \left\{  a\right\}  |_{U}\left\{  b^{\prime
}\right\}  \right\rangle  & \left\langle \left\{  a\right\}  |_{U}\left\{
c^{\prime}\right\}  \right\rangle \\
\left\langle \left\{  b\right\}  |_{U}\left\{  a^{\prime}\right\}
\right\rangle  & \left\langle \left\{  b\right\}  |_{U}\left\{  b^{\prime
}\right\}  \right\rangle  & \left\langle \left\{  b\right\}  |_{U}\left\{
c^{\prime}\right\}  \right\rangle \\
\left\langle \left\{  c\right\}  |_{U}\left\{  a^{\prime}\right\}
\right\rangle  & \left\langle \left\{  c\right\}  |_{U}\left\{  b^{\prime
}\right\}  \right\rangle  & \left\langle \left\{  c\right\}  |_{U}\left\{
c^{\prime}\right\}  \right\rangle
\end{bmatrix}
=%
\begin{bmatrix}
1 & 0 & 1\\
1 & 1 & 1\\
0 & 1 & 1
\end{bmatrix}
_{U\leftarrow U^{\prime}}$.
\end{center}

The conversion the other way is given by the inverse matrix (remember
$\operatorname{mod}\left(  2\right)  $ arithmetic):

\begin{center}
$\mathcal{C}_{U^{\prime}\leftarrow U}=%
\begin{bmatrix}
0 & 1 & 1\\
1 & 1 & 0\\
1 & 1 & 1
\end{bmatrix}
_{U^{\prime}\leftarrow U}=\mathcal{C}_{U\leftarrow U^{\prime}}^{-1}$
\end{center}

\noindent which could also be directly seen from the ket table since $\left\{
a\right\}  =\left\{  b^{\prime},c^{\prime}\right\}  $, $\left\{  b\right\}
=\left\{  a^{\prime},b^{\prime},c^{\prime}\right\}  $, and $\left\{
c\right\}  =\left\{  a^{\prime},c^{\prime}\right\}  $.

The projection matrix for $\left\{  a^{\prime},b^{\prime}\right\}  \cap()$ in
the $U^{\prime}$-basis can be converted to the $U$-basis by computing the
matrix that starting with any $U$-basis vector will convert it to the
$U^{\prime}$-basis, then apply the projection matrix in that $U^{\prime}%
$-basis and then convert the result back to the $U$-basis:

\begin{center}
$\mathcal{C}_{U\leftarrow U^{\prime}}%
\begin{bmatrix}
1 & 0 & 0\\
0 & 1 & 0\\
0 & 0 & 0
\end{bmatrix}
_{U^{\prime}}\mathcal{C}_{U^{\prime}\leftarrow U}$

$=%
\begin{bmatrix}
1 & 0 & 1\\
1 & 1 & 1\\
0 & 1 & 1
\end{bmatrix}
_{U\leftarrow U^{\prime}}%
\begin{bmatrix}
1 & 0 & 0\\
0 & 1 & 0\\
0 & 0 & 0
\end{bmatrix}
_{U^{\prime}}%
\begin{bmatrix}
0 & 1 & 1\\
1 & 1 & 0\\
1 & 1 & 1
\end{bmatrix}
_{U^{\prime}\leftarrow U}$

$=\allowbreak%
\begin{bmatrix}
0 & 1 & 1\\
1 & 0 & 1\\
1 & 1 & 0
\end{bmatrix}
_{U}$

$\left\{  a^{\prime},b^{\prime}\right\}  \cap()$ projection operator in the
$U$-basis.
\end{center}

Now the two projection operators are represented as projection matrices in the
same $U$-basis so they can be multiplied to see if they commute:

\begin{center}
$g\upharpoonright f\upharpoonright\left(  {}\right)  =%
\begin{bmatrix}
0 & 1 & 1\\
1 & 0 & 1\\
1 & 1 & 0
\end{bmatrix}
_{U}%
\begin{bmatrix}
0 & 0 & 0\\
0 & 1 & 0\\
0 & 0 & 1
\end{bmatrix}
_{U}=\allowbreak%
\begin{bmatrix}
0 & 1 & 1\\
0 & 0 & 1\\
0 & 1 & 0
\end{bmatrix}
_{U}$

$f\upharpoonright g\upharpoonright\left(  {}\right)  =%
\begin{bmatrix}
0 & 0 & 0\\
0 & 1 & 0\\
0 & 0 & 1
\end{bmatrix}
_{U}%
\begin{bmatrix}
0 & 1 & 1\\
1 & 0 & 1\\
1 & 1 & 0
\end{bmatrix}
_{U}=\allowbreak%
\begin{bmatrix}
0 & 0 & 0\\
1 & 0 & 1\\
1 & 1 & 0
\end{bmatrix}
_{U}$
\end{center}

\noindent so the two projection matrices do not commute, as we previously saw
in the table computation.

There is a standard theorem of linear algebra:

\begin{proposition}
For two diagonalizable (i.e., eigenvectors span the space) linear operators on
a finite dimensional space: the operators commute if and only if there is a
basis of simultaneous eigenvectors \cite[p. 177]{hk:la}.
\end{proposition}

In the above example of non-commuting projection operators, there is no basis
of simultaneous eigenvectors (in fact $\left\{  b,c\right\}  =\left\{
b^{\prime}\right\}  $ is the only common eigenvector).

In the following example of a third $U^{\prime\prime}$-basis where
$U^{\prime\prime}=\left\{  a^{\prime\prime},b^{\prime\prime},c^{\prime\prime
}\right\}  $ with the set attributes $f=\chi_{\left\{  b,c\right\}
}:U\rightarrow\left\{  0,1\right\}  $ and $g=\chi_{\left\{  a^{\prime\prime
},b^{\prime\prime}\right\}  }:U^{\prime\prime}\rightarrow\left\{  0,1\right\}
$, the projections $\left\{  b,c\right\}  \cap()$ and $\left\{  a^{\prime
\prime},b^{\prime\prime}\right\}  \cap()$ commute as we see from the last two columns.

\begin{center}%
\begin{tabular}
[c]{|c|c|c|c|c|c|}\hline
$U$ & $U^{\prime\prime}$ & $f\upharpoonright=\left\{  b,c\right\}  \cap()$ &
$g\upharpoonright=\left\{  a^{\prime\prime},b^{\prime\prime}\right\}  \cap()$
& $g\upharpoonright f\upharpoonright$ & $f\upharpoonright g\upharpoonright
$\\\hline\hline
$\left\{  a,b,c\right\}  $ & $\left\{  a^{\prime\prime},b^{\prime\prime
},c^{\prime\prime}\right\}  $ & $\left\{  b,c\right\}  $ & $\left\{
a^{\prime\prime},b^{\prime\prime}\right\}  =\left\{  a,c\right\}  $ &
$\left\{  c\right\}  $ & $\left\{  c\right\}  $\\\hline
$\left\{  a,b\right\}  $ & $\left\{  b^{\prime\prime},c^{\prime\prime
}\right\}  $ & $\left\{  b\right\}  $ & $\left\{  b^{\prime\prime}\right\}
=\{a\}$ & $\emptyset$ & $\emptyset$\\\hline
$\left\{  b,c\right\}  $ & $\left\{  a^{\prime\prime},c^{\prime\prime
}\right\}  $ & $\left\{  b,c\right\}  $ & $\left\{  a^{\prime\prime}\right\}
=\left\{  c\right\}  $ & $\left\{  c\right\}  $ & $\left\{  c\right\}
$\\\hline
$\left\{  a,c\right\}  $ & $\left\{  a^{\prime\prime},b^{\prime\prime
}\right\}  $ & $\left\{  c\right\}  $ & $\left\{  a^{\prime\prime}%
,b^{\prime\prime}\right\}  =\left\{  a,c\right\}  $ & $\left\{  c\right\}  $ &
$\left\{  c\right\}  $\\\hline
$\left\{  a\right\}  $ & $\left\{  b^{\prime\prime}\right\}  $ & $\emptyset$ &
$\left\{  b^{\prime\prime}\right\}  =\{a\}$ & $\emptyset$ & $\emptyset
$\\\hline
$\left\{  b\right\}  $ & $\left\{  c^{\prime\prime}\right\}  $ & $\left\{
b\right\}  $ & $\emptyset$ & $\emptyset$ & $\emptyset$\\\hline
$\left\{  c\right\}  $ & $\left\{  a^{\prime\prime}\right\}  $ & $\left\{
c\right\}  $ & $\left\{  a^{\prime\prime}\right\}  =\left\{  c\right\}  $ &
$\left\{  c\right\}  $ & $\left\{  c\right\}  $\\\hline
$\emptyset$ & $\emptyset$ & $\emptyset$ & $\emptyset$ & $\emptyset$ &
$\emptyset$\\\hline
\end{tabular}

Commuting projection operators $\left\{  b,c\right\}  \cap()$ and $\left\{
a^{\prime\prime},b^{\prime\prime}\right\}  \cap()$.
\end{center}

\noindent Hence in this case, there is a basis of simultaneous eigenvectors
$\left\{  a\right\}  =\left\{  b^{\prime\prime}\right\}  $, $\left\{
b\right\}  =\left\{  c^{\prime\prime}\right\}  $, and $\left\{  c\right\}
=\left\{  a^{\prime\prime}\right\}  $, so that $f$ and $g$ are defined on the
same set (which we could take to be either $U$ or $U^{\prime\prime}$).

Returning to the two basis sets $\left\{  \left\{  a\right\}  ,\left\{
b\right\}  ,...\mid a,b,...\in U\right\}  $ and $\left\{  \left\{  a^{\prime
}\right\}  ,\left\{  b^{\prime}\right\}  ,...\mid a^{\prime},b^{\prime},...\in
U^{\prime}\right\}  $ for $%
%TCIMACRO{\U{2124} }%
%BeginExpansion
\mathbb{Z}
%EndExpansion
_{2}^{n}$ with two real-valued set attributes $f:U\rightarrow%
%TCIMACRO{\U{211d} }%
%BeginExpansion
\mathbb{R}
%EndExpansion
$ and $g:U^{\prime}\rightarrow%
%TCIMACRO{\U{211d} }%
%BeginExpansion
\mathbb{R}
%EndExpansion
$, the attributes cannot be represented as operators on $%
%TCIMACRO{\U{2124} }%
%BeginExpansion
\mathbb{Z}
%EndExpansion
_{2}^{n}$ but each block $f^{-1}\left(  r\right)  $ and $g^{-1}\left(
s\right)  $ can be analyzed using the projection operators $f^{-1}\left(
r\right)  \cap()$ and $g^{-1}\left(  s\right)  \cap()$ for those subsets. Thus
instead of the criterion of operators commuting, we define that attributes $f$
and $g$ \textit{"commute"} if all their projection operators $f^{-1}\left(
r\right)  \cap()$ and $g^{-1}\left(  s\right)  \cap()$ commute. Then the above
proposition about commuting operators can be applied to the commuting
operators to yield the result: 

\begin{quote}
set attributes $f:U\rightarrow%
%TCIMACRO{\U{211d} }%
%BeginExpansion
\mathbb{R}
%EndExpansion
$ and $g:U^{\prime}\rightarrow%
%TCIMACRO{\U{211d} }%
%BeginExpansion
\mathbb{R}
%EndExpansion
$ "commute" if and only if they are \textit{compatible} in the sense that
there is a basis set $\left\{  \left\{  a^{\prime\prime}\right\}  ,\left\{
b^{\prime\prime}\right\}  ,...\right\}  $ for $%
%TCIMACRO{\U{2124} }%
%BeginExpansion
\mathbb{Z}
%EndExpansion
_{2}^{n}$ whose subsets (vectors) are "simultaneous eigenvectors" for all the
projection operators--so that $f$ and $g$ can be taken as being defined on the
same basis set of $n$ vectors. 
\end{quote}

\noindent This result also justifies our earlier simplification that $f$ and
$g$ were defined as compatible if they were defined on the same set
$U=U^{\prime}$.

If the two set attributes $f$ and $g$ could be defined on the same set, then
they could have definite values at the same time, and that holds if and only
if the attributes "commute." But in the non-commutative case, $f$ and $g$
cannot always have definite values in any state. A definite value for one
means an indefinite value for the other. In the first example, we have
$f=\chi_{\left\{  b,c\right\}  }$ and $g=\chi_{\left\{  a^{\prime},b^{\prime
}\right\}  }$ so, for example, in the state $\left\{  c\right\}  =\left\{
a^{\prime},c^{\prime}\right\}  $, $f$ has the definite value $f\left(
c\right)  =1$ while $g$ is indefinite between the values of $g\left(
a^{\prime}\right)  =1$ and $g\left(  c^{\prime}\right)  =0$. In this manner,
we see how the essential points (but not the numerical formulas) of
Heisenberg's indeterminacy principle, i.e., when two observables can or cannot
have simultaneous definite values, are evidenced in the model of "quantum
mechanics" on sets.

\subsection{Entanglement in "quantum mechanics" on sets}

Another QM concept that also generates much mystery is entanglement. Hence it
might be useful to consider entanglement in "quantum mechanics" on sets.

First we need to lift the set notion of the direct (or Cartesian) product
$X\times Y$ of two sets $X$ and $Y$. Using the basis principle, we apply the
set concept to the two basis sets $\left\{  v_{1},...,v_{m}\right\}  $ and
$\left\{  w_{1},...,w_{n}\right\}  $ of two vector spaces $V$ and $W$ (over
the same base field) and then we see what it generates. The set direct product
of the two basis sets is the set of all ordered pairs $\left(  v_{i}%
,w_{j}\right)  $, which we will write as $v_{i}\otimes w_{j}$, and then we
generate the vector space, denoted $V\otimes W$, over the same base field from
those basis elements $v_{i}\otimes w_{j}$. That vector space is the
\textit{tensor product}, and it not the direct product $V\times W$ of the
vector spaces. The cardinality of $X\times Y$ is the product of the
cardinalities of the two sets, and the dimension of the tensor product
$V\otimes W$ is the product of the dimensions of the two spaces (while the
dimension of the direct product $V\times W$ is the sum of the two dimensions).

A vector $z\in V\otimes W$ is said to be \textit{separated} if there are
vectors $v\in V$ and $w\in W$ such that $z=v\otimes w$; otherwise, $z$ is said
to be \textit{entangled}. Since vectors delift to subsets, a subset
$S\subseteq X\times Y$ is said to be \textit{"separated"} or a
\textit{product} if there exists subsets $S_{X}\subseteq X$ and $S_{Y}%
\subseteq Y$ such that $S=S_{X}\times S_{Y}$; otherwise $S\subseteq X\times Y$
is said to be \textit{"entangled."} In general, let $S_{X}$ be the support or
projection of $S$ on $X$, i.e., $S_{X}=\left\{  x:\exists y\in Y,\left(
x,y\right)  \in S\right\}  $ and similarly for $S_{Y}$. Then $S$ is
"separated" iff $S=S_{X}\times S_{Y}$.

For any subset $S\subseteq X\times Y$, where $X$ and $Y$ are finite sets, a
natural measure of its "entanglement" can be constructed by first viewing $S$
as the support of the equiprobable or Laplacian joint probability distribution
on $S$. If $\left\vert S\right\vert =N$, then define $\Pr\left(  x,y\right)
=\frac{1}{N}$ if $\left(  x,y\right)  \in S$ and $\Pr\left(  x,y\right)  =0$ otherwise.

The marginal distributions\footnote{The marginal distributions are the set
versions of the reduced density matrices of QM.} are defined in the usual way:

\begin{center}
$\Pr\left(  x\right)  =\sum_{y}\Pr\left(  x,y\right)  $

$\Pr\left(  y\right)  =\sum_{x}\Pr\left(  x,y\right)  $.
\end{center}

\noindent A joint probability distribution $\Pr\left(  x,y\right)  $ on
$X\times Y$ is \textit{independent} if for all $\left(  x,y\right)  \in
X\times Y$,

\begin{center}
$\Pr\left(  x,y\right)  =\Pr\left(  x\right)  \Pr\left(  y\right)  $.

Independent distribution
\end{center}

\noindent Otherwise $\Pr\left(  x,y\right)  $ is said to be
\textit{correlated}.

\begin{proposition}
A subset $S\subseteq X\times Y$ is "entangled" iff the equiprobable
distribution on $S$ is correlated.
\end{proposition}

Proof: If $S$ is "separated", i.e., $S=S_{X}\times S_{Y}$, then $\Pr\left(
x\right)  =|S_{Y}|/N$ for $x\in S_{X}$ and $\Pr\left(  y\right)  =\left\vert
S_{X}\right\vert /N$ for $y\in S_{Y}$ where $\left\vert S_{X}\right\vert
\left\vert S_{Y}\right\vert =N$. Then for $\left(  x,y\right)  \in S$,

\begin{center}
$\Pr\left(  x,y\right)  =\frac{1}{N}=\frac{N}{N^{2}}=\frac{\left\vert
S_{X}\right\vert \left\vert S_{Y}\right\vert }{N^{2}}=\Pr\left(  x\right)
\Pr\left(  y\right)  $
\end{center}

\noindent and $\Pr(x,y)=0=\Pr\left(  x\right)  \Pr\left(  y\right)  $ for
$\left(  x,y\right)  \notin S$ so the equiprobable distribution is
independent. If $S$ is "entangled," i.e., $S\neq S_{X}\times S_{Y}$, then
$S\subsetneqq S_{X}\times S_{Y}$ so let $\left(  x,y\right)  \in S_{X}\times
S_{Y}-S$. Then $\Pr\left(  x\right)  ,\Pr\left(  y\right)  >0$ but $\Pr\left(
x,y\right)  =0$ so it is not independent, i.e., is correlated. $\square$

Consider the set version of one qubit space where $U=\left\{  a,b\right\}  $.
The product set $U\times U$ has $15$ nonempty subsets. Each factor $U$ and $U$
has $3$ nonempty subsets so $3\times3=9$ of the $15$ subsets are "separated"
subsets leaving $6$ "entangled" subsets.

\begin{center}%
\begin{tabular}
[c]{|c|}\hline
$S\subseteq U\times U$\\\hline\hline
$\left\{  \left(  a,a\right)  ,\left(  b,b\right)  \right\}  $\\\hline
$\left\{  \left(  a,b\right)  ,\left(  b,a\right)  \right\}  $\\\hline
$\left\{  \left(  a,a\right)  ,(a,b),\left(  b,a\right)  \right\}  $\\\hline
$\left\{  \left(  a,a\right)  ,(a,b),\left(  b,b\right)  \right\}  $\\\hline
$\left\{  (a,b),\left(  b,a\right)  ,\left(  b,b\right)  \right\}  $\\\hline
$\left\{  (a,a),\left(  b,a\right)  ,\left(  b,b\right)  \right\}  $\\\hline
\end{tabular}

The six entangled subsets
\end{center}

The first two are the "Bell states" which are the two graphs of bijections
$U\longleftrightarrow U$ and have the maximum entanglement if entanglement is
measured by the logical divergence $d\left(  \Pr(x,y)||\Pr\left(  x\right)
\Pr\left(  y\right)  \right)  $\cite{ell:distinctions}. All the $9$
"separated" states have zero "entanglement" by the same measure.

For an "entangled" subset $S$, a sampling $x$ of left-hand system will change
the probability distribution for a sampling of the right-hand system $y$,
$\Pr\left(  y|x\right)  \neq\Pr\left(  y\right)  $. In the case of maximal
"entanglement" (e.g., the "Bell states"), when $S$ is the graph of a bijection
between $U$ and $U$, the value of $y$ is determined by the value of $x$ (and vice-versa).

In this manner, we see that many of the basic ideas and relationships of
quantum mechanical entanglement (e.g., "entangled states," "reduced density
matrices," maximally "entangled states," and "Bell states"), can be reproduced
in "quantum mechanics" on sets.

The two-slit experiment and the Bell inequality for "quantum mechanics" on
sets are developed in Appendices 2 and 3.

\section{Waving good-by to waves}

\subsection{Wave-particle duality = indistinct-distinct particle duality}

States that are indistinct for an observable are represented as weighted
vector sums or superpositions of the eigenstates that might be actualized by
further distinctions. This indistinctness-represented-as-superpositions is
usually interpreted as "wave-like aspects" of the particles in the indefinite
state. Hence the distinction-making measurements take away the
indistinctness--which is usually interpreted as taking away the "wave-like
aspects," i.e., "collapse of the wave packet." But there are no actual
physical waves in quantum mechanics (e.g., the "wave amplitudes" are complex
numbers); only particles with indistinct attributes for certain observables.
Thus the "collapse of the wave packet" is better described as the "collapse of
indefiniteness" to achieve definiteness. And the "wave-particle duality" is
actually the \textit{indistinct-distinct particle duality }or\textit{
complementarity}.

We have provided the back-story to objective indefiniteness by building the
notion of distinctions from the ground up starting with partition logic and
logical information theory. But the importance of distinctions and
indistinguishability has been there all along in quantum mechanics.

Consider the standard double-slit experiment. When there is no distinction
between the two slits, then the position attribute of the traversing particle
is indefinite, neither top slit nor bottom slit (not "going through both
slits"), which is usually interpreted as the "wave-like aspects" that show
interference. But when a distinction is made between the slits, e.g.,
inserting a detector in one slit or closing one slit, then the distinction
reduces the indefiniteness to definiteness so the indefiniteness disappears,
i.e., the "wave-like aspects" disappear. For instance, Feynman makes this
point about distinctions in terms of \textit{distinguishing} the alternative
final states (such as hitting the wall after traversing top slit or hitting
the wall after traversing bottom slit).

\begin{quotation}
\noindent If you could, \textit{in principle}, distinguish the alternative
\textit{final} states (even though you do not bother to do so), the total,
final probability is obtained by calculating the \textit{probability} for each
state (not the amplitude) and then adding them together. If you
\textit{cannot} distinguish the final states \textit{even in principle}, then
the probability amplitudes must be summed before taking the absolute square to
find the actual probability.\cite[p. 3-9]{feynman:vIII}
\end{quotation}

Moreover, when the properties of entities are carved out by distinctions
(starting at the blob), then it is perfectly possible to have two entities
that result from the same distinctions but with no other distinctions so they
are \textit{in principle} indistinguishable (unlike two twins who are "hard to
tell apart"). In QM, this has enormous consequences as in the distinction
between bosons and fermions, the Pauli exclusion principle, and the chemical
properties of the elements. This sort of in-principle indistinguishability is
a feature of the micro-reality envisaged by partition logic, but is not
possible under the "properties all the way down" vision of subset logic.

\subsection{Wave math without waves = indistinctness-preserving mathematics}

What about the Schr\"{o}dinger \textit{wave} equation? Since measurements, or,
more generally, interactions between a quantum system and the environment, may
make distinctions (measurement and decoherence), we might ask the following
question. What is the evolution of a quantum system that is isolated so that
not only are no distinctions made, but even the degree of indistinctness
between state vectors is not changed? Two states $\psi$ and $\varphi$ in a
Hilbert space are \textit{fully distinct} if they are orthogonal, i.e.,
$\left\langle \psi|\varphi\right\rangle =0$. Two states are \textit{fully
indistinct} if $\left\langle \psi|\varphi\right\rangle =1$. In between, the
\textit{degree of indistinctness} can be measured by the \textit{overlap}
$\left\langle \psi|\varphi\right\rangle $, the inner product of the state
vectors. Hence the evolution of an isolated quantum system where the degree of
indistinctness does not change is described by a linear transformation that
preserves inner products, i.e., a unitary transformation.

The connection between unitary transformations and the solutions to the
Schr\"{o}dinger "wave" equation is given by Stone's Theorem \cite{stone:thm}:
there is a one-to-one correspondence between strongly continuous $1$-parameter
unitary groups $\left\{  U_{t}\right\}  _{t\in%
%TCIMACRO{\U{211d} }%
%BeginExpansion
\mathbb{R}
%EndExpansion
}$ and Hermitian operators $A$ on the Hilbert space so that $U_{t}=e^{itA}$.

In simplest terms, a unitary transformation describes a rotation such as the
rotation of the unit vector in the complex plane.%

%TCIMACRO{\FRAME{dtbpF}{1.3334in}{1.351in}{0in}{}{}{fig11-rotate-arrow.eps}%
%width 1.3334in;  height 1.351in;  depth 0in;  original-width 1.2747in;
%original-height 1.2915in;  cropleft "0";  croptop "1";  cropright "1";
%cropbottom "0";  filename 'fig11-rotate-arrow.eps';file-properties "XNPEU";}}
%}%
%BeginExpansion
\begin{center}
\includegraphics[
height=1.351in,
width=1.3334in
]%
{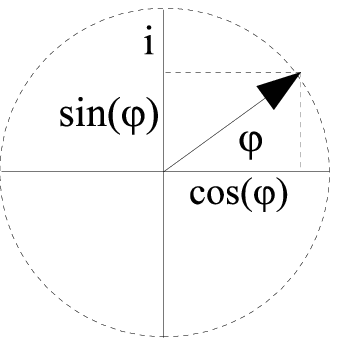}%
\end{center}
%EndExpansion

\begin{center}
Figure 11: Rotating vector
\end{center}

\noindent The rotating unit vector traces out the cosine and sine functions on
the two axes, and the position of the arrow can be compactly described as a
function of $\varphi$ using Euler's formula:

\begin{center}
$e^{i\varphi}=\cos\left(  \varphi\right)  +i\sin\left(  \varphi\right)  $.
\end{center}

\noindent Such complex exponentials and their superpositions are the "wave
functions" of QM. The "wave functions" describe the evolution of particles in
indefinite states in isolated systems where there are no interactions to
change the degree of indistinctness between states, i.e., the context where
Schr\"{o}dinger's equation holds. Previously it has been assumed that the
mathematics of waves must describe physical waves of some sort, and thus the
puzzlement about the "waves" of QM having complex amplitudes and no
corresponding physical waves. But we have supplied \textit{another}
interpretation; wave mathematics is the mathematics of indefiniteness, e.g.,
superposition represents indefiniteness and unitary evolution represents the
indistinctness-preserving evolution of an isolated system. Feynman's addition
of shrinking and turning arrows \cite{fey:qed} builds an imagery different
from the usual wave imagery.

Thus the objective indefiniteness approach to interpreting QM provides an
explanation for the appearance of the wave mathematics (which implies
interference as well as the quantized solutions to the "wave" equation that
gave QM its name) when, in fact, there are no actual physical waves involved.

\section{Logical entropy measures measurement}

\subsection{Logical entropy as the total distinction probability}

The notion of logical entropy of a probability distribution $p=\left(
p_{1},...,p_{n}\right)  $, $h\left(  p\right)  =1-\sum_{i}p_{i}^{2}$,
generalizes to the \textit{quantum logical entropy} of a density matrix $\rho$
\cite{fano:density},

\begin{center}
$h\left(  \rho\right)  =1-\operatorname*{tr}\left[  \rho^{2}\right]  $.
\end{center}

Given a state vector $\left\vert \psi\right\rangle =\sum_{i}\alpha
_{i}\left\vert i\right\rangle $ expressed in the orthonormal basis $\left\{
\left\vert i\right\rangle \right\}  _{i=1,...,n}$, the density matrix

\begin{center}
$\rho=\left\vert \psi\right\rangle \left\langle \psi\right\vert =\left[
\rho_{ij}\right]  =\left[  \alpha_{i}\alpha_{j}^{\ast}\right]  $
\end{center}

\noindent(where $\alpha_{j}^{\ast}$ is the complex conjugate of $\alpha_{j}$)
is a \textit{pure state} density matrix. For a pure state density matrix:

\begin{center}
$h\left(  \rho\right)  =1-\operatorname*{tr}\left[  \rho^{2}\right]
=1-\sum_{i}\sum_{j}\alpha_{i}\alpha_{j}^{\ast}\alpha_{j}\alpha_{i}^{\ast
}=1-\sum_{i}\alpha_{i}\alpha_{i}^{\ast}\sum_{j}\alpha_{j}\alpha_{j}^{\ast
}=1-1=0$.
\end{center}

\noindent Otherwise, a density matrix $\rho$ is said to represent a
\textit{mixed state}, and its logical entropy is positive.

In the set case, the logical entropy $h\left(  \pi\right)  $ of a partition
$\pi$ was interpreted as the probability that two independent draws from $U$
(equiprobable elements) would give a distinction of $\pi$. For a probability
distribution $p=\left(  p_{1},...,p_{n}\right)  $, the logical entropy
$h\left(  p\right)  =1-\sum_{i}p_{i}^{2}$ is the probability that two
independent samples from the distribution will give distinct outcomes $i\neq
j$. The probability of the distinct outcomes $\left(  i,j\right)  $ for $i\neq
j$ is $p_{i}p_{j}$. Since $1=\left(  p_{1}+...+p_{n}\right)  \left(
p_{1}+...+p_{n}\right)  =\sum_{i,j}p_{i}p_{j}$, we have:

\begin{center}
$h\left(  p\right)  =1-\sum_{i}p_{i}^{2}=\sum_{i,j}p_{i}p_{j}-\sum_{i}%
p_{i}^{2}=\sum_{i\neq j}p_{i}p_{j}$
\end{center}

\noindent which is the sum of all the distinction (i.e., distinct indices) probabilities.

This interpretation generalizes to the quantum logical entropy $h\left(
\rho\right)  $. The diagonal terms $\left\{  p_{i}\right\}  $ in a density matrix:

\begin{center}
$\rho=%
\begin{bmatrix}
p_{1} & \rho_{12} & \cdots & \rho_{1n}\\
\rho_{21} & p_{2} & \cdots & \rho_{2n}\\
\vdots & \vdots & \ddots & \vdots\\
\rho_{n1} & \rho_{n2} & \cdots & p_{n}%
\end{bmatrix}
$
\end{center}

\noindent are the probabilities of getting the $i^{th}$ eigenvector
$\left\vert i\right\rangle $ in a projective measurement of a system in the
state $\rho$ (using $\left\{  \left\vert i\right\rangle \right\}  $ as the
measurement basis). The off-diagonal terms $\rho_{ij}$ give the amplitude that
the eigenstates $\left\vert i\right\rangle $ and $\left\vert j\right\rangle $
cohere, i.e., are indistinct, in the state $\rho$ so the absolute square
$\left\vert \rho_{ij}\right\vert ^{2}$ is the \textit{indistinction
probability}. Since $p_{i}p_{j}$ is the probability of getting $\left\vert
i\right\rangle $ and $\left\vert j\right\rangle $ in two independent
measurements, the difference $p_{i}p_{j}-\left\vert \rho_{ij}\right\vert ^{2}%
$, is the \textit{distinction probability}. But $1=\sum_{i,j}p_{i}p_{j}$ so we
see that the interpretation of the logical entropy as the total distinction
probability carries over to the quantum case:

\begin{center}
$h\left(  \rho\right)  =1-\operatorname*{tr}\left[  \rho^{2}\right]
=1-\sum_{ij}\left\vert \rho_{ij}\right\vert ^{2}=\sum_{ij}\left[  p_{i}%
p_{j}-\left\vert \rho_{ij}\right\vert ^{2}\right]  =\sum_{i\neq j}\left[
p_{i}p_{j}-\left\vert \rho_{ij}\right\vert ^{2}\right]  $

Quantum logical entropy = sum of distinction probabilities
\end{center}

\noindent where the last step follows since $p_{i}p_{i}-\left\vert \rho
_{ii}\right\vert ^{2}=0$.

\subsection{Measuring measurement}

Since $h\left(  \rho\right)  =0$ for a pure state $\rho$, that means that all
the eigenstates $\left\vert i\right\rangle $ and $\left\vert j\right\rangle $
cohere together, i.e., are indistinct, in a pure state, like the indiscrete
partition in the set case. For set partitions, the transition, $\mathbf{0}%
\rightarrow\mathbf{1}$, from the indiscrete to the discrete partition turns
all the indistinctions $\left(  i,j\right)  $ (where $i\neq j$) into
distinctions, and the logical entropy increases from $0$ to $1-\sum_{i}%
p_{i}^{2}=1-\frac{1}{n}$ where $p_{i}=\frac{1}{n}$ for $\left\vert
U\right\vert =n$.

Similarly in quantum mechanics, a nondegenerate measurement turns a pure state
density matrix $\rho$ into the mixed state diagonal matrix $\hat{\rho}$ with
the same diagonal entries $p_{1},...,p_{n}$:

\begin{center}
$\rho=%
\begin{bmatrix}
p_{1} & \rho_{12} & \cdots & \rho_{1n}\\
\rho_{21} & p_{2} & \cdots & \rho_{2n}\\
\vdots & \vdots & \ddots & \vdots\\
\rho_{n1} & \rho_{n2} & \cdots & p_{n}%
\end{bmatrix}
\overset{measurement}{\Rightarrow}\hat{\rho}=%
\begin{bmatrix}
p_{1} & 0 & \cdots & 0\\
0 & p_{2} & \cdots & 0\\
\vdots & \vdots & \ddots & \vdots\\
0 & 0 & \cdots & p_{n}%
\end{bmatrix}
$.
\end{center}

\noindent Hence the quantum logical entropy similarly goes from $h\left(
\rho\right)  =0$ to $h\left(  \hat{\rho}\right)  =1-\sum_{i}p_{i}^{2}$. This
is usually described by saying that all the off-diagonal coherence terms are
decohered in a nondegenerate measurement--which means that all the
indistinctions $\left(  \left\vert i\right\rangle ,\left\vert j\right\rangle
\right)  $ where $\left\vert i\right\rangle \neq\left\vert j\right\rangle $ of
the pure state are distinguished by the measurement. And the sum of all those
new distinction probabilities for the decohered off-diagonal terms is
precisely the quantum logical entropy since $h\left(  \hat{\rho}\right)
=\sum_{i\neq j}\left[  p_{i}p_{j}-\left\vert \hat{\rho}_{ij}\right\vert
^{2}\right]  =\sum_{i\neq j}p_{i}p_{j}$. For any measurement (degenerate or
not), the increase in logical entropy

\begin{center}
$h\left(  \hat{\rho}\right)  -h\left(  \rho\right)  =\sum_{new}\left\vert
\rho_{ij}\right\vert ^{2}$ = sum of \textit{new} distinction probabilities
\end{center}

\noindent where the sum is over the zeroed or decohered coherence terms
$\left\vert \rho_{ij}\right\vert ^{2}$ that gave indistinction probabilities
in the pre-measurement state $\rho$. Thus we see how quantum logical entropy
interprets the off-diagonal entries in the state density matrices and how the
change in the quantum logical entropy measures precisely the decoherence,
i.e., the distinctions, made by a measurement.\footnote{In contrast, the
standard notion of entropy currently used in quantum information theory, the
von Neumann entropy, is only qualitatively related to measurement, i.e.,
projective measurement increases von Neumann entropy \cite[p. 515]{nc:qcqi}.}

\section{Lifting to the axioms of quantum mechanics}

We have now reached the point where the program of lifting partition logic and
logical information theory to the quantum concepts of Hilbert spaces
essentially yields the axioms of quantum mechanics.

Using axioms based on \cite{nc:qcqi}, the first axiom gives the vector space
endpoint of the lifting program.

\textbf{Axiom 1:} \textit{An isolated system is represented by a complex inner
product vector space (i.e., a Hilbert space) where the complete description of
a state of the system is given by a state vector, a unit vector in the
system's space.}

Two fully distinct states would be orthogonal (thinking of them as eigenstates
of an observable), and a state indefinite between them would be represented as
a weighted vector sum or superposition of the two states. Taking a
superposition state as a "complete description" is essentially the same as
saying that the indefiniteness is objective.

We previously saw that the evolution of a closed system that preserves the
degree of indistinction between states would be a unitary transformation.

\textbf{Axiom 2:} \textit{The evolution of a closed quantum system is
described by a unitary transformation}.

In the last section, we saw how a projective measurement would zero some or
all of the off-diagonal coherence terms in a pure state $\rho$ to give a mixed
state $\hat{\rho}$ (and how the sum of the absolute squares of the zeroed
coherence terms gave the change in quantum logical entropy).

\textbf{Axiom 3:} \textit{A projective measurement for an observable
(Hermitian operator) }$M=\sum_{m}mP_{m}$\textit{ (spectral decomposition using
projection operators }$P_{m}$\textit{) on a pure state }$\rho$\textit{ has an
outcome }$m$\textit{ with probability }$p_{m}=\rho_{mm}$\textit{ giving the
mixed state }$\hat{\rho}=\sum_{m}P_{m}\rho P_{m}$.

And finally we saw how the basis principle lifted the notion of combining sets
with the direct product of sets $X\times Y$ to the notion of representing
combined quantum systems with the vector space generated by the direct product
of two bases of the state spaces.

\textbf{Axiom 4:} \textit{The state space of a composite system is the tensor
product of the state spaces of the component systems}.

\section{Conclusion}

The objective indefiniteness interpretation of quantum mechanics is based on
using partition logic, logical information theory, and the lifting program to
fill out the back story to the old notion of "objective indefiniteness"
(\cite{shim:reality}, \cite{shim:concept}). In Appendix 1, the lifting program
is further applied to lift set representations of groups to vector space
representations, and thus to explain the fundamental importance of group
representation theory in quantum mechanics (not to mention particle physics).
In Appendices 2 and 3, the two-slit experiment and Bell's Theorem are treated
in "quantum mechanics" on sets--which lays out the bare logical structure in
both cases. 

At the level of sets, if we start with a universe set $U$ as representing our
common-sense macroscopic world, then there are \textit{only two} logics, the
logics of subsets and quotient sets (i.e., partitions), to envisage the
"creation story" for $U$. Increase the size of subsets or increase the
refinement of quotient sets until reaching the universe $U$. That is, starting
with the empty subset of $U$, take larger and larger subsets of well-defined
fully definite elements until finally reaching all the fully definite elements
of $U$. Or starting with the indiscrete partition on $U$, take more and more
refined partitions, each block interpreted as an indefinite element, until
finally reaching all the fully definite elements of $U$. Those are the two
dual options.

Classical physics was compatible with the subset creation story in the sense
that the elements were always fully propertied ("properties all the way
down"). But almost from the beginning, quantum mechanics was seen not to be
compatible with that world view of always fully definite entities; QM seems to
envisage entities at the micro-level that are objectively indefinite. Within
the framework of the two logics given by subset-partition duality, the
"obvious" thing to do is to elaborate on the dual creation story to try to
build \textit{the} other interpretation of QM.

With the development of the logic of partitions (dual to the logic of subsets)
and logical information theory built on top of it, the foundation was in place
to lift those set concepts to the richer mathematical environment of vector
spaces (Hilbert spaces in particular). In that manner, \textit{the} other
interpretation of QM was constructed. Unlike the interpretation based on
entities with fully definite properties expressed by Boolean subset logic,
\textit{the} dual interpretation works. That is, the result reproduces the
basic ideas and mathematical machinery of quantum mechanics, e.g., as
expressed in four axioms given above. That completes an outline of the vision
of micro-reality that provides the objective indefiniteness interpretation of
quantum mechanics.

\section{Appendix 1: Lifting in group representation theory}

\subsection{Group representations define partitions}

Given a \textit{set} $G$ of mappings $R=\left\{  R_{g}:U\rightarrow U\right\}
_{g\in G}$ on a set $U$, what are the conditions on the set of mappings so
that it is a set representation of a group? Define the binary relation on
$U\times U$:

\begin{center}
$u\thicksim u^{\prime}$ if $\exists g\in G$ such that $R_{g}\left(  u\right)
=u^{\prime}$.
\end{center}

\noindent Then the conditions that make $R$ into a group representation are
the conditions that imply $u\thicksim u^{\prime}$ is an equivalence relation:

\begin{enumerate}
\item existence of the identity $1_{U}\in U$ implies reflexivity of
$\thicksim$;

\item existence of inverses implies symmetry of $\thicksim$; and

\item closure under products, i.e., for $g,g^{\prime}\in G$, $\exists
g^{\prime\prime}\in G$ such that $R_{g^{\prime\prime}}=R_{g^{\prime}}R_{g}$,
implies transitivity of $\thicksim$.
\end{enumerate}

Hence a set representation of a group might be seen as a "dynamic" way to
define an equivalence relation and thus a partition on the set. A symmetry
group defines indistinctions. For instance, if linear translations form a
symmetry group for a quantum system, then the system behavior before a linear
translation is indistinct from the behavior of the translated system. Given
this intimate connection between groups and partitions, it is then no surprise
that group representation theory has a basic role to play in quantum mechanics
and in the partition-based objective indefiniteness interpretation of QM.

\subsection{Where do the fully distinct eigen-alternatives come from?}

In the vector space case, we may be \textit{given} the observable with its
distinct eigenstates so the indefinite states are linear combinations of those eigenstates.

In the set case, we are \textit{given} the universe $U$ of distinct
eigen-alternatives $u\in U$, and then the indistinct entities are the subsets
such as the blocks $B\in\pi$ in a partition of $U$. A "measurement" is some
distinction-making operation that reduces an indistinct state $B$ down to a
more distinct state $B^{\prime}\subseteq B$ or, in the nondegenerate case, to
a fully distinct singleton $\left\{  u\right\}  $ for some $u\in B$. But where
do the fully distinct elements come from?

The \textit{basic idea} is that a symmetry group defines indistinctions, so
what are all the ways that there can be distinct eigen-elements that are
consistent with those indistinctions? In a representation of a group by
permutations on a set $U$, the answer is:

\begin{center}
distinct eigen-elements consistent with symmetry group $\approx$ orbits of
group representation.
\end{center}

\noindent Two elements of $U$ inside the same orbit cannot be considered
distinct in a way consistent with the indistinction-making action of the group
since they are, by definition, mapped from one to the other by a group
operation. Intuitively, this is how the partition ideas mesh with group
representation theory. First we consider the set version, and then we lift to
the vector space version of group representation theory.

Let $U$ be a set and $S\left(  U\right)  $ the group of all permutations of
$U$. Then a \textit{set representation} of a group $G$ is an assignment
$R:G\rightarrow S\left(  U\right)  $ where for $g\in G$, $g\longmapsto
R_{g}\in S\left(  U\right)  $ such that $R_{1}$ is the identity on $U$ and for
any $g,g^{\prime}\in G$, $R_{g^{\prime}}R_{g}=R_{g^{\prime}g}$. Equivalently,
a \textit{group action} is a binary operation $G\times U\rightarrow U$ such
that $1u=u$ and $g^{\prime}\left(  gu\right)  =\left(  g^{\prime}g\right)  u$
for all $u\in U$.

Defining $u\thicksim u^{\prime}$ if $\exists g\in G$ such that $R_{g}\left(
u\right)  =u^{\prime}$ [or $gu=u^{\prime}$ using the group action notation],
we have an equivalence relation on $U$ where the blocks are called the
\textit{orbits}.

How are the ultimate distinct eigen-alternatives, the distinct "eigen-forms"
of "substance," defined in the set case? Instead of just assuming $U$ as the
set of eigen-alternatives, we start with $U$ as the carrier for a set
representation of the group $G$ as a group of symmetries. What are the
smallest subsets (forming the blocks $B$ in a set partition) that respect the
symmetries, i.e., that are \textit{invariant} in the sense that $R_{g}%
(B)\subseteq B$ for all $g\in G$? Those minimal invariant subsets are the
orbits, and all invariant subsets are unions of orbits. Thus the orbits,
thought of as points in the quotient set $U/G$ (set of orbits), are the
eigen-alternatives, the "eigen-forms" of "substance," defined by the symmetry
group $G$ in the set case.

\textbf{Example 1}: Let $U=\left\{  0,1,2,3,4,5\right\}  $ and let
$G=S_{2}=\left\{  1,\sigma\right\}  $ (symmetric group on two elements) where
$R_{1}=1_{U}$ and $R_{\sigma}(u)=u+3$ $\operatorname{mod}6$.%

%TCIMACRO{\FRAME{dtbpFU}{4.3054in}{1.4365in}{0pt}{\Qcb{Figure 12: Action of
%$S_{2}$ on $U=\left\{  0,1,2,3,4,5\right\}  $}}{}{fig12-s2-rep.eps}%
%width 4.3054in;  height 1.4365in;  depth 0pt;  original-width 4.1729in;
%original-height 1.3753in;  cropleft "0";  croptop "1";  cropright "1";
%cropbottom "0";  filename 'fig12-s2-rep.eps';file-properties "XNPEU";}} }%
%BeginExpansion
\begin{center}
\includegraphics[
height=1.4365in,
width=4.3054in
]%
{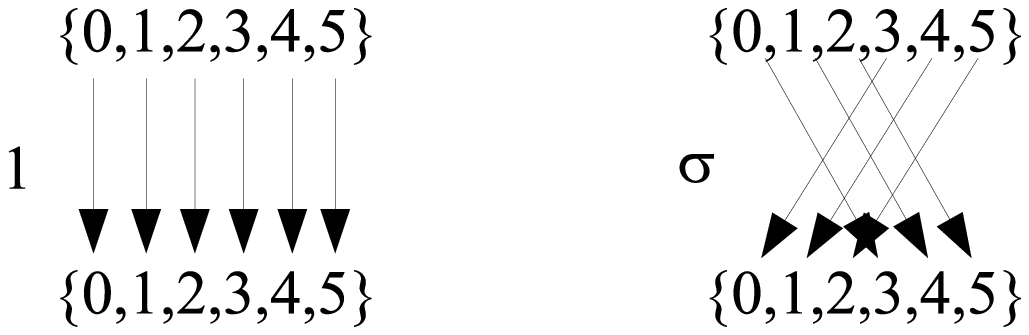}%
\\
Figure 12: Action of $S_{2}$ on $U=\left\{  0,1,2,3,4,5\right\}  $%
\end{center}
%EndExpansion

There are $3$ orbits: $\left\{  0,3\right\}  ,\left\{  1,4\right\}  $, and
$\left\{  2,5\right\}  $, and they partition $U$. Those three orbits are the
"points" in the quotient set $U/G$, i.e., they are the distinct
eigen-alternatives defined by the symmetry group's $S_{2}$ action on $U$.

A \textit{vector space representation} of a group $G$ on a vector space $V$ is
a mapping $g\longmapsto R_{g}:V\rightarrow V$ from $G$ to invertible linear
transformations on $V$ such that $R_{g^{\prime}}R_{g}=R_{g^{\prime}g}$.

The lifts to the vector space representations of groups are;

\begin{itemize}
\item minimal invariant subset = orbits $\overset{Lifts}{\longrightarrow}$
minimal invariant subspaces = \textit{irreducible subspaces},

\item representation restricted to orbits $\overset{Lifts}{\longrightarrow}$
representation restricted to irreducible subspaces which gives the
\textit{irreducible representations} (the eigen-forms of substance in the
vector space case\footnote{For a certain group of particle physics, "an
elementary particle `is' an irreducible unitary representation of the
group."\cite[p. 149]{stern:group} In Heisenberg's philosophical terms, the
irreducible representations of certain symmetry groups of particle physics
determine the fundamental eigen-forms that the substance (energy) can take.
\par
\begin{quotation}
\noindent The elementary particles are therefore the fundamental forms that
the substance energy must take in order to become matter, and these basic
forms must in some way be determined by a fundamental law expressible in
mathematical terms. ... The real conceptual core of the fundamental law must,
however, be formed by the mathematical properties of the symmetry it
represents.\cite[pp. 16-17]{heisen:modphy}
\end{quotation}
}), and

\item set partition of orbits $\overset{Lifts}{\longrightarrow}$ vector space
partition of irreducible subspaces.
\end{itemize}

The "irreducible representations" in the set case are just the restrictions of
the representation to the orbits, e.g., $R\upharpoonright\left\{  0,3\right\}
:S_{2}\rightarrow S\left(  \left\{  0,3\right\}  \right)  $, as their
carriers. A set representation is said to be \textit{transitive}, if for any
$u,u^{\prime}\in U$, $\exists g\in G$ such that $R_{g}\left(  u\right)
=u^{\prime}$. A transitive set representation has only one orbit, all of $U$.
Any set "irreducible representation" is transitive.

We are accustomed to thinking of some distinction-making operation as reducing
a whole partition to a more refined partition, and thus breaking up a block
$B$ into distinguishable non-overlapping subsets $B^{\prime},B^{\prime\prime
},...\subseteq B$. Now we are working at the more basic level of determining
the distinct eigen-alternatives, i.e., the orbits of a set representation of a
symmetry group. Here we might also consider how distinctions are made to move
to a more refined partition of orbits. Since the group operations identify
elements, $u\thicksim u^{\prime}$ if $\exists g\in G$ such that $R_{g}\left(
u\right)  =u^{\prime}$, we would further \textit{distinguish} elements by
moving to a subgroup. The symmetry operations in the larger group are
"broken," so the remaining group of symmetries is a subgroup. Breaking
symmetries makes distinctions.

\textbf{Example 1 revisited}: the group $S_{2}$ has only one subgroup, the
trivial subgroup of the identity operation, and its orbits are clearly the
singletons $\left\{  u\right\}  $ for $u\in U$. That is the simplest example
of \textit{symmetry-breaking} that gives a more distinct set of eigen-alternatives.

In any set representation, the maximum distinctions are made by the smallest
symmetry subgroup which is always the identity subgroup, so that is always the
waste case that takes us back to the singleton orbits in $U$, i.e., the
distinct elements of $U$.

Thus we see that symmetry-breaking is analogous to measurement but at this
more fundamental level where the distinct eigen-forms are determined in the
first place by symmetries.

\subsection{Attributes and observables}

An (real-valued) \textit{attribute} on a set $U$ is a function $f:U\rightarrow%
%TCIMACRO{\U{211d} }%
%BeginExpansion
\mathbb{R}
%EndExpansion
$. An attribute induces a set partition $\left\{  f^{-1}(r)\right\}  $ on $U$.
An attribute $f:U\rightarrow%
%TCIMACRO{\U{211d} }%
%BeginExpansion
\mathbb{R}
%EndExpansion
$ \textit{commutes} with a set representation $R:G\rightarrow S\left(
U\right)  $ if for any $R_{g}$, the following diagram commutes in the sense
that $fR_{g}=f$:

\begin{center}
$%
\begin{array}
[c]{ccc}%
U & \overset{R_{g}}{\longrightarrow} & U\\
& \searrow^{f} & \;\downarrow^{f}\\
&  &
%TCIMACRO{\U{211d}}%
%BeginExpansion
\mathbb{R}%
%EndExpansion
\end{array}
$

Commuting attribute.
\end{center}

The lifts to vector space representations are immediate:

\begin{itemize}
\item a real-valued attribute on a set $\overset{Lifts}{\longrightarrow}$ an
observable represented by a Hermitian operator on a complex vector space; and

\item the commutativity condition on a set-attribute
$\overset{Lifts}{\longrightarrow}$ an observable operator $H$ (like the
Hamiltonian) commuting with a symmetry group in the sense that $HR_{g}=R_{g}H$
for all $g\in G$.
\end{itemize}

The commutativity-condition in the set case means that whenever $R_{g}\left(
u\right)  =u^{\prime}$ then $f\left(  u\right)  =f\left(  u^{\prime}\right)
$, i.e., that $f$ is an \textit{invariant} of the group. Recall that each
orbit of a set representation is transitive so for any $u,u^{\prime}$ in the
same orbit, $\exists R_{g}$ such that $R_{g}\left(  u\right)  =u^{\prime}$ so
$f\left(  u\right)  =f\left(  u^{\prime}\right)  $ for any two $u,u^{\prime}$
in the same orbit. In other words:

"\textbf{Schur's Lemma"} (set version): a commuting attribute restricted to an
orbit is constant.

The lift to vector space representations is one version of the usual

\textbf{Schur's Lemma }(vector space version): An operator $H$ commuting with
$G$ restricted to irreducible subspace is a constant operator.

This also means that the inverse-image partition $\left\{  f^{-1}\left(
r\right)  \right\}  $ of a commuting attribute is refined by the orbit
partition. If an orbit $B\subseteq f^{-1}\left(  r\right)  $, then the
"eigenvalue" $r$ of the attribute $f$ is associated with that orbit. Every
commuting attribute $f:U\rightarrow%
%TCIMACRO{\U{211d} }%
%BeginExpansion
\mathbb{R}
%EndExpansion
$ can be uniquely expressed as a decomposition:

\begin{center}
$f=\sum_{o\in Orbits}r_{o}\chi_{o}$,
\end{center}

\noindent where $r_{o}$ is the constant value on the orbit $o\subseteq U$ and
$\chi_{o}:U\rightarrow%
%TCIMACRO{\U{211d} }%
%BeginExpansion
\mathbb{R}
%EndExpansion
$ is the characteristic function of the orbit $o$.

There may be other orbits with the same "eigenvalue." Then we would need
another commuting attribute $g:U\rightarrow%
%TCIMACRO{\U{211d} }%
%BeginExpansion
\mathbb{R}
%EndExpansion
$ so that for each orbit $B$, there is an "eigenvalue" $s$ of the attribute
$g$ such that $B\subseteq g^{-1}\left(  s\right)  $. Then the
eigen-alternative $B$ may be characterized by the ordered pair $\left\vert
r,s\right\rangle $ if $B=f^{-1}\left(  r\right)  \cap g^{-1}\left(  s\right)
$. If not, we continue until we have a Complete Set of Commuting Attributes
(CSCA) whose ordered $n$-tuples of "eigenvalues" would characterize the
eigen-alternatives, the orbits of the set representation $R:G\rightarrow
S\left(  U\right)  $.

Obviously, we are just spelling out the set version whose lift is the use of a
Complete Set of Commuting Operators (CSCO) to characterize the eigenstates by
kets of ordered $n$-tuples $\left\vert \lambda,\mu,...\right\rangle $ of
eigenvalues of the commuting operators.\footnote{For a presentation of group
representation theory that uses a CSCO approach to characterizing the
irreducible representations, see \cite{chen:text}.} But these "eigenstates"
are not the singletons $\left\{  u\right\}  $ but are the minimal invariant
subsets or orbits of the set representation of the symmetry group $G$.

\textbf{Example 1 again}: Consider the attribute $f:U=\left\{
0,1,2,3,4,5\right\}  \rightarrow%
%TCIMACRO{\U{211d} }%
%BeginExpansion
\mathbb{R}
%EndExpansion
$ where $f\left(  n\right)  =n$ $\operatorname{mod}3$. This attribute commutes
with the previous set representation of $S_{2}$, namely $R_{1}=1_{U}$ and
$R_{\sigma}(u)=u+3$ $\operatorname{mod}6$, and accordingly by "Schur's Lemma"
(set version), the attribute is constant on each orbit $\left\{  0,3\right\}
,\left\{  1,4\right\}  $, and $\left\{  2,5\right\}  $. In this case, the
blocks of the inverse-image partition $\left\{  f^{-1}\left(  0\right)
,f^{-1}\left(  1\right)  ,f^{-1}\left(  2\right)  \right\}  $ equal the blocks
of the orbit partition, so this attribute is the set version of a
"nondegenerate measurement" in that its "eigenvalues" suffice to characterize
the eigen-alternatives, i.e., the orbits. By itself, it forms a complete set
of attributes.

\textbf{Example 2: }Let $U=\left\{  0,1,...,11\right\}  $ where $S_{2}%
=\left\{  1,\sigma\right\}  $ is represented by the operations $R_{1}=1_{U}$
and $R_{\sigma}\left(  n\right)  =n+6$ $\operatorname{mod}\left(  12\right)
$. Then the orbits are $\left\{  0,6\right\}  ,\left\{  1,7\right\}  ,\left\{
2,8\right\}  ,\left\{  3,9\right\}  ,\left\{  4,10\right\}  ,$ and $\left\{
5,11\right\}  $. Consider the attribute $f:U\rightarrow%
%TCIMACRO{\U{211d} }%
%BeginExpansion
\mathbb{R}
%EndExpansion
$ where $f\left(  n\right)  =n$ $\operatorname{mod}\left(  2\right)  $. This
attribute commutes with the symmetry group and is thus constant on the orbits.
But the blocks in the inverse-image partition are now larger than the orbits,
i.e., $f^{-1}\left(  0\right)  =\left\{  0,2,4,6,8,10\right\}  $ and
$f^{-1}\left(  1\right)  =\left\{  1,3,5,7,9,11\right\}  $ so the orbit
partition strictly refines $\left\{  f^{-1}\left(  r\right)  \right\}  $. Thus
this attribute corresponds to a degenerate measurement in that the two
"eigenvalues" do not suffice to characterize the orbits.

Consider the attribute $g:U\rightarrow%
%TCIMACRO{\U{211d} }%
%BeginExpansion
\mathbb{R}
%EndExpansion
$ where $g\left(  n\right)  =n$ $\operatorname{mod}\left(  3\right)  $. This
attribute commutes with the symmetry group and is thus constant on the orbits.
The blocks in the inverse-image partition are: $g^{-1}\left(  0\right)
=\left\{  0,3,6,9\right\}  $, $g^{-1}\left(  1\right)  =\left\{
1,4,7,10\right\}  $, and $g^{-1}\left(  2\right)  =\left\{  2,5,8,11\right\}
$. The blocks in the join of the two partitions $\left\{  f^{-1}\left(
r\right)  \right\}  $ and $\left\{  g^{-1}\left(  s\right)  \right\}  $ are
the non-empty intersections of the blocks:

\begin{center}%
\begin{tabular}
[c]{|c|c|c|c|}\hline
$f^{-1}\left(  r\right)  $ & $g^{-1}\left(  s\right)  $ & $f^{-1}\left(
r\right)  \cap g^{-1}\left(  s\right)  $ & $\left\vert r,s\right\rangle
$\\\hline\hline
$\left\{  0,2,4,6,8,10\right\}  $ & $\left\{  0,3,6,9\right\}  $ & $\left\{
0,6\right\}  $ & $\left\vert 0,0\right\rangle $\\\hline
$\left\{  0,2,4,6,8,10\right\}  $ & $\left\{  1,4,7,10\right\}  $ & $\left\{
4,10\right\}  $ & $\left\vert 0,1\right\rangle $\\\hline
$\left\{  0,2,4,6,8,10\right\}  $ & $\left\{  2,5,8,11\right\}  $ & $\left\{
2,8\right\}  $ & $\left\vert 0,2\right\rangle $\\\hline
$\left\{  1,3,5,7,9,11\right\}  $ & $\left\{  0,3,6,9\right\}  $ & $\left\{
3,9\right\}  $ & $\left\vert 1,0\right\rangle $\\\hline
$\left\{  1,3,5,7,9,11\right\}  $ & $\left\{  1,4,7,10\right\}  $ & $\left\{
1,7\right\}  $ & $\left\vert 1,1\right\rangle $\\\hline
$\left\{  1,3,5,7,9,11\right\}  $ & $\left\{  2,5,8,11\right\}  $ & $\left\{
5,11\right\}  $ & $\left\vert 1,2\right\rangle $\\\hline
\end{tabular}

$f$ and $g$ as a complete set of commuting attributes
\end{center}

Thus $f$ and $g$ form a Complete Set of Commuting Attributes to characterize
the eigen-alternatives, the orbits, by the "kets" of ordered pairs of their "eigenvalues."

\textbf{Example 3:} Let $U=%
%TCIMACRO{\U{211d} }%
%BeginExpansion
\mathbb{R}
%EndExpansion
^{2}$ as a set and let $G$ be the special orthogonal matrix group $SO\left(
2,%
%TCIMACRO{\U{211d} }%
%BeginExpansion
\mathbb{R}
%EndExpansion
\right)  $ of matrices of the form;

\begin{center}
$%
\begin{bmatrix}
\cos\varphi & -\sin\varphi\\
\sin\varphi & \cos\varphi
\end{bmatrix}
$ for $0\leq\varphi<2\pi$.
\end{center}

\noindent This group is trivially represented by the rotations in $U=%
%TCIMACRO{\U{211d} }%
%BeginExpansion
\mathbb{R}
%EndExpansion
^{2}$:

\begin{center}
$%
\begin{bmatrix}
x^{\prime}\\
y^{\prime}%
\end{bmatrix}
=%
\begin{bmatrix}
\cos\varphi & -\sin\varphi\\
\sin\varphi & \cos\varphi
\end{bmatrix}%
\begin{bmatrix}
x\\
y
\end{bmatrix}
$.
\end{center}

\noindent The orbits are the circular orbits around the origin. The attribute
"radius" $f:%
%TCIMACRO{\U{211d} }%
%BeginExpansion
\mathbb{R}
%EndExpansion
^{2}\rightarrow%
%TCIMACRO{\U{211d} }%
%BeginExpansion
\mathbb{R}
%EndExpansion
$ where $f\left(  x,y\right)  =\sqrt{x^{2}+y^{2}}$ commutes with the
representation since:

$f\left(  x^{\prime},y^{\prime}\right)  =\sqrt{\left(  x^{\prime}\right)
^{2}+\left(  y^{\prime}\right)  ^{2}}$

$=\sqrt{\left(  x\cos\varphi-y\sin\varphi\right)  ^{2}+\left(  x\sin
\varphi+y\cos\varphi\right)  ^{2}}$

$=\sqrt{x^{2}\left(  \cos^{2}\varphi+\sin^{2}\varphi\right)  +y^{2}\left(
\cos^{2}\varphi+\sin^{2}\varphi\right)  }$

$=f\left(  x,y\right)  $.

\noindent That means that "radius" is an invariant of the rotation symmetry
group. The blocks in the set partition $\left\{  f^{-1}\left(  r\right)
:0\leq r\right\}  $ of $%
%TCIMACRO{\U{211d} }%
%BeginExpansion
\mathbb{R}
%EndExpansion
^{2}$ coincide with the orbits so the "eigenvalues" of the radius attribute
suffice to characterize the orbits.

\textbf{Example 4:} The Cayley set representation of any group $G$ is given by
permutations on $U=G$ itself defined by $R_{g}(g^{\prime})=gg^{\prime}$, which
is also called the \textit{left regular representation}. Given any
$g,g^{\prime}\in G$, $R_{g^{\prime}g^{-1}}\left(  g\right)  =g^{\prime}$ so
the Cayley representation is always transitive, i.e., has only one orbit
consisting of all of $U=G$. Since any commuting attribute $f:U=G\rightarrow%
%TCIMACRO{\U{211d} }%
%BeginExpansion
\mathbb{R}
%EndExpansion
$ is constant on each orbit, it can only be a constant function such as
$\chi_{G}$.

Thus the Cayley set representation is rather simple, but we could break some
symmetry by considering a proper subgroup $H\subseteq G$. Then using only the
$R_{h}$ for $h\in H$, we have a representation $H\rightarrow S\left(
G\right)  $. The orbit-defining equivalence relation is $g\thicksim g^{\prime
}$ if $\exists h\in H$ such that $hg=g^{\prime}$, i.e., the orbits are the
\textit{right cosets} of the form $Hg$.%

%TCIMACRO{\FRAME{dtbpFU}{5.6723in}{3.3855in}{0pt}{\Qcb{Figure 12: Action of
%$S_{2}$ on $U=\left\{  0,1,2,3,4,5\right\}  $}}{}{fig13-grpreptable.eps}%
%width 5.6723in;  height 3.3855in;  depth 0pt;  original-width 7.4753in;
%original-height 4.448in;  cropleft "0";  croptop "1";  cropright "1";
%cropbottom "0";  filename 'fig13-grpreptable.eps';file-properties "XNPEU";}}
%}%
%BeginExpansion
\begin{center}
\includegraphics[
height=3.3855in,
width=5.6723in
]%
{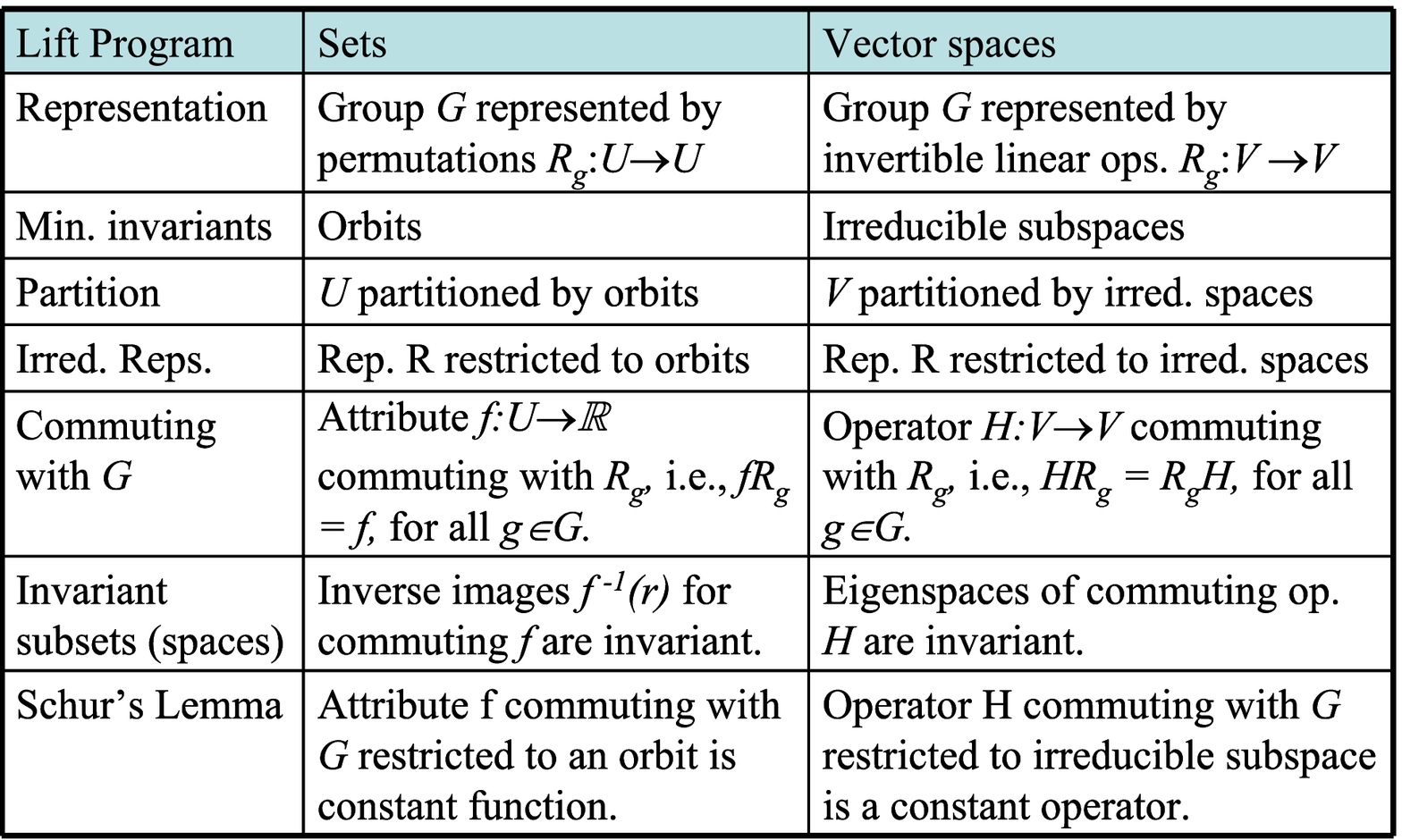}%
\\
Figure 12: Action of $S_{2}$ on $U=\left\{  0,1,2,3,4,5\right\}  $%
\end{center}
%EndExpansion

\section{Appendix 2: "Unitary evolution" and the two-slit experiment in
"quantum mechanics" on sets}

To illustrate a two-slit experiment in "quantum mechanics" on sets, we need to
introduce some "dynamics." In quantum mechanics, the requirement was that the
linear transformation had to preserve the degree of indistinctness
$\left\langle \psi|\varphi\right\rangle $, i.e., that it preserved the inner
product. Where two states are fully distinct if $\left\langle \psi
|\varphi\right\rangle =0$ and fully indistinct if $\left\langle \psi
|\varphi\right\rangle =1$, it is also sufficient to just require that full
distinctness and indistinctness be preserved since that would imply
orthonormal bases are preserved and that is equivalent to being unitary. In
"quantum mechanics" on sets, we have no inner product but the idea of a linear
transformation $A:%
%TCIMACRO{\U{2124} }%
%BeginExpansion
\mathbb{Z}
%EndExpansion
_{2}^{n}\rightarrow%
%TCIMACRO{\U{2124} }%
%BeginExpansion
\mathbb{Z}
%EndExpansion
_{2}^{n}$ preserving distinctness would simply mean being
non-singular.\footnote{Moreover, it might be noted that since the "brackets"
are basis-dependent, the condition analogous to preserving inner product would
be $\left\langle S|_{U}T\right\rangle =\left\langle A\left(  S\right)
|_{A\left(  U\right)  }A\left(  T\right)  \right\rangle $ where $A\left(
U\right)  =U^{\prime}$ is defined by $A\left(  \left\{  u\right\}  \right)
=\left\{  u^{\prime}\right\}  $. When $A:%
%TCIMACRO{\U{2124} }%
%BeginExpansion
\mathbb{Z}
%EndExpansion
_{2}^{\left\vert U\right\vert }\rightarrow%
%TCIMACRO{\U{2124} }%
%BeginExpansion
\mathbb{Z}
%EndExpansion
_{2}^{\left\vert U\right\vert }$ is a linear isomorphism (i.e., non-singular),
then the image $A\left(  U\right)  $ of the $U$-basis is a basis, i.e., the
$U^{\prime}$-basis, and the "bracket-preserving" condition holds since
$\left\vert S\cap T\right\vert =\left\vert A\left(  S\right)  \cap A\left(
T\right)  \right\vert $ for $A\left(  S\right)  ,A\left(  T\right)  \subseteq
A\left(  U\right)  =U^{\prime}$.}

Hence our only requirement on the "dynamics" is that the change-of-state
matrix is non-singular (so states are not merged). Consider the dynamics given
in terms of the $U$-basis where: $\left\{  a\right\}  \rightarrow\left\{
a,b\right\}  $; $\left\{  b\right\}  \rightarrow\left\{  a,b,c\right\}  $; and
$\left\{  c\right\}  \rightarrow\left\{  b,c\right\}  $ in one time period.
This is represented by the non-singular one-period change of state matrix:

\begin{center}
$A=%
\begin{bmatrix}
1 & 1 & 0\\
1 & 1 & 1\\
0 & 1 & 1
\end{bmatrix}
$.
\end{center}

The seven nonzero vectors in the vector space are divided by this "dynamics"
into a $4$ -orbit: $\left\{  a\right\}  \rightarrow\left\{  a,b\right\}
\rightarrow\left\{  c\right\}  \rightarrow\left\{  b,c\right\}  \rightarrow
\left\{  a\right\}  $, a $2$-orbit: $\left\{  b\right\}  \rightarrow\left\{
a,b,c\right\}  \rightarrow\left\{  b\right\}  $, and a $1$-orbit: $\left\{
a,c\right\}  \rightarrow\left\{  a,c\right\}  $.

If we take the $U$-basis vectors as "vertical position" eigenstates, we can
device a "quantum mechanics" version of the "two-slit experiment" which models
"all of the mystery of quantum mechanics" \cite[p. 130]{fey-phylaw}. Taking
$a$, $b$, and $c$ as three vertical positions, we have a vertical diaphragm
with slits at $a$ and $c$. Then there is a screen or wall to the right of the
slits so that a "particle" will travel from the diaphragm to the wall in one
time period according to the $A$-dynamics.%

%TCIMACRO{\FRAME{dtbpF}{2.5108in}{1.8156in}{0in}{}{}{fig14-two-slit-setup.eps}%
%width 2.5108in;  height 1.8156in;  depth 0in;  original-width 2.4236in;
%original-height 1.7451in;  cropleft "0";  croptop "1";  cropright "1";
%cropbottom "0";  filename 'fig14-two-slit-setup.eps';file-properties "XNPEU";}%
%} }%
%BeginExpansion
\begin{center}
\includegraphics[
height=1.8156in,
width=2.5108in
]%
{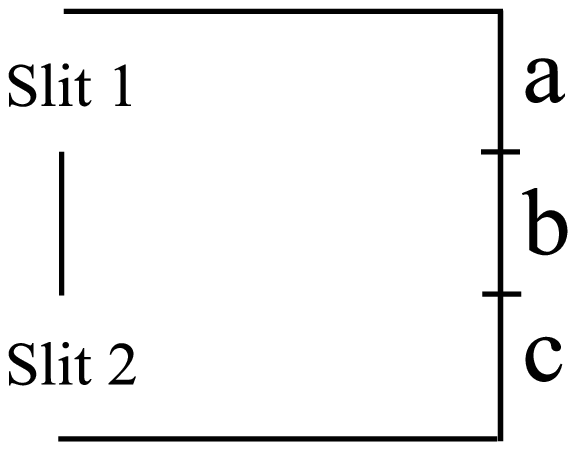}%
\end{center}
%EndExpansion

\begin{center}
Figure 14: Two-slit setup
\end{center}

We start with or "prepare" the state of a particle being at the slits in the
indefinite position state $\left\{  a,c\right\}  $. Then there are two cases.

\textbf{First case of distinctions at slits}: The first case is where we
measure the $U$-state at the slits and then let the resultant position
eigenstate evolve by the $A$-dynamics to hit the wall at the right where the
position is measured again. The probability that the particle is at slit 1 or
at slit 2 is:

\begin{center}
$\Pr\left(  \left\{  a\right\}  \text{ at slits }|\left\{  a,c\right\}  \text{
at slits}\right)  =\frac{\left\langle \left\{  a\right\}  |_{U}\left\{
a,c\right\}  \right\rangle ^{2}}{\left\Vert \left\{  a,c\right\}  \right\Vert
_{U}^{2}}=\frac{\left\vert \left\{  a\right\}  \cap\left\{  a,c\right\}
\right\vert }{\left\vert \left\{  a,c\right\}  \right\vert }=\frac{1}{2}$;

$\Pr\left(  \left\{  c\right\}  \text{ at slits }|\left\{  a,c\right\}  \text{
at slits}\right)  =\frac{\left\langle \left\{  c\right\}  |_{U}\left\{
a,c\right\}  \right\rangle ^{2}}{\left\Vert \left\{  a,c\right\}  \right\Vert
_{U}^{2}}=\frac{\left\vert \left\{  c\right\}  \cap\left\{  a,c\right\}
\right\vert }{\left\vert \left\{  a,c\right\}  \right\vert }=\frac{1}{2}$.
\end{center}

If the particle was at slit 1, i.e., was in eigenstate $\left\{  a\right\}  $,
then it evolves in one time period by the $A$-dynamics to $\left\{
a,b\right\}  $ where the position measurements yield the probabilities of
being at $a$ or at $b$ as:%

\begin{align*}
\Pr\left(  \left\{  a\right\}  \text{ at wall }|\left\{  a,b\right\}  \text{
at wall}\right)   &  =\frac{\left\langle \left\{  a\right\}  |_{U}\left\{
a,b\right\}  \right\rangle ^{2}}{\left\Vert \left\{  a,b\right\}  \right\Vert
_{U}^{2}}=\frac{\left\vert \left\{  a\right\}  \cap\left\{  a,b\right\}
\right\vert }{\left\vert \left\{  a,b\right\}  \right\vert }=\frac{1}{2}\\
\Pr\left(  \left\{  b\right\}  \text{ at wall }|\left\{  a,b\right\}  \text{
at wall}\right)   &  =\frac{\left\langle \left\{  b\right\}  |_{U}\left\{
a,b\right\}  \right\rangle ^{2}}{\left\Vert \left\{  a,b\right\}  \right\Vert
_{U}^{2}}=\frac{\left\vert \left\{  b\right\}  \cap\left\{  a,b\right\}
\right\vert }{\left\vert \left\{  a,b\right\}  \right\vert }=\frac{1}%
{2}\text{.}%
\end{align*}

If on the other hand the particle was found in the first measurement to be at
slit 2, i.e., was in eigenstate $\left\{  c\right\}  $, then it evolved in one
time period by the $A$-dynamics to $\left\{  b,c\right\}  $ where the position
measurements yield the probabilities of being at $b$ or at $c$ as:

\begin{center}
$\Pr\left(  \left\{  b\right\}  \text{ at wall }|\left\{  b,c\right\}  \text{
at wall}\right)  =\frac{\left\vert \left\{  b\right\}  \cap\left\{
b,c\right\}  \right\vert }{\left\vert \left\{  b,c\right\}  \right\vert
}=\frac{1}{2}$

$\Pr\left(  \left\{  c\right\}  \text{ at wall }|\left\{  b,c\right\}  \text{
at wall}\right)  =\frac{\left\vert \left\{  c\right\}  \cap\left\{
b,c\right\}  \right\vert }{\left\vert \left\{  b,c\right\}  \right\vert
}=\frac{1}{2}$.
\end{center}

Hence we can use the laws of probability theory to compute the probabilities
of the particle being measured at the three positions on the wall at the right
if it starts at the slits in the superposition state $\left\{  a,c\right\}  $
\textit{and} the measurements were made at the slits:

\begin{center}%
\begin{tabular}
[c]{l}%
$\Pr(\left\{  a\right\}  $ at wall $|\left\{  a,c\right\}  $ at slits$)=\frac
{1}{2}\frac{1}{2}=\frac{1}{4}$;\\
$\Pr(\left\{  b\right\}  $ at wall $|\left\{  a,c\right\}  $ at slits$)=\frac
{1}{2}\frac{1}{2}+\frac{1}{2}\frac{1}{2}=\frac{1}{2}$;\\
$\Pr(\left\{  c\right\}  $ at wall $|\left\{  a,c\right\}  $ at slits$)=\frac
{1}{2}\frac{1}{2}=\frac{1}{4}$.
\end{tabular}
%

%TCIMACRO{\FRAME{dtbpF}{2.5108in}{1.8013in}{0in}{}{}{fig15-case1-two-slit.eps}%
%width 2.5108in;  height 1.8013in;  depth 0in;  original-width 2.4236in;
%original-height 1.7309in;  cropleft "0";  croptop "1";  cropright "1";
%cropbottom "0";  filename 'fig15-case1-two-slit.eps';file-properties "XNPEU";}%
%} }%
%BeginExpansion
\begin{center}
\includegraphics[
height=1.8013in,
width=2.5108in
]%
{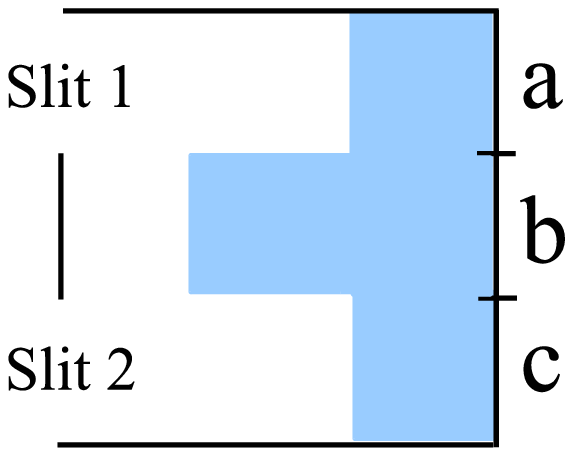}%
\end{center}
%EndExpansion

Figure 15: Final probability distribution with measurements at slits
\end{center}

\textbf{Second case of no distinctions at slits}: The second case is when no
measurements are made at the slits and then the superposition state $\left\{
a,c\right\}  $ evolves by the $A$-dynamics to $\left\{  a,b\right\}
+\left\langle b,c\right\rangle =\left\{  a,c\right\}  $ where the
superposition at $\left\{  b\right\}  $ cancels out. Then the final
probabilities will just be probabilities of finding $\left\{  a\right\}  $,
$\left\{  b\right\}  $, or $\left\{  c\right\}  $ when the measurement is made
only at the wall on the right is:

\begin{center}%
\begin{tabular}
[c]{l}%
$\Pr(\left\{  a\right\}  $ at wall $|\left\{  a,c\right\}  $ at slits$)=\Pr
\left(  \left\{  a\right\}  |\left\{  a,c\right\}  \right)  =\frac{\left\vert
\left\{  a\right\}  \cap\left\{  a,c\right\}  \right\vert }{\left\vert
\left\{  a,c\right\}  \right\vert }=\frac{1}{2}$;\\
$\Pr(\left\{  b\right\}  $ at wall $|\left\{  a,c\right\}  $ at slits$)=\Pr
\left(  \left\{  b\right\}  |\left\{  a,c\right\}  \right)  =\frac{\left\vert
\left\{  b\right\}  \cap\left\{  a,c\right\}  \right\vert }{\left\vert
\left\{  a,c\right\}  \right\vert }=0$;\\
$\Pr(\left\{  c\right\}  $ at wall $|\left\{  a,c\right\}  $ at slits$)=\Pr
\left(  \left\{  c\right\}  |\left\{  a,c\right\}  \right)  =\frac{\left\vert
\left\{  c\right\}  \cap\left\{  a,c\right\}  \right\vert }{\left\vert
\left\{  a,c\right\}  \right\vert }=\frac{1}{2}$.
\end{tabular}
%

%TCIMACRO{\FRAME{dtbpF}{2.5108in}{1.8013in}{0in}{}{}{fig16-case2-two-slit.eps}%
%width 2.5108in;  height 1.8013in;  depth 0in;  original-width 2.4236in;
%original-height 1.7309in;  cropleft "0";  croptop "1";  cropright "1";
%cropbottom "0";  filename 'fig16-case2-two-slit.eps';file-properties "XNPEU";}%
%} }%
%BeginExpansion
\begin{center}
\includegraphics[
height=1.8013in,
width=2.5108in
]%
{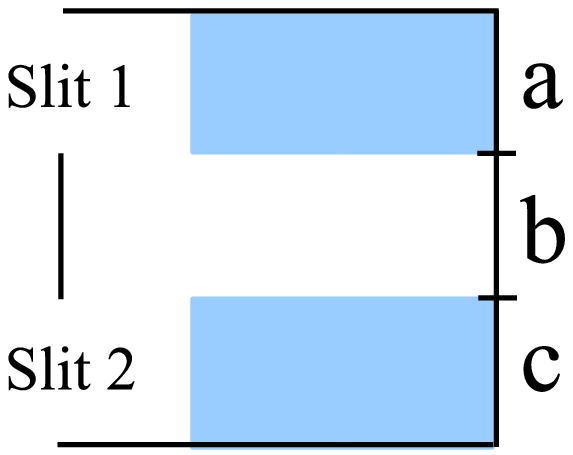}%
\end{center}
%EndExpansion

Figure 16: Final probability distribution with no measurement at slits
\end{center}

Since no "collapse" took place at the slits due to no distinctions being made
there, the indistinct element $\left\{  a,c\right\}  $ evolved (rather than
one or the other of the distinct elements $\left\{  a\right\}  $ or $\left\{
c\right\}  $). The action of $A$ is the same on $\left\{  a\right\}  $ and
$\left\{  c\right\}  $ as when they evolve separately since $A$ is a linear
operator but the two results are now added together \textit{as part of the
evolution}. This allows the "interference" of the two results and thus the
cancellation of the $\left\{  b\right\}  $ term in $\left\{  a,b\right\}
+\left\langle b,c\right\rangle =\left\{  a,c\right\}  $. The addition is, of
course, mod $2$ (where $-1=+1$) so, in "wave language," the two "wave crests"
that add at the location $\left\{  b\right\}  $ cancel out. When this
indistinct element $\left\{  a,c\right\}  $ "hits the wall" on the right,
there is an equal probability of that distinction yielding either of those eigenstates.

\section{Appendix 3: Bell inequality in "quantum mechanics" on sets}

A simple version of a Bell inequality can be derived in the case of $%
%TCIMACRO{\U{2124} }%
%BeginExpansion
\mathbb{Z}
%EndExpansion
_{2}^{2}$ with three bases $U=\left\{  a,b\right\}  $, $U^{\prime}=\left\{
a^{\prime},b^{\prime}\right\}  $, and $U^{\prime\prime}=\left\{
a^{\prime\prime},b^{\prime\prime}\right\}  $, and where the kets are:

\begin{center}%
\begin{tabular}
[c]{|c|c|c|c|}\hline
kets & $U$-basis & $U^{\prime}$-basis & $U^{\prime\prime}$-basis\\\hline\hline
$\left\vert 1\right\rangle $ & $\left\{  a,b\right\}  $ & $\left\{  a^{\prime
}\right\}  $ & $\left\{  a^{\prime\prime}\right\}  $\\\hline
$\left\vert 2\right\rangle $ & $\left\{  b\right\}  $ & $\left\{  b^{\prime
}\right\}  $ & $\left\{  a^{\prime\prime},b^{\prime\prime}\right\}  $\\\hline
$\left\vert 3\right\rangle $ & $\left\{  a\right\}  $ & $\left\{  a^{\prime
},b^{\prime}\right\}  $ & $\left\{  b^{\prime\prime}\right\}  $\\\hline
$\left\vert 4\right\rangle $ & $\emptyset$ & $\emptyset$ & $\emptyset$\\\hline
\end{tabular}

Ket table for $\wp\left(  U\right)  \cong\wp\left(  U^{\prime}\right)
\cong\wp\left(  U^{\prime\prime}\right)  \cong%
%TCIMACRO{\U{2124} }%
%BeginExpansion
\mathbb{Z}
%EndExpansion
_{2}^{2}$.
\end{center}

Attributes defined on the three universe sets $U$, $U^{\prime}$, and
$U^{\prime\prime}$, such as say $\chi_{\left\{  a\right\}  }$, $\chi_{\left\{
b^{\prime}\right\}  }$, and $\chi_{\left\{  a^{\prime\prime}\right\}  }$, are
incompatible as can be seen in several ways. For instance the set partitions
defined on $U$ and $U^{\prime}$, namely $\left\{  \left\{  a\right\}
,\left\{  b\right\}  \right\}  $ and $\left\{  \left\{  a^{\prime}\right\}
,\left\{  b^{\prime}\right\}  \right\}  $, cannot be obtained as two different
ways to partition the same set since $\left\{  a\right\}  =\left\{  a^{\prime
},b^{\prime}\right\}  $ and $\left\{  a^{\prime}\right\}  =\left\{
a,b\right\}  $, i.e., an "eigenstate" in one basis is a superposition in the
other. The same holds in the other pairwise comparison of $U$ and
$U^{\prime\prime}$ and of $U^{\prime}$ and $U^{\prime\prime}$.

A more technical way to show incompatibility is to exploit the vector space
structure of $%
%TCIMACRO{\U{2124} }%
%BeginExpansion
\mathbb{Z}
%EndExpansion
_{2}^{2}$ and to see if the projection matrices for $\left\{  a\right\}
\cap()$ and $\left\{  b^{\prime}\right\}  \cap()$ commute. The basis
conversion matrices between the $U$-basis and $U^{\prime}$-basis are:

\begin{center}
$\mathcal{C}_{U\leftarrow U^{\prime}}=%
\begin{bmatrix}
1 & 0\\
1 & 1
\end{bmatrix}
$ and $\mathcal{C}_{U^{\prime}\leftarrow U}=%
\begin{bmatrix}
1 & 0\\
1 & 1
\end{bmatrix}
$.
\end{center}

\noindent The projection matrix for $\left\{  a\right\}  \cap()$ in the
$U$-basis is, of course, $%
\begin{bmatrix}
1 & 0\\
0 & 0
\end{bmatrix}
$ and the projection matrix for $\left\{  b^{\prime}\right\}  \cap()$ in the
$U^{\prime}$-basis is $%
\begin{bmatrix}
0 & 0\\
0 & 1
\end{bmatrix}
$. Converting the latter to the $U$-basis to check commutativity gives:

\begin{center}
$\left[  \left\{  b^{\prime}\right\}  \cap()\right]  _{U}=\mathcal{C}%
_{U\leftarrow U^{\prime}}%
\begin{bmatrix}
0 & 0\\
0 & 1
\end{bmatrix}
$ $\mathcal{C}_{U^{\prime}\leftarrow U}$

$=%
\begin{bmatrix}
1 & 0\\
1 & 1
\end{bmatrix}%
\begin{bmatrix}
0 & 0\\
0 & 1
\end{bmatrix}%
\begin{bmatrix}
1 & 0\\
1 & 1
\end{bmatrix}
=\allowbreak%
\begin{bmatrix}
0 & 0\\
1 & 1
\end{bmatrix}
$.
\end{center}

Hence the commutativity check is:

\begin{center}
$\qquad\left[  \left\{  a\right\}  \cap()\right]  _{U}\left[  \left\{
b^{\prime}\right\}  \cap()\right]  _{U}=$ $%
\begin{bmatrix}
1 & 0\\
0 & 0
\end{bmatrix}%
\begin{bmatrix}
0 & 0\\
1 & 1
\end{bmatrix}
=%
\begin{bmatrix}
0 & 0\\
0 & 0
\end{bmatrix}
\neq$

$\left[  \left\{  b^{\prime}\right\}  \cap()\right]  _{U}\left[  \left\{
a\right\}  \cap()\right]  _{U}=%
\begin{bmatrix}
0 & 0\\
1 & 1
\end{bmatrix}%
\begin{bmatrix}
1 & 0\\
0 & 0
\end{bmatrix}
=%
\begin{bmatrix}
0 & 0\\
1 & 0
\end{bmatrix}
$
\end{center}

\noindent so the two operators for the "observables" $\chi_{\left\{
a\right\}  }$ and $\chi_{\left\{  b^{\prime}\right\}  }$ do not commute. In a
similar manner, it is seen that the three "observables" are mutually incompatible.

Given a ket in $%
%TCIMACRO{\U{2124} }%
%BeginExpansion
\mathbb{Z}
%EndExpansion
_{2}^{2}\cong\wp\left(  U\right)  \cong\wp\left(  U^{\prime}\right)  \cong%
\wp\left(  U^{\prime\prime}\right)  $, and using the usual equiprobability
assumption on sets, the probabilities of getting the different outcomes for
the various "observables" in the different given states are given in the
following table.

\begin{center}%
\begin{tabular}
[c]{|l||c|c||c|c||c|c|}\hline
Given state
%TCIMACRO{\TEXTsymbol{\backslash} }%
%BeginExpansion
$\backslash$
%EndExpansion
Outcome of test & $a$ & $b$ & $a^{\prime}$ & $b^{\prime}$ & $a^{\prime\prime}$
& $b^{\prime\prime}$\\\hline\hline
$\left\{  a,b\right\}  =\left\{  a^{\prime}\right\}  =\left\{  a^{\prime
\prime}\right\}  $ & $\frac{1}{2}$ & $\frac{1}{2}$ & $1$ & $0$ & $1$ &
$0$\\\hline
$\left\{  b\right\}  =\left\{  b^{\prime}\right\}  =\left\{  a^{\prime\prime
},b^{\prime\prime}\right\}  $ & $0$ & $1$ & $0$ & $1$ & $\frac{1}{2}$ &
$\frac{1}{2}$\\\hline
$\left\{  a\right\}  =\left\{  a^{\prime},b^{\prime}\right\}  =\left\{
b^{\prime\prime}\right\}  $ & $1$ & $0$ & $\frac{1}{2}$ & $\frac{1}{2}$ & $0$
& $1$\\\hline
\end{tabular}

State-outcome table.
\end{center}

The delift of the tensor product of vector spaces is the Cartesian or direct
product of sets, and the delift of the vectors in the tensor product are the
subsets of direct product of sets (as seen in the above treatment of
entanglement in "quantum mechanics" on sets). Thus in the $U$-basis, the basis
elements are the elements of $U\times U$ and the "vectors" are all the subsets
in $\wp\left(  U\times U\right)  $. But we could obtain the same "space" as
$\wp\left(  U^{\prime}\times U^{\prime}\right)  $ and $\wp\left(
U^{\prime\prime}\times U^{\prime\prime}\right)  $, and we can construct a ket
table where each row is a ket expressed in the different bases. And these
calculations in terms of sets could also be carried out in terms of vector
spaces over $%
%TCIMACRO{\U{2124} }%
%BeginExpansion
\mathbb{Z}
%EndExpansion
_{2}$ where the rows of the ket table are the kets in the tensor product:

\begin{center}
$%
%TCIMACRO{\U{2124} }%
%BeginExpansion
\mathbb{Z}
%EndExpansion
_{2}^{2}\otimes%
%TCIMACRO{\U{2124} }%
%BeginExpansion
\mathbb{Z}
%EndExpansion
_{2}^{2}\cong\wp\left(  U\times U\right)  \cong\wp\left(  U^{\prime}\times
U^{\prime}\right)  \cong\wp\left(  U^{\prime\prime}\times U^{\prime\prime
}\right)  $.
\end{center}

Since $\left\{  a\right\}  =\left\{  a^{\prime},b^{\prime}\right\}  =\left\{
b^{\prime\prime}\right\}  $ and $\left\{  b\right\}  =\left\{  b^{\prime
}\right\}  =\left\{  a^{\prime\prime},b^{\prime\prime}\right\}  $, the subset
$\left\{  a\right\}  \times\left\{  b\right\}  =\left\{  \left(  a,b\right)
\right\}  \subseteq U\times U$ is expressed in the $U^{\prime}\times
U^{\prime}$-basis as $\left\{  a^{\prime},b^{\prime}\right\}  \times\left\{
b^{\prime}\right\}  =\left\{  \left(  a^{\prime},b^{\prime}\right)  ,\left(
b^{\prime},b^{\prime}\right)  \right\}  $, and in the $U^{\prime\prime}\times
U^{\prime\prime}$-basis it is $\left\{  b^{\prime\prime}\right\}
\times\left\{  a^{\prime\prime},b^{\prime\prime}\right\}  =\left\{  \left(
b^{\prime\prime},a^{\prime\prime}\right)  ,\left(  b^{\prime\prime}%
,b^{\prime\prime}\right)  \right\}  $. Hence one row in the ket table has:

\begin{center}
$\left\{  \left(  a,b\right)  \right\}  =\left\{  \left(  a^{\prime}%
,b^{\prime}\right)  ,\left(  b^{\prime},b^{\prime}\right)  \right\}  =\left\{
\left(  b^{\prime\prime},a^{\prime\prime}\right)  ,\left(  b^{\prime\prime
},b^{\prime\prime}\right)  \right\}  $.
\end{center}

\noindent Since the full ket table has $16$ rows, we will just give a partial
table that suffices for our calculations.

\begin{center}%
\begin{tabular}
[c]{|c|c|c|}\hline
$U\times U$ & \multicolumn{1}{|c|}{$U^{\prime}\times U^{\prime}$} &
$U^{\prime\prime}\times U^{\prime\prime}$\\\hline\hline
$\left\{  \left(  a,a\right)  \right\}  $ & $\left\{  \left(  a^{\prime
},a^{\prime}\right)  ,\left(  a^{\prime},b^{\prime}\right)  ,\left(
b^{\prime},a^{\prime}\right)  ,\left(  b^{\prime},b^{\prime}\right)  \right\}
$ & $\left\{  \left(  b^{\prime\prime},b^{\prime\prime}\right)  \right\}
$\\\hline
$\left\{  \left(  a,b\right)  \right\}  $ & $\left\{  \left(  a^{\prime
},b^{\prime}\right)  ,\left(  b^{\prime},b^{\prime}\right)  \right\}  $ &
$\left\{  \left(  b^{\prime\prime},a^{\prime\prime}\right)  ,\left(
b^{\prime\prime},b^{\prime\prime}\right)  \right\}  $\\\hline
$\left\{  \left(  b,a\right)  \right\}  $ & $\left\{  \left(  b^{\prime
},a^{\prime}\right)  ,\left(  b^{\prime},b^{\prime}\right)  \right\}  $ &
$\left\{  \left(  a^{\prime\prime},b^{\prime\prime}\right)  ,\left(
b^{\prime\prime},b^{\prime\prime}\right)  \right\}  $\\\hline
$\left\{  b,b\right\}  $ & $\left\{  \left(  b^{\prime},b^{\prime}\right)
\right\}  $ & $\left\{  \left(  a^{\prime\prime},a^{\prime\prime}\right)
,\left(  a^{\prime\prime},b^{\prime\prime}\right)  ,\left(  b^{\prime\prime
},a^{\prime\prime}\right)  ,\left(  b^{\prime\prime},b^{\prime\prime}\right)
\right\}  $\\\hline
$\left\{  \left(  a,a\right)  ,\left(  a,b\right)  \right\}  $ & $\left\{
\left(  a^{\prime},a^{\prime}\right)  ,\left(  b^{\prime},a^{\prime}\right)
\right\}  $ & $\left\{  \left(  b^{\prime\prime},a^{\prime\prime}\right)
\right\}  $\\\hline
$\left\{  \left(  a,a\right)  ,\left(  b,a\right)  \right\}  $ & $\left\{
\left(  a^{\prime},a^{\prime}\right)  ,\left(  a^{\prime},b^{\prime}\right)
\right\}  $ & $\left\{  \left(  a^{\prime\prime},b^{\prime\prime}\right)
\right\}  $\\\hline
$\left\{  \left(  a,a\right)  ,\left(  b,b\right)  \right\}  $ & $\left\{
\left(  a^{\prime},a^{\prime}\right)  ,\left(  a^{\prime},b^{\prime}\right)
,\left(  b^{\prime},a^{\prime}\right)  \right\}  $ & $\left\{  \left(
a^{\prime\prime},a^{\prime\prime}\right)  ,\left(  a^{\prime\prime}%
,b^{\prime\prime}\right)  ,\left(  b^{\prime\prime},a^{\prime\prime}\right)
\right\}  $\\\hline
$\left\{  \left(  a,b\right)  ,\left(  b,a\right)  \right\}  $ & $\left\{
\left(  a^{\prime},b^{\prime}\right)  ,\left(  b^{\prime},a^{\prime}\right)
\right\}  $ & $\left\{  \left(  a^{\prime\prime},b^{\prime\prime}\right)
,\left(  b^{\prime\prime},a^{\prime\prime}\right)  \right\}  $\\\hline
\end{tabular}

Partial ket table for $\wp\left(  U\times U\right)  \cong\wp\left(  U^{\prime
}\times U^{\prime}\right)  \cong\wp\left(  U^{\prime\prime}\times
U^{\prime\prime}\right)  $
\end{center}

As before, we can classify each "vector" or subset as "separated" or
"entangled" and we can furthermore see how that is independent of the basis.
For instance $\left\{  \left(  a,a\right)  ,\left(  a,b\right)  \right\}  $ is
"separated" since:

\begin{center}
$\left\{  \left(  a,a\right)  ,\left(  a,b\right)  \right\}  =\left\{
a\right\}  \times\left\{  a,b\right\}  =\left\{  \left(  a^{\prime},a^{\prime
}\right)  ,\left(  b^{\prime},a^{\prime}\right)  \right\}  =\left\{
a^{\prime},b^{\prime}\right\}  \times\left\{  a^{\prime}\right\}  =\left\{
\left(  b^{\prime\prime},a^{\prime\prime}\right)  \right\}  =\left\{
b^{\prime\prime}\right\}  \times\left\{  a^{\prime\prime}\right\}  $.
\end{center}

An example of an "entangled state" is:

\begin{center}
$\left\{  \left(  a,a\right)  ,\left(  b,b\right)  \right\}  =\left\{  \left(
a^{\prime},a^{\prime}\right)  ,\left(  a^{\prime},b^{\prime}\right)  ,\left(
b^{\prime},a^{\prime}\right)  \right\}  =\left\{  \left(  a^{\prime\prime
},a^{\prime\prime}\right)  ,\left(  a^{\prime\prime},b^{\prime\prime}\right)
,\left(  b^{\prime\prime},a^{\prime\prime}\right)  \right\}  $.
\end{center}

\noindent Taking this "entangled state" as the initial "state," there is a
probability distribution on $U\times U^{\prime}\times U^{\prime\prime}$ where
$\Pr\left(  a,a^{\prime},a^{\prime\prime}\right)  $ (for instance) is defined
as the probability of getting the result $\left\{  a\right\}  $ if a
$U$-measurement is performed on the left-hand system, and if instead a
$U^{\prime}$-measurement is performed on the left-hand system then $\left\{
a^{\prime}\right\}  $ is obtained, and if instead a $U^{\prime\prime}%
$-measurement is performed on the left-hand system then $\left\{
a^{\prime\prime}\right\}  $ is obtained. Thus we would have $\Pr\left(
a,a^{\prime},a^{\prime\prime}\right)  =\frac{1}{2}\frac{2}{3}\frac{2}{3}%
=\frac{2}{9}$. In this way the probability distribution $\Pr\left(
x,y,z\right)  $ is defined on $U\times U^{\prime}\times U^{\prime\prime}$.

A Bell inequality can be obtained from this joint probability distribution
over the outcomes $U\times U^{\prime}\times U^{\prime\prime}$ of measuring
these three incompatible attributes \cite{d'esp:sciam}. Consider the following marginals:%

\begin{align*}
\Pr\left(  a,a^{\prime}\right)   &  =\Pr\left(  a,a^{\prime},a^{\prime\prime
}\right)  +\Pr\left(  a,a^{\prime},b^{\prime\prime}\right)  \checkmark\\
\Pr\left(  b^{\prime},b^{\prime\prime}\right)   &  =\Pr\left(  a,b^{\prime
},b^{\prime\prime}\right)  \checkmark+\Pr\left(  b,b^{\prime},b^{\prime\prime
}\right) \\
\Pr\left(  a,b^{\prime\prime}\right)   &  =\Pr\left(  a,a^{\prime}%
,b^{\prime\prime}\right)  \checkmark+\Pr\left(  a,b^{\prime},b^{\prime\prime
}\right)  \checkmark\text{.}%
\end{align*}

\noindent The two terms in the last marginal are each contained in one of the
two previous marginals (as indicated by the check marks) and all the
probabilities are non-negative, so we have the following inequality:

\begin{center}
$\Pr\left(  a,a^{\prime}\right)  +\Pr\left(  b^{\prime},b^{\prime\prime
}\right)  \geq\Pr\left(  a,b^{\prime\prime}\right)  $

Bell inequality.
\end{center}

All this has to do with measurements on the left-hand system. But there is an
alternative interpretation to the probabilities $\Pr\left(  x,y\right)  $,
$\Pr\left(  y,z\right)  $, and $\Pr\left(  x,z\right)  $ \textit{if }we assume
that the outcome of a measurement on the right-hand system is
\textit{independent} of the outcome of the same measurement on the left-hand
system. Then $\Pr\left(  a,a^{\prime}\right)  $ is the probability of a
$U$-measurement on the left-hand system giving $\left\{  a\right\}  $ and then
a $U^{\prime}$-measurement on the right-hand system giving $\left\{
a^{\prime}\right\}  $, and so forth. Under that \textit{independence
assumption} and for this initially prepared "Bell state" (which is left-right
symmetrical in each basis),

\begin{center}
$\left\{  \left(  a,a\right)  ,\left(  b,b\right)  \right\}  =\left\{  \left(
a^{\prime},a^{\prime}\right)  ,\left(  a^{\prime},b^{\prime}\right)  ,\left(
b^{\prime},a^{\prime}\right)  \right\}  =\left\{  \left(  a^{\prime\prime
},a^{\prime\prime}\right)  ,\left(  a^{\prime\prime},b^{\prime\prime}\right)
,\left(  b^{\prime\prime},a^{\prime\prime}\right)  \right\}  $,
\end{center}

\noindent the probabilities would be the same.\footnote{The same holds for the
other "Bell state": $\left\{  \left(  a,b\right)  ,\left(  b,a\right)
\right\}  $.} That is, under that assumption, the probabilities, $\Pr\left(
a\right)  =\frac{1}{2}=\Pr\left(  b\right)  $, $\Pr\left(  a^{\prime}\right)
=\frac{2}{3}=\Pr\left(  a^{\prime\prime}\right)  $, and $\Pr\left(  b^{\prime
}\right)  =\frac{1}{3}=\Pr\left(  b^{\prime\prime}\right)  $ are the same
regardless of whether we are measuring the left-hand or right-hand system of
that composite state. Hence the above Bell inequality would still hold. But we
can use "quantum mechanics" on sets to compute the probabilities for those
different measurements on the two systems to see if the independence
assumption is compatible with "QM" on sets.

To compute $\Pr\left(  a,a^{\prime}\right)  $, we first measure the left-hand
component in the $U$-basis. Since $\left\{  \left(  a,a\right)  ,\left(
b,b\right)  \right\}  $ is the given state, and $\left(  a,a\right)  $ and
$\left(  b,b\right)  $ are equiprobable, the probability of getting $\left\{
a\right\}  $ (i.e., the "eigenvalue" $1$ for the "observable $\chi_{\left\{
a\right\}  }$) is $\frac{1}{2}$. But the right-hand system is then in the
state $\left\{  a\right\}  $ and the probability of getting $\left\{
a^{\prime}\right\}  $ (i.e., "eigenvalue" $0$ for the "observable"
$\chi_{\left\{  b^{\prime}\right\}  }$) is $\frac{1}{2}$ (as seen in the
state-outcome table). Thus the probability is $\Pr\left(  a,a^{\prime}\right)
=\frac{1}{2}\frac{1}{2}=\frac{1}{4}$.

To compute $\Pr\left(  b^{\prime},b^{\prime\prime}\right)  $, we first perform
a $U^{\prime}$-basis "measurement" on the left-hand component of the given
state $\left\{  \left(  a,a\right)  ,\left(  b,b\right)  \right\}  =\left\{
\left(  a^{\prime},a^{\prime}\right)  ,\left(  a^{\prime},b^{\prime}\right)
,\left(  b^{\prime},a^{\prime}\right)  \right\}  $, and we see that the
probability of getting $\left\{  b^{\prime}\right\}  $ is $\frac{1}{3}$. Then
the right-hand system is in the state $\left\{  a^{\prime}\right\}  $ and the
probability of getting $\left\{  b^{\prime\prime}\right\}  $ in a
$U^{\prime\prime}$-basis "measurement" of the right-hand system in the state
$\left\{  a^{\prime}\right\}  $ is $0$ (as seen from the state-outcome table).
Hence the probability is $\Pr\left(  b^{\prime},b^{\prime\prime}\right)  =0$.

Finally we compute $\Pr\left(  a,b^{\prime\prime}\right)  $ by first making a
$U$-measurement on the left-hand component of the given state $\left\{
\left(  a,a\right)  ,\left(  b,b\right)  \right\}  $ and get the result
$\left\{  a\right\}  $ with probability $\frac{1}{2}$. Then the state of the
second system is $\left\{  a\right\}  $ so a $U^{\prime\prime}$-measurement
will give the $\left\{  b^{\prime\prime}\right\}  $ result with probability
$1$ so the probability is $\Pr\left(  a,b^{\prime\prime}\right)  =\frac{1}{2}$.

Then we plug the probabilities into the Bell inequality:

\begin{center}
$\Pr\left(  a,a^{\prime}\right)  +\Pr\left(  b^{\prime},b^{\prime\prime
}\right)  \geq\Pr\left(  a,b^{\prime\prime}\right)  $

$\frac{1}{4}+0\ngeq\frac{1}{2}$

Violation of Bell inequality.
\end{center}

\noindent The violation of the Bell inequality shows that the independence
assumption about the measurement outcomes on the left-hand and right-hand
systems is incompatible with "QM" on sets so the effects of the "QM" on sets
measurements are "nonlocal."

\end{document}